\documentclass[12pt]{article}
\usepackage{publication}
\usepackage{array}
\usepackage[utf8]{inputenc}
\usepackage{amsmath}
\usepackage{natbib}
\usepackage{epstopdf, epsfig}
\usepackage{graphicx}   
\usepackage{subcaption} 
\usepackage{tikz,pgfplots,pgfplotstable} 
\usetikzlibrary{spy, shapes, decorations.pathmorphing, plotmarks, arrows, arrows.meta, automata, patterns, decorations.text, calc, backgrounds, shadows, positioning,  matrix, calendar, fit, svg.path, pgfplots.colorbrewer, pgfplots.colormaps, colorbrewer, decorations.markings, external, bending, shapes.geometric, patterns.meta}
\usepgfplotslibrary{groupplots}
\pgfplotsset{compat=newest}
\pgfplotsset{plot coordinates/math parser=false}
\usepgfplotslibrary{patchplots,fillbetween,groupplots,units,statistics}
\usetikzlibrary{spy}
\usetikzlibrary{fit,calc,positioning}

\title{A Surfactant Prediction Model for Rising Bubbles}

\author{Lim Chi Tung James$^{1}$}

\author{Ivo R. Peters$^1$}

\author{Swathi Krishna$^{1*}$}
\contact{$^*$s.b.krishna@soton.ac.uk}

\affil{$^1$Department of Aeronautics and Astronautics, University of Southampton, Southampton, United Kingdom, S016 7QF}

\date{\today}

\definecolor{white}{rgb}{1.000,1.000,1.000}
\definecolor{snow}{rgb}{1.000,0.979,0.979}
\definecolor{honeydew}{rgb}{0.938,1.000,0.938}
\definecolor{mintcream}{rgb}{0.958,1.000,0.979}
\definecolor{azure}{rgb}{0.938,1.000,1.000}
\definecolor{aliceblue}{rgb}{0.938,0.971,1.000}
\definecolor{ghostwhite}{rgb}{0.971,0.971,1.000}
\definecolor{whitesmoke}{rgb}{0.958,0.958,0.958}
\definecolor{seashell}{rgb}{1.000,0.958,0.930}
\definecolor{beige}{rgb}{0.958,0.958,0.859}
\definecolor{oldlace}{rgb}{0.992,0.958,0.898}
\definecolor{floralwhite}{rgb}{1.000,0.979,0.938}
\definecolor{ivory}{rgb}{1.000,1.000,0.938}
\definecolor{antiquewhite}{rgb}{0.979,0.918,0.840}
\definecolor{linen}{rgb}{0.979,0.938,0.898}
\definecolor{lavenderblush}{rgb}{1.000,0.938,0.958}
\definecolor{mistyrose}{rgb}{1.000,0.891,0.879}
\definecolor{gray}{rgb}{0.500,0.500,0.500}
\definecolor{sgray}{rgb}{0.211,0.211,0.211}
\definecolor{gainsboro}{rgb}{0.859,0.859,0.859}
\definecolor{lightgray}{rgb}{0.824,0.824,0.824}
\definecolor{silver}{rgb}{0.750,0.750,0.750}
\definecolor{darkgray}{rgb}{0.660,0.660,0.660}
\definecolor{dimgray}{rgb}{0.410,0.410,0.410}
\definecolor{lightslategray}{rgb}{0.465,0.531,0.598}
\definecolor{slategray}{rgb}{0.438,0.500,0.562}
\definecolor{darkslategray}{rgb}{0.184,0.309,0.309}
\definecolor{black}{rgb}{0.000,0.000,0.000}
\definecolor{red}{rgb}{1.000,0.000,0.000}
\definecolor{lightsalmon}{rgb}{1.000,0.625,0.477}
\definecolor{salmon}{rgb}{0.979,0.500,0.445}
\definecolor{darksalmon}{rgb}{0.910,0.586,0.477}
\definecolor{lightcoral}{rgb}{0.938,0.500,0.500}
\definecolor{indianred}{rgb}{0.801,0.359,0.359}
\definecolor{crimson}{rgb}{0.859,0.078,0.234}
\definecolor{firebrick}{rgb}{0.695,0.133,0.133}
\definecolor{darkred}{rgb}{0.543,0.000,0.000}
\definecolor{pink}{rgb}{1.000,0.750,0.793}
\definecolor{lightpink}{rgb}{1.000,0.711,0.754}
\definecolor{hotpink}{rgb}{1.000,0.410,0.703}
\definecolor{deeppink}{rgb}{1.000,0.078,0.574}
\definecolor{palevioletred}{rgb}{0.855,0.438,0.574}
\definecolor{mediumvioletred}{rgb}{0.777,0.082,0.520}
\definecolor{orange}{rgb}{1.000,0.645,0.000}
\definecolor{darkorange}{rgb}{1.000,0.547,0.000}
\definecolor{coral}{rgb}{1.000,0.496,0.312}
\definecolor{tomato}{rgb}{1.000,0.387,0.277}
\definecolor{orangered}{rgb}{1.000,0.270,0.000}
\definecolor{yellow}{rgb}{1.000,1.000,0.000}
\definecolor{lightyellow}{rgb}{1.000,1.000,0.875}
\definecolor{lemonchiffon}{rgb}{1.000,0.979,0.801}
\definecolor{lightgoldenrodyellow}{rgb}{0.979,0.979,0.820}
\definecolor{papayawhip}{rgb}{1.000,0.934,0.832}
\definecolor{moccasin}{rgb}{1.000,0.891,0.707}
\definecolor{peachpuff}{rgb}{1.000,0.852,0.723}
\definecolor{palegoldenrod}{rgb}{0.930,0.906,0.664}
\definecolor{khaki}{rgb}{0.938,0.898,0.547}
\definecolor{darkkhaki}{rgb}{0.738,0.715,0.418}
\definecolor{gold}{rgb}{1.000,0.840,0.000}
\definecolor{brown}{rgb}{0.645,0.164,0.164}
\definecolor{cornsilk}{rgb}{1.000,0.971,0.859}
\definecolor{blanchedalmond}{rgb}{1.000,0.918,0.801}
\definecolor{bisque}{rgb}{1.000,0.891,0.766}
\definecolor{navajowhite}{rgb}{1.000,0.867,0.676}
\definecolor{wheat}{rgb}{0.958,0.867,0.699}
\definecolor{burlywood}{rgb}{0.867,0.719,0.527}
\definecolor{tan}{rgb}{0.820,0.703,0.547}
\definecolor{rosybrown}{rgb}{0.734,0.559,0.559}
\definecolor{sandybrown}{rgb}{0.954,0.641,0.375}
\definecolor{goldenrod}{rgb}{0.852,0.645,0.125}
\definecolor{darkgoldenrod}{rgb}{0.719,0.523,0.043}
\definecolor{peru}{rgb}{0.801,0.520,0.246}
\definecolor{chocolate}{rgb}{0.820,0.410,0.117}
\definecolor{saddlebrown}{rgb}{0.543,0.270,0.074}
\definecolor{sienna}{rgb}{0.625,0.320,0.176}
\definecolor{maroon}{rgb}{0.500,0.000,0.000}
\definecolor{green}{rgb}{0.000,0.500,0.000}
\definecolor{palegreen}{rgb}{0.594,0.983,0.594}
\definecolor{lightgreen}{rgb}{0.562,0.930,0.562}
\definecolor{yellowgreen}{rgb}{0.602,0.801,0.195}
\definecolor{greenyellow}{rgb}{0.676,1.000,0.184}
\definecolor{chartreuse}{rgb}{0.496,1.000,0.000}
\definecolor{lawngreen}{rgb}{0.484,0.988,0.000}
\definecolor{lime}{rgb}{0.000,1.000,0.000}
\definecolor{limegreen}{rgb}{0.195,0.801,0.195}
\definecolor{mediumspringgreen}{rgb}{0.000,0.979,0.602}
\definecolor{springgreen}{rgb}{0.000,1.000,0.496}
\definecolor{mediumaquamarine}{rgb}{0.398,0.801,0.664}
\definecolor{aquamarine}{rgb}{0.496,1.000,0.828}
\definecolor{lightseagreen}{rgb}{0.125,0.695,0.664}
\definecolor{mediumseagreen}{rgb}{0.234,0.699,0.441}
\definecolor{seagreen}{rgb}{0.180,0.543,0.340}
\definecolor{darkseagreen}{rgb}{0.559,0.734,0.559}
\definecolor{forestgreen}{rgb}{0.133,0.543,0.133}
\definecolor{darkgreen}{rgb}{0.000,0.391,0.000}
\definecolor{olivedrab}{rgb}{0.418,0.555,0.137}
\definecolor{olive}{rgb}{0.500,0.500,0.000}
\definecolor{darkolivegreen}{rgb}{0.332,0.418,0.184}
\definecolor{teal}{rgb}{0.000,0.500,0.500}
\definecolor{blue}{rgb}{0.000,0.000,1.000}
\definecolor{lightblue}{rgb}{0.676,0.844,0.898}
\definecolor{powderblue}{rgb}{0.688,0.875,0.898}
\definecolor{paleturquoise}{rgb}{0.684,0.930,0.930}
\definecolor{turquoise}{rgb}{0.250,0.875,0.812}
\definecolor{mediumturquoise}{rgb}{0.281,0.816,0.797}
\definecolor{darkturquoise}{rgb}{0.000,0.805,0.816}
\definecolor{lightcyan}{rgb}{0.875,1.000,1.000}
\definecolor{cyan}{rgb}{0.000,1.000,1.000}
\definecolor{aqua}{rgb}{0.000,1.000,1.000}
\definecolor{darkcyan}{rgb}{0.000,0.543,0.543}
\definecolor{cadetblue}{rgb}{0.371,0.617,0.625}
\definecolor{lightsteelblue}{rgb}{0.688,0.766,0.867}
\definecolor{steelblue}{rgb}{0.273,0.508,0.703}
\definecolor{lightskyblue}{rgb}{0.527,0.805,0.979}
\definecolor{skyblue}{rgb}{0.527,0.805,0.918}
\definecolor{deepskyblue}{rgb}{0.000,0.746,1.000}
\definecolor{dodgerblue}{rgb}{0.117,0.562,1.000}
\definecolor{cornflowerblue}{rgb}{0.391,0.582,0.926}
\definecolor{royalblue}{rgb}{0.254,0.410,0.879}
\definecolor{mediumblue}{rgb}{0.000,0.000,0.801}
\definecolor{darkblue}{rgb}{0.000,0.000,0.543}
\definecolor{navy}{rgb}{0.000,0.000,0.500}
\definecolor{midnightblue}{rgb}{0.098,0.098,0.438}
\definecolor{purple}{rgb}{0.500,0.000,0.500}
\definecolor{lavender}{rgb}{0.898,0.898,0.979}
\definecolor{thistle}{rgb}{0.844,0.746,0.844}
\definecolor{plum}{rgb}{0.863,0.625,0.863}
\definecolor{violet}{rgb}{0.930,0.508,0.930}
\definecolor{orchid}{rgb}{0.852,0.438,0.836}
\definecolor{fuchsia}{rgb}{1.000,0.000,1.000}
\definecolor{magenta}{rgb}{1.000,0.000,1.000}
\definecolor{mediumorchid}{rgb}{0.727,0.332,0.824}
\definecolor{mediumpurple}{rgb}{0.574,0.438,0.855}
\definecolor{amethyst}{rgb}{0.598,0.398,0.797}
\definecolor{blueviolet}{rgb}{0.539,0.168,0.883}
\definecolor{darkviolet}{rgb}{0.578,0.000,0.824}
\definecolor{darkorchid}{rgb}{0.598,0.195,0.797}
\definecolor{darkmagenta}{rgb}{0.543,0.000,0.543}
\definecolor{slateblue}{rgb}{0.414,0.352,0.801}
\definecolor{darkslateblue}{rgb}{0.281,0.238,0.543}
\definecolor{mediumslateblue}{rgb}{0.480,0.406,0.930}
\definecolor{indigo}{rgb}{0.293,0.000,0.508}
\definecolor{grey}{rgb}{0.500,0.500,0.500}
\definecolor{lightgrey}{rgb}{0.824,0.824,0.824}
\definecolor{darkgrey}{rgb}{0.660,0.660,0.660}
\definecolor{dimgrey}{rgb}{0.410,0.410,0.410}
\definecolor{lightslategrey}{rgb}{0.465,0.531,0.598}
\definecolor{slategrey}{rgb}{0.438,0.500,0.562}
\definecolor{darkslategrey}{rgb}{0.184,0.309,0.309}
\definecolor{rosa}{rgb}{1,0.5,0.5}

\definecolor{parula-1}{rgb}{0.2081,0.1663,0.5292}
\definecolor{parula-2}{rgb}{0.0146,0.3845,0.8813}
\definecolor{parula-3}{rgb}{0.0795,0.5159,0.8328}
\definecolor{parula-4}{rgb}{0.0228,0.6492,0.7823}
\definecolor{parula-5}{rgb}{0.1986,0.7214,0.6310}
\definecolor{parula-6}{rgb}{0.5456,0.7490,0.4597}
\definecolor{parula-7}{rgb}{0.8266,0.7320,0.3464}
\definecolor{parula-8}{rgb}{0.9948,0.7886,0.1943}
\definecolor{parula-9}{rgb}{0.9763,0.9831,0.0538}
\definecolor{YlGn-3-1}{RGB}{247,252,185}
\definecolor{YlGn-3-C}{RGB}{247,252,185}
\definecolor{YlGn-3-2}{RGB}{173,221,142}
\definecolor{YlGn-3-F}{RGB}{173,221,142}
\definecolor{YlGn-3-3}{RGB}{49,163,84}
\definecolor{YlGn-3-I}{RGB}{49,163,84}
\definecolor{YlGn-4-1}{RGB}{255,255,204}
\definecolor{YlGn-4-B}{RGB}{255,255,204}
\definecolor{YlGn-4-2}{RGB}{194,230,153}
\definecolor{YlGn-4-E}{RGB}{194,230,153}
\definecolor{YlGn-4-3}{RGB}{120,198,121}
\definecolor{YlGn-4-G}{RGB}{120,198,121}
\definecolor{YlGn-4-4}{RGB}{35,132,67}
\definecolor{YlGn-4-J}{RGB}{35,132,67}
\definecolor{YlGn-5-1}{RGB}{255,255,204}
\definecolor{YlGn-5-B}{RGB}{255,255,204}
\definecolor{YlGn-5-2}{RGB}{194,230,153}
\definecolor{YlGn-5-E}{RGB}{194,230,153}
\definecolor{YlGn-5-3}{RGB}{120,198,121}
\definecolor{YlGn-5-G}{RGB}{120,198,121}
\definecolor{YlGn-5-4}{RGB}{49,163,84}
\definecolor{YlGn-5-I}{RGB}{49,163,84}
\definecolor{YlGn-5-5}{RGB}{0,104,55}
\definecolor{YlGn-5-K}{RGB}{0,104,55}
\definecolor{YlGn-6-1}{RGB}{255,255,204}
\definecolor{YlGn-6-B}{RGB}{255,255,204}
\definecolor{YlGn-6-2}{RGB}{217,240,163}
\definecolor{YlGn-6-D}{RGB}{217,240,163}
\definecolor{YlGn-6-3}{RGB}{173,221,142}
\definecolor{YlGn-6-F}{RGB}{173,221,142}
\definecolor{YlGn-6-4}{RGB}{120,198,121}
\definecolor{YlGn-6-G}{RGB}{120,198,121}
\definecolor{YlGn-6-5}{RGB}{49,163,84}
\definecolor{YlGn-6-I}{RGB}{49,163,84}
\definecolor{YlGn-6-6}{RGB}{0,104,55}
\definecolor{YlGn-6-K}{RGB}{0,104,55}
\definecolor{YlGn-7-1}{RGB}{255,255,204}
\definecolor{YlGn-7-B}{RGB}{255,255,204}
\definecolor{YlGn-7-2}{RGB}{217,240,163}
\definecolor{YlGn-7-D}{RGB}{217,240,163}
\definecolor{YlGn-7-3}{RGB}{173,221,142}
\definecolor{YlGn-7-F}{RGB}{173,221,142}
\definecolor{YlGn-7-4}{RGB}{120,198,121}
\definecolor{YlGn-7-G}{RGB}{120,198,121}
\definecolor{YlGn-7-5}{RGB}{65,171,93}
\definecolor{YlGn-7-H}{RGB}{65,171,93}
\definecolor{YlGn-7-6}{RGB}{35,132,67}
\definecolor{YlGn-7-J}{RGB}{35,132,67}
\definecolor{YlGn-7-7}{RGB}{0,90,50}
\definecolor{YlGn-7-L}{RGB}{0,90,50}
\definecolor{YlGn-8-1}{RGB}{255,255,229}
\definecolor{YlGn-8-A}{RGB}{255,255,229}
\definecolor{YlGn-8-2}{RGB}{247,252,185}
\definecolor{YlGn-8-C}{RGB}{247,252,185}
\definecolor{YlGn-8-3}{RGB}{217,240,163}
\definecolor{YlGn-8-D}{RGB}{217,240,163}
\definecolor{YlGn-8-4}{RGB}{173,221,142}
\definecolor{YlGn-8-F}{RGB}{173,221,142}
\definecolor{YlGn-8-5}{RGB}{120,198,121}
\definecolor{YlGn-8-G}{RGB}{120,198,121}
\definecolor{YlGn-8-6}{RGB}{65,171,93}
\definecolor{YlGn-8-H}{RGB}{65,171,93}
\definecolor{YlGn-8-7}{RGB}{35,132,67}
\definecolor{YlGn-8-J}{RGB}{35,132,67}
\definecolor{YlGn-8-8}{RGB}{0,90,50}
\definecolor{YlGn-8-L}{RGB}{0,90,50}
\definecolor{YlGn-9-1}{RGB}{255,255,229}
\definecolor{YlGn-9-A}{RGB}{255,255,229}
\definecolor{YlGn-9-2}{RGB}{247,252,185}
\definecolor{YlGn-9-C}{RGB}{247,252,185}
\definecolor{YlGn-9-3}{RGB}{217,240,163}
\definecolor{YlGn-9-D}{RGB}{217,240,163}
\definecolor{YlGn-9-4}{RGB}{173,221,142}
\definecolor{YlGn-9-F}{RGB}{173,221,142}
\definecolor{YlGn-9-5}{RGB}{120,198,121}
\definecolor{YlGn-9-G}{RGB}{120,198,121}
\definecolor{YlGn-9-6}{RGB}{65,171,93}
\definecolor{YlGn-9-H}{RGB}{65,171,93}
\definecolor{YlGn-9-7}{RGB}{35,132,67}
\definecolor{YlGn-9-J}{RGB}{35,132,67}
\definecolor{YlGn-9-8}{RGB}{0,104,55}
\definecolor{YlGn-9-K}{RGB}{0,104,55}
\definecolor{YlGn-9-9}{RGB}{0,69,41}
\definecolor{YlGn-9-M}{RGB}{0,69,41}
\definecolor{YlGnBu-3-1}{RGB}{237,248,177}
\definecolor{YlGnBu-3-C}{RGB}{237,248,177}
\definecolor{YlGnBu-3-2}{RGB}{127,205,187}
\definecolor{YlGnBu-3-F}{RGB}{127,205,187}
\definecolor{YlGnBu-3-3}{RGB}{44,127,184}
\definecolor{YlGnBu-3-I}{RGB}{44,127,184}
\definecolor{YlGnBu-4-1}{RGB}{255,255,204}
\definecolor{YlGnBu-4-B}{RGB}{255,255,204}
\definecolor{YlGnBu-4-2}{RGB}{161,218,180}
\definecolor{YlGnBu-4-E}{RGB}{161,218,180}
\definecolor{YlGnBu-4-3}{RGB}{65,182,196}
\definecolor{YlGnBu-4-G}{RGB}{65,182,196}
\definecolor{YlGnBu-4-4}{RGB}{34,94,168}
\definecolor{YlGnBu-4-J}{RGB}{34,94,168}
\definecolor{YlGnBu-5-1}{RGB}{255,255,204}
\definecolor{YlGnBu-5-B}{RGB}{255,255,204}
\definecolor{YlGnBu-5-2}{RGB}{161,218,180}
\definecolor{YlGnBu-5-E}{RGB}{161,218,180}
\definecolor{YlGnBu-5-3}{RGB}{65,182,196}
\definecolor{YlGnBu-5-G}{RGB}{65,182,196}
\definecolor{YlGnBu-5-4}{RGB}{44,127,184}
\definecolor{YlGnBu-5-I}{RGB}{44,127,184}
\definecolor{YlGnBu-5-5}{RGB}{37,52,148}
\definecolor{YlGnBu-5-K}{RGB}{37,52,148}
\definecolor{YlGnBu-6-1}{RGB}{255,255,204}
\definecolor{YlGnBu-6-B}{RGB}{255,255,204}
\definecolor{YlGnBu-6-2}{RGB}{199,233,180}
\definecolor{YlGnBu-6-D}{RGB}{199,233,180}
\definecolor{YlGnBu-6-3}{RGB}{127,205,187}
\definecolor{YlGnBu-6-F}{RGB}{127,205,187}
\definecolor{YlGnBu-6-4}{RGB}{65,182,196}
\definecolor{YlGnBu-6-G}{RGB}{65,182,196}
\definecolor{YlGnBu-6-5}{RGB}{44,127,184}
\definecolor{YlGnBu-6-I}{RGB}{44,127,184}
\definecolor{YlGnBu-6-6}{RGB}{37,52,148}
\definecolor{YlGnBu-6-K}{RGB}{37,52,148}
\definecolor{YlGnBu-7-1}{RGB}{255,255,204}
\definecolor{YlGnBu-7-B}{RGB}{255,255,204}
\definecolor{YlGnBu-7-2}{RGB}{199,233,180}
\definecolor{YlGnBu-7-D}{RGB}{199,233,180}
\definecolor{YlGnBu-7-3}{RGB}{127,205,187}
\definecolor{YlGnBu-7-F}{RGB}{127,205,187}
\definecolor{YlGnBu-7-4}{RGB}{65,182,196}
\definecolor{YlGnBu-7-G}{RGB}{65,182,196}
\definecolor{YlGnBu-7-5}{RGB}{29,145,192}
\definecolor{YlGnBu-7-H}{RGB}{29,145,192}
\definecolor{YlGnBu-7-6}{RGB}{34,94,168}
\definecolor{YlGnBu-7-J}{RGB}{34,94,168}
\definecolor{YlGnBu-7-7}{RGB}{12,44,132}
\definecolor{YlGnBu-7-L}{RGB}{12,44,132}
\definecolor{YlGnBu-8-1}{RGB}{255,255,217}
\definecolor{YlGnBu-8-A}{RGB}{255,255,217}
\definecolor{YlGnBu-8-2}{RGB}{237,248,177}
\definecolor{YlGnBu-8-C}{RGB}{237,248,177}
\definecolor{YlGnBu-8-3}{RGB}{199,233,180}
\definecolor{YlGnBu-8-D}{RGB}{199,233,180}
\definecolor{YlGnBu-8-4}{RGB}{127,205,187}
\definecolor{YlGnBu-8-F}{RGB}{127,205,187}
\definecolor{YlGnBu-8-5}{RGB}{65,182,196}
\definecolor{YlGnBu-8-G}{RGB}{65,182,196}
\definecolor{YlGnBu-8-6}{RGB}{29,145,192}
\definecolor{YlGnBu-8-H}{RGB}{29,145,192}
\definecolor{YlGnBu-8-7}{RGB}{34,94,168}
\definecolor{YlGnBu-8-J}{RGB}{34,94,168}
\definecolor{YlGnBu-8-8}{RGB}{12,44,132}
\definecolor{YlGnBu-8-L}{RGB}{12,44,132}
\definecolor{YlGnBu-9-1}{RGB}{255,255,217}
\definecolor{YlGnBu-9-A}{RGB}{255,255,217}
\definecolor{YlGnBu-9-2}{RGB}{237,248,177}
\definecolor{YlGnBu-9-C}{RGB}{237,248,177}
\definecolor{YlGnBu-9-3}{RGB}{199,233,180}
\definecolor{YlGnBu-9-D}{RGB}{199,233,180}
\definecolor{YlGnBu-9-4}{RGB}{127,205,187}
\definecolor{YlGnBu-9-F}{RGB}{127,205,187}
\definecolor{YlGnBu-9-5}{RGB}{65,182,196}
\definecolor{YlGnBu-9-G}{RGB}{65,182,196}
\definecolor{YlGnBu-9-6}{RGB}{29,145,192}
\definecolor{YlGnBu-9-H}{RGB}{29,145,192}
\definecolor{YlGnBu-9-7}{RGB}{34,94,168}
\definecolor{YlGnBu-9-J}{RGB}{34,94,168}
\definecolor{YlGnBu-9-8}{RGB}{37,52,148}
\definecolor{YlGnBu-9-K}{RGB}{37,52,148}
\definecolor{YlGnBu-9-9}{RGB}{8,29,88}
\definecolor{YlGnBu-9-M}{RGB}{8,29,88}
\definecolor{GnBu-3-1}{RGB}{224,243,219}
\definecolor{GnBu-3-C}{RGB}{224,243,219}
\definecolor{GnBu-3-2}{RGB}{168,221,181}
\definecolor{GnBu-3-F}{RGB}{168,221,181}
\definecolor{GnBu-3-3}{RGB}{67,162,202}
\definecolor{GnBu-3-I}{RGB}{67,162,202}
\definecolor{GnBu-4-1}{RGB}{240,249,232}
\definecolor{GnBu-4-B}{RGB}{240,249,232}
\definecolor{GnBu-4-2}{RGB}{186,228,188}
\definecolor{GnBu-4-E}{RGB}{186,228,188}
\definecolor{GnBu-4-3}{RGB}{123,204,196}
\definecolor{GnBu-4-G}{RGB}{123,204,196}
\definecolor{GnBu-4-4}{RGB}{43,140,190}
\definecolor{GnBu-4-J}{RGB}{43,140,190}
\definecolor{GnBu-5-1}{RGB}{240,249,232}
\definecolor{GnBu-5-B}{RGB}{240,249,232}
\definecolor{GnBu-5-2}{RGB}{186,228,188}
\definecolor{GnBu-5-E}{RGB}{186,228,188}
\definecolor{GnBu-5-3}{RGB}{123,204,196}
\definecolor{GnBu-5-G}{RGB}{123,204,196}
\definecolor{GnBu-5-4}{RGB}{67,162,202}
\definecolor{GnBu-5-I}{RGB}{67,162,202}
\definecolor{GnBu-5-5}{RGB}{8,104,172}
\definecolor{GnBu-5-K}{RGB}{8,104,172}
\definecolor{GnBu-6-1}{RGB}{240,249,232}
\definecolor{GnBu-6-B}{RGB}{240,249,232}
\definecolor{GnBu-6-2}{RGB}{204,235,197}
\definecolor{GnBu-6-D}{RGB}{204,235,197}
\definecolor{GnBu-6-3}{RGB}{168,221,181}
\definecolor{GnBu-6-F}{RGB}{168,221,181}
\definecolor{GnBu-6-4}{RGB}{123,204,196}
\definecolor{GnBu-6-G}{RGB}{123,204,196}
\definecolor{GnBu-6-5}{RGB}{67,162,202}
\definecolor{GnBu-6-I}{RGB}{67,162,202}
\definecolor{GnBu-6-6}{RGB}{8,104,172}
\definecolor{GnBu-6-K}{RGB}{8,104,172}
\definecolor{GnBu-7-1}{RGB}{240,249,232}
\definecolor{GnBu-7-B}{RGB}{240,249,232}
\definecolor{GnBu-7-2}{RGB}{204,235,197}
\definecolor{GnBu-7-D}{RGB}{204,235,197}
\definecolor{GnBu-7-3}{RGB}{168,221,181}
\definecolor{GnBu-7-F}{RGB}{168,221,181}
\definecolor{GnBu-7-4}{RGB}{123,204,196}
\definecolor{GnBu-7-G}{RGB}{123,204,196}
\definecolor{GnBu-7-5}{RGB}{78,179,211}
\definecolor{GnBu-7-H}{RGB}{78,179,211}
\definecolor{GnBu-7-6}{RGB}{43,140,190}
\definecolor{GnBu-7-J}{RGB}{43,140,190}
\definecolor{GnBu-7-7}{RGB}{8,88,158}
\definecolor{GnBu-7-L}{RGB}{8,88,158}
\definecolor{GnBu-8-1}{RGB}{247,252,240}
\definecolor{GnBu-8-A}{RGB}{247,252,240}
\definecolor{GnBu-8-2}{RGB}{224,243,219}
\definecolor{GnBu-8-C}{RGB}{224,243,219}
\definecolor{GnBu-8-3}{RGB}{204,235,197}
\definecolor{GnBu-8-D}{RGB}{204,235,197}
\definecolor{GnBu-8-4}{RGB}{168,221,181}
\definecolor{GnBu-8-F}{RGB}{168,221,181}
\definecolor{GnBu-8-5}{RGB}{123,204,196}
\definecolor{GnBu-8-G}{RGB}{123,204,196}
\definecolor{GnBu-8-6}{RGB}{78,179,211}
\definecolor{GnBu-8-H}{RGB}{78,179,211}
\definecolor{GnBu-8-7}{RGB}{43,140,190}
\definecolor{GnBu-8-J}{RGB}{43,140,190}
\definecolor{GnBu-8-8}{RGB}{8,88,158}
\definecolor{GnBu-8-L}{RGB}{8,88,158}
\definecolor{GnBu-9-1}{RGB}{247,252,240}
\definecolor{GnBu-9-A}{RGB}{247,252,240}
\definecolor{GnBu-9-2}{RGB}{224,243,219}
\definecolor{GnBu-9-C}{RGB}{224,243,219}
\definecolor{GnBu-9-3}{RGB}{204,235,197}
\definecolor{GnBu-9-D}{RGB}{204,235,197}
\definecolor{GnBu-9-4}{RGB}{168,221,181}
\definecolor{GnBu-9-F}{RGB}{168,221,181}
\definecolor{GnBu-9-5}{RGB}{123,204,196}
\definecolor{GnBu-9-G}{RGB}{123,204,196}
\definecolor{GnBu-9-6}{RGB}{78,179,211}
\definecolor{GnBu-9-H}{RGB}{78,179,211}
\definecolor{GnBu-9-7}{RGB}{43,140,190}
\definecolor{GnBu-9-J}{RGB}{43,140,190}
\definecolor{GnBu-9-8}{RGB}{8,104,172}
\definecolor{GnBu-9-K}{RGB}{8,104,172}
\definecolor{GnBu-9-9}{RGB}{8,64,129}
\definecolor{GnBu-9-M}{RGB}{8,64,129}
\definecolor{BuGn-3-1}{RGB}{229,245,249}
\definecolor{BuGn-3-C}{RGB}{229,245,249}
\definecolor{BuGn-3-2}{RGB}{153,216,201}
\definecolor{BuGn-3-F}{RGB}{153,216,201}
\definecolor{BuGn-3-3}{RGB}{44,162,95}
\definecolor{BuGn-3-I}{RGB}{44,162,95}
\definecolor{BuGn-4-1}{RGB}{237,248,251}
\definecolor{BuGn-4-B}{RGB}{237,248,251}
\definecolor{BuGn-4-2}{RGB}{178,226,226}
\definecolor{BuGn-4-E}{RGB}{178,226,226}
\definecolor{BuGn-4-3}{RGB}{102,194,164}
\definecolor{BuGn-4-G}{RGB}{102,194,164}
\definecolor{BuGn-4-4}{RGB}{35,139,69}
\definecolor{BuGn-4-J}{RGB}{35,139,69}
\definecolor{BuGn-5-1}{RGB}{237,248,251}
\definecolor{BuGn-5-B}{RGB}{237,248,251}
\definecolor{BuGn-5-2}{RGB}{178,226,226}
\definecolor{BuGn-5-E}{RGB}{178,226,226}
\definecolor{BuGn-5-3}{RGB}{102,194,164}
\definecolor{BuGn-5-G}{RGB}{102,194,164}
\definecolor{BuGn-5-4}{RGB}{44,162,95}
\definecolor{BuGn-5-I}{RGB}{44,162,95}
\definecolor{BuGn-5-5}{RGB}{0,109,44}
\definecolor{BuGn-5-K}{RGB}{0,109,44}
\definecolor{BuGn-6-1}{RGB}{237,248,251}
\definecolor{BuGn-6-B}{RGB}{237,248,251}
\definecolor{BuGn-6-2}{RGB}{204,236,230}
\definecolor{BuGn-6-D}{RGB}{204,236,230}
\definecolor{BuGn-6-3}{RGB}{153,216,201}
\definecolor{BuGn-6-F}{RGB}{153,216,201}
\definecolor{BuGn-6-4}{RGB}{102,194,164}
\definecolor{BuGn-6-G}{RGB}{102,194,164}
\definecolor{BuGn-6-5}{RGB}{44,162,95}
\definecolor{BuGn-6-I}{RGB}{44,162,95}
\definecolor{BuGn-6-6}{RGB}{0,109,44}
\definecolor{BuGn-6-K}{RGB}{0,109,44}
\definecolor{BuGn-7-1}{RGB}{237,248,251}
\definecolor{BuGn-7-B}{RGB}{237,248,251}
\definecolor{BuGn-7-2}{RGB}{204,236,230}
\definecolor{BuGn-7-D}{RGB}{204,236,230}
\definecolor{BuGn-7-3}{RGB}{153,216,201}
\definecolor{BuGn-7-F}{RGB}{153,216,201}
\definecolor{BuGn-7-4}{RGB}{102,194,164}
\definecolor{BuGn-7-G}{RGB}{102,194,164}
\definecolor{BuGn-7-5}{RGB}{65,174,118}
\definecolor{BuGn-7-H}{RGB}{65,174,118}
\definecolor{BuGn-7-6}{RGB}{35,139,69}
\definecolor{BuGn-7-J}{RGB}{35,139,69}
\definecolor{BuGn-7-7}{RGB}{0,88,36}
\definecolor{BuGn-7-L}{RGB}{0,88,36}
\definecolor{BuGn-8-1}{RGB}{247,252,253}
\definecolor{BuGn-8-A}{RGB}{247,252,253}
\definecolor{BuGn-8-2}{RGB}{229,245,249}
\definecolor{BuGn-8-C}{RGB}{229,245,249}
\definecolor{BuGn-8-3}{RGB}{204,236,230}
\definecolor{BuGn-8-D}{RGB}{204,236,230}
\definecolor{BuGn-8-4}{RGB}{153,216,201}
\definecolor{BuGn-8-F}{RGB}{153,216,201}
\definecolor{BuGn-8-5}{RGB}{102,194,164}
\definecolor{BuGn-8-G}{RGB}{102,194,164}
\definecolor{BuGn-8-6}{RGB}{65,174,118}
\definecolor{BuGn-8-H}{RGB}{65,174,118}
\definecolor{BuGn-8-7}{RGB}{35,139,69}
\definecolor{BuGn-8-J}{RGB}{35,139,69}
\definecolor{BuGn-8-8}{RGB}{0,88,36}
\definecolor{BuGn-8-L}{RGB}{0,88,36}
\definecolor{BuGn-9-1}{RGB}{247,252,253}
\definecolor{BuGn-9-A}{RGB}{247,252,253}
\definecolor{BuGn-9-2}{RGB}{229,245,249}
\definecolor{BuGn-9-C}{RGB}{229,245,249}
\definecolor{BuGn-9-3}{RGB}{204,236,230}
\definecolor{BuGn-9-D}{RGB}{204,236,230}
\definecolor{BuGn-9-4}{RGB}{153,216,201}
\definecolor{BuGn-9-F}{RGB}{153,216,201}
\definecolor{BuGn-9-5}{RGB}{102,194,164}
\definecolor{BuGn-9-G}{RGB}{102,194,164}
\definecolor{BuGn-9-6}{RGB}{65,174,118}
\definecolor{BuGn-9-H}{RGB}{65,174,118}
\definecolor{BuGn-9-7}{RGB}{35,139,69}
\definecolor{BuGn-9-J}{RGB}{35,139,69}
\definecolor{BuGn-9-8}{RGB}{0,109,44}
\definecolor{BuGn-9-K}{RGB}{0,109,44}
\definecolor{BuGn-9-9}{RGB}{0,68,27}
\definecolor{BuGn-9-M}{RGB}{0,68,27}
\definecolor{PuBuGn-3-1}{RGB}{236,226,240}
\definecolor{PuBuGn-3-C}{RGB}{236,226,240}
\definecolor{PuBuGn-3-2}{RGB}{166,189,219}
\definecolor{PuBuGn-3-F}{RGB}{166,189,219}
\definecolor{PuBuGn-3-3}{RGB}{28,144,153}
\definecolor{PuBuGn-3-I}{RGB}{28,144,153}
\definecolor{PuBuGn-4-1}{RGB}{246,239,247}
\definecolor{PuBuGn-4-B}{RGB}{246,239,247}
\definecolor{PuBuGn-4-2}{RGB}{189,201,225}
\definecolor{PuBuGn-4-E}{RGB}{189,201,225}
\definecolor{PuBuGn-4-3}{RGB}{103,169,207}
\definecolor{PuBuGn-4-G}{RGB}{103,169,207}
\definecolor{PuBuGn-4-4}{RGB}{2,129,138}
\definecolor{PuBuGn-4-J}{RGB}{2,129,138}
\definecolor{PuBuGn-5-1}{RGB}{246,239,247}
\definecolor{PuBuGn-5-B}{RGB}{246,239,247}
\definecolor{PuBuGn-5-2}{RGB}{189,201,225}
\definecolor{PuBuGn-5-E}{RGB}{189,201,225}
\definecolor{PuBuGn-5-3}{RGB}{103,169,207}
\definecolor{PuBuGn-5-G}{RGB}{103,169,207}
\definecolor{PuBuGn-5-4}{RGB}{28,144,153}
\definecolor{PuBuGn-5-I}{RGB}{28,144,153}
\definecolor{PuBuGn-5-5}{RGB}{1,108,89}
\definecolor{PuBuGn-5-K}{RGB}{1,108,89}
\definecolor{PuBuGn-6-1}{RGB}{246,239,247}
\definecolor{PuBuGn-6-B}{RGB}{246,239,247}
\definecolor{PuBuGn-6-2}{RGB}{208,209,230}
\definecolor{PuBuGn-6-D}{RGB}{208,209,230}
\definecolor{PuBuGn-6-3}{RGB}{166,189,219}
\definecolor{PuBuGn-6-F}{RGB}{166,189,219}
\definecolor{PuBuGn-6-4}{RGB}{103,169,207}
\definecolor{PuBuGn-6-G}{RGB}{103,169,207}
\definecolor{PuBuGn-6-5}{RGB}{28,144,153}
\definecolor{PuBuGn-6-I}{RGB}{28,144,153}
\definecolor{PuBuGn-6-6}{RGB}{1,108,89}
\definecolor{PuBuGn-6-K}{RGB}{1,108,89}
\definecolor{PuBuGn-7-1}{RGB}{246,239,247}
\definecolor{PuBuGn-7-B}{RGB}{246,239,247}
\definecolor{PuBuGn-7-2}{RGB}{208,209,230}
\definecolor{PuBuGn-7-D}{RGB}{208,209,230}
\definecolor{PuBuGn-7-3}{RGB}{166,189,219}
\definecolor{PuBuGn-7-F}{RGB}{166,189,219}
\definecolor{PuBuGn-7-4}{RGB}{103,169,207}
\definecolor{PuBuGn-7-G}{RGB}{103,169,207}
\definecolor{PuBuGn-7-5}{RGB}{54,144,192}
\definecolor{PuBuGn-7-H}{RGB}{54,144,192}
\definecolor{PuBuGn-7-6}{RGB}{2,129,138}
\definecolor{PuBuGn-7-J}{RGB}{2,129,138}
\definecolor{PuBuGn-7-7}{RGB}{1,100,80}
\definecolor{PuBuGn-7-L}{RGB}{1,100,80}
\definecolor{PuBuGn-8-1}{RGB}{255,247,251}
\definecolor{PuBuGn-8-A}{RGB}{255,247,251}
\definecolor{PuBuGn-8-2}{RGB}{236,226,240}
\definecolor{PuBuGn-8-C}{RGB}{236,226,240}
\definecolor{PuBuGn-8-3}{RGB}{208,209,230}
\definecolor{PuBuGn-8-D}{RGB}{208,209,230}
\definecolor{PuBuGn-8-4}{RGB}{166,189,219}
\definecolor{PuBuGn-8-F}{RGB}{166,189,219}
\definecolor{PuBuGn-8-5}{RGB}{103,169,207}
\definecolor{PuBuGn-8-G}{RGB}{103,169,207}
\definecolor{PuBuGn-8-6}{RGB}{54,144,192}
\definecolor{PuBuGn-8-H}{RGB}{54,144,192}
\definecolor{PuBuGn-8-7}{RGB}{2,129,138}
\definecolor{PuBuGn-8-J}{RGB}{2,129,138}
\definecolor{PuBuGn-8-8}{RGB}{1,100,80}
\definecolor{PuBuGn-8-L}{RGB}{1,100,80}
\definecolor{PuBuGn-9-1}{RGB}{255,247,251}
\definecolor{PuBuGn-9-A}{RGB}{255,247,251}
\definecolor{PuBuGn-9-2}{RGB}{236,226,240}
\definecolor{PuBuGn-9-C}{RGB}{236,226,240}
\definecolor{PuBuGn-9-3}{RGB}{208,209,230}
\definecolor{PuBuGn-9-D}{RGB}{208,209,230}
\definecolor{PuBuGn-9-4}{RGB}{166,189,219}
\definecolor{PuBuGn-9-F}{RGB}{166,189,219}
\definecolor{PuBuGn-9-5}{RGB}{103,169,207}
\definecolor{PuBuGn-9-G}{RGB}{103,169,207}
\definecolor{PuBuGn-9-6}{RGB}{54,144,192}
\definecolor{PuBuGn-9-H}{RGB}{54,144,192}
\definecolor{PuBuGn-9-7}{RGB}{2,129,138}
\definecolor{PuBuGn-9-J}{RGB}{2,129,138}
\definecolor{PuBuGn-9-8}{RGB}{1,108,89}
\definecolor{PuBuGn-9-K}{RGB}{1,108,89}
\definecolor{PuBuGn-9-9}{RGB}{1,70,54}
\definecolor{PuBuGn-9-M}{RGB}{1,70,54}
\definecolor{PuBu-3-1}{RGB}{236,231,242}
\definecolor{PuBu-3-C}{RGB}{236,231,242}
\definecolor{PuBu-3-2}{RGB}{166,189,219}
\definecolor{PuBu-3-F}{RGB}{166,189,219}
\definecolor{PuBu-3-3}{RGB}{43,140,190}
\definecolor{PuBu-3-I}{RGB}{43,140,190}
\definecolor{PuBu-4-1}{RGB}{241,238,246}
\definecolor{PuBu-4-B}{RGB}{241,238,246}
\definecolor{PuBu-4-2}{RGB}{189,201,225}
\definecolor{PuBu-4-E}{RGB}{189,201,225}
\definecolor{PuBu-4-3}{RGB}{116,169,207}
\definecolor{PuBu-4-G}{RGB}{116,169,207}
\definecolor{PuBu-4-4}{RGB}{5,112,176}
\definecolor{PuBu-4-J}{RGB}{5,112,176}
\definecolor{PuBu-5-1}{RGB}{241,238,246}
\definecolor{PuBu-5-B}{RGB}{241,238,246}
\definecolor{PuBu-5-2}{RGB}{189,201,225}
\definecolor{PuBu-5-E}{RGB}{189,201,225}
\definecolor{PuBu-5-3}{RGB}{116,169,207}
\definecolor{PuBu-5-G}{RGB}{116,169,207}
\definecolor{PuBu-5-4}{RGB}{43,140,190}
\definecolor{PuBu-5-I}{RGB}{43,140,190}
\definecolor{PuBu-5-5}{RGB}{4,90,141}
\definecolor{PuBu-5-K}{RGB}{4,90,141}
\definecolor{PuBu-6-1}{RGB}{241,238,246}
\definecolor{PuBu-6-B}{RGB}{241,238,246}
\definecolor{PuBu-6-2}{RGB}{208,209,230}
\definecolor{PuBu-6-D}{RGB}{208,209,230}
\definecolor{PuBu-6-3}{RGB}{166,189,219}
\definecolor{PuBu-6-F}{RGB}{166,189,219}
\definecolor{PuBu-6-4}{RGB}{116,169,207}
\definecolor{PuBu-6-G}{RGB}{116,169,207}
\definecolor{PuBu-6-5}{RGB}{43,140,190}
\definecolor{PuBu-6-I}{RGB}{43,140,190}
\definecolor{PuBu-6-6}{RGB}{4,90,141}
\definecolor{PuBu-6-K}{RGB}{4,90,141}
\definecolor{PuBu-7-1}{RGB}{241,238,246}
\definecolor{PuBu-7-B}{RGB}{241,238,246}
\definecolor{PuBu-7-2}{RGB}{208,209,230}
\definecolor{PuBu-7-D}{RGB}{208,209,230}
\definecolor{PuBu-7-3}{RGB}{166,189,219}
\definecolor{PuBu-7-F}{RGB}{166,189,219}
\definecolor{PuBu-7-4}{RGB}{116,169,207}
\definecolor{PuBu-7-G}{RGB}{116,169,207}
\definecolor{PuBu-7-5}{RGB}{54,144,192}
\definecolor{PuBu-7-H}{RGB}{54,144,192}
\definecolor{PuBu-7-6}{RGB}{5,112,176}
\definecolor{PuBu-7-J}{RGB}{5,112,176}
\definecolor{PuBu-7-7}{RGB}{3,78,123}
\definecolor{PuBu-7-L}{RGB}{3,78,123}
\definecolor{PuBu-8-1}{RGB}{255,247,251}
\definecolor{PuBu-8-A}{RGB}{255,247,251}
\definecolor{PuBu-8-2}{RGB}{236,231,242}
\definecolor{PuBu-8-C}{RGB}{236,231,242}
\definecolor{PuBu-8-3}{RGB}{208,209,230}
\definecolor{PuBu-8-D}{RGB}{208,209,230}
\definecolor{PuBu-8-4}{RGB}{166,189,219}
\definecolor{PuBu-8-F}{RGB}{166,189,219}
\definecolor{PuBu-8-5}{RGB}{116,169,207}
\definecolor{PuBu-8-G}{RGB}{116,169,207}
\definecolor{PuBu-8-6}{RGB}{54,144,192}
\definecolor{PuBu-8-H}{RGB}{54,144,192}
\definecolor{PuBu-8-7}{RGB}{5,112,176}
\definecolor{PuBu-8-J}{RGB}{5,112,176}
\definecolor{PuBu-8-8}{RGB}{3,78,123}
\definecolor{PuBu-8-L}{RGB}{3,78,123}
\definecolor{PuBu-9-1}{RGB}{255,247,251}
\definecolor{PuBu-9-A}{RGB}{255,247,251}
\definecolor{PuBu-9-2}{RGB}{236,231,242}
\definecolor{PuBu-9-C}{RGB}{236,231,242}
\definecolor{PuBu-9-3}{RGB}{208,209,230}
\definecolor{PuBu-9-D}{RGB}{208,209,230}
\definecolor{PuBu-9-4}{RGB}{166,189,219}
\definecolor{PuBu-9-F}{RGB}{166,189,219}
\definecolor{PuBu-9-5}{RGB}{116,169,207}
\definecolor{PuBu-9-G}{RGB}{116,169,207}
\definecolor{PuBu-9-6}{RGB}{54,144,192}
\definecolor{PuBu-9-H}{RGB}{54,144,192}
\definecolor{PuBu-9-7}{RGB}{5,112,176}
\definecolor{PuBu-9-J}{RGB}{5,112,176}
\definecolor{PuBu-9-8}{RGB}{4,90,141}
\definecolor{PuBu-9-K}{RGB}{4,90,141}
\definecolor{PuBu-9-9}{RGB}{2,56,88}
\definecolor{PuBu-9-M}{RGB}{2,56,88}
\definecolor{BuPu-3-1}{RGB}{224,236,244}
\definecolor{BuPu-3-C}{RGB}{224,236,244}
\definecolor{BuPu-3-2}{RGB}{158,188,218}
\definecolor{BuPu-3-F}{RGB}{158,188,218}
\definecolor{BuPu-3-3}{RGB}{136,86,167}
\definecolor{BuPu-3-I}{RGB}{136,86,167}
\definecolor{BuPu-4-1}{RGB}{237,248,251}
\definecolor{BuPu-4-B}{RGB}{237,248,251}
\definecolor{BuPu-4-2}{RGB}{179,205,227}
\definecolor{BuPu-4-E}{RGB}{179,205,227}
\definecolor{BuPu-4-3}{RGB}{140,150,198}
\definecolor{BuPu-4-G}{RGB}{140,150,198}
\definecolor{BuPu-4-4}{RGB}{136,65,157}
\definecolor{BuPu-4-J}{RGB}{136,65,157}
\definecolor{BuPu-5-1}{RGB}{237,248,251}
\definecolor{BuPu-5-B}{RGB}{237,248,251}
\definecolor{BuPu-5-2}{RGB}{179,205,227}
\definecolor{BuPu-5-E}{RGB}{179,205,227}
\definecolor{BuPu-5-3}{RGB}{140,150,198}
\definecolor{BuPu-5-G}{RGB}{140,150,198}
\definecolor{BuPu-5-4}{RGB}{136,86,167}
\definecolor{BuPu-5-I}{RGB}{136,86,167}
\definecolor{BuPu-5-5}{RGB}{129,15,124}
\definecolor{BuPu-5-K}{RGB}{129,15,124}
\definecolor{BuPu-6-1}{RGB}{237,248,251}
\definecolor{BuPu-6-B}{RGB}{237,248,251}
\definecolor{BuPu-6-2}{RGB}{191,211,230}
\definecolor{BuPu-6-D}{RGB}{191,211,230}
\definecolor{BuPu-6-3}{RGB}{158,188,218}
\definecolor{BuPu-6-F}{RGB}{158,188,218}
\definecolor{BuPu-6-4}{RGB}{140,150,198}
\definecolor{BuPu-6-G}{RGB}{140,150,198}
\definecolor{BuPu-6-5}{RGB}{136,86,167}
\definecolor{BuPu-6-I}{RGB}{136,86,167}
\definecolor{BuPu-6-6}{RGB}{129,15,124}
\definecolor{BuPu-6-K}{RGB}{129,15,124}
\definecolor{BuPu-7-1}{RGB}{237,248,251}
\definecolor{BuPu-7-B}{RGB}{237,248,251}
\definecolor{BuPu-7-2}{RGB}{191,211,230}
\definecolor{BuPu-7-D}{RGB}{191,211,230}
\definecolor{BuPu-7-3}{RGB}{158,188,218}
\definecolor{BuPu-7-F}{RGB}{158,188,218}
\definecolor{BuPu-7-4}{RGB}{140,150,198}
\definecolor{BuPu-7-G}{RGB}{140,150,198}
\definecolor{BuPu-7-5}{RGB}{140,107,177}
\definecolor{BuPu-7-H}{RGB}{140,107,177}
\definecolor{BuPu-7-6}{RGB}{136,65,157}
\definecolor{BuPu-7-J}{RGB}{136,65,157}
\definecolor{BuPu-7-7}{RGB}{110,1,107}
\definecolor{BuPu-7-L}{RGB}{110,1,107}
\definecolor{BuPu-8-1}{RGB}{247,252,253}
\definecolor{BuPu-8-A}{RGB}{247,252,253}
\definecolor{BuPu-8-2}{RGB}{224,236,244}
\definecolor{BuPu-8-C}{RGB}{224,236,244}
\definecolor{BuPu-8-3}{RGB}{191,211,230}
\definecolor{BuPu-8-D}{RGB}{191,211,230}
\definecolor{BuPu-8-4}{RGB}{158,188,218}
\definecolor{BuPu-8-F}{RGB}{158,188,218}
\definecolor{BuPu-8-5}{RGB}{140,150,198}
\definecolor{BuPu-8-G}{RGB}{140,150,198}
\definecolor{BuPu-8-6}{RGB}{140,107,177}
\definecolor{BuPu-8-H}{RGB}{140,107,177}
\definecolor{BuPu-8-7}{RGB}{136,65,157}
\definecolor{BuPu-8-J}{RGB}{136,65,157}
\definecolor{BuPu-8-8}{RGB}{110,1,107}
\definecolor{BuPu-8-L}{RGB}{110,1,107}
\definecolor{BuPu-9-1}{RGB}{247,252,253}
\definecolor{BuPu-9-A}{RGB}{247,252,253}
\definecolor{BuPu-9-2}{RGB}{224,236,244}
\definecolor{BuPu-9-C}{RGB}{224,236,244}
\definecolor{BuPu-9-3}{RGB}{191,211,230}
\definecolor{BuPu-9-D}{RGB}{191,211,230}
\definecolor{BuPu-9-4}{RGB}{158,188,218}
\definecolor{BuPu-9-F}{RGB}{158,188,218}
\definecolor{BuPu-9-5}{RGB}{140,150,198}
\definecolor{BuPu-9-G}{RGB}{140,150,198}
\definecolor{BuPu-9-6}{RGB}{140,107,177}
\definecolor{BuPu-9-H}{RGB}{140,107,177}
\definecolor{BuPu-9-7}{RGB}{136,65,157}
\definecolor{BuPu-9-J}{RGB}{136,65,157}
\definecolor{BuPu-9-8}{RGB}{129,15,124}
\definecolor{BuPu-9-K}{RGB}{129,15,124}
\definecolor{BuPu-9-9}{RGB}{77,0,75}
\definecolor{BuPu-9-M}{RGB}{77,0,75}
\definecolor{RdPu-3-1}{RGB}{253,224,221}
\definecolor{RdPu-3-C}{RGB}{253,224,221}
\definecolor{RdPu-3-2}{RGB}{250,159,181}
\definecolor{RdPu-3-F}{RGB}{250,159,181}
\definecolor{RdPu-3-3}{RGB}{197,27,138}
\definecolor{RdPu-3-I}{RGB}{197,27,138}
\definecolor{RdPu-4-1}{RGB}{254,235,226}
\definecolor{RdPu-4-B}{RGB}{254,235,226}
\definecolor{RdPu-4-2}{RGB}{251,180,185}
\definecolor{RdPu-4-E}{RGB}{251,180,185}
\definecolor{RdPu-4-3}{RGB}{247,104,161}
\definecolor{RdPu-4-G}{RGB}{247,104,161}
\definecolor{RdPu-4-4}{RGB}{174,1,126}
\definecolor{RdPu-4-J}{RGB}{174,1,126}
\definecolor{RdPu-5-1}{RGB}{254,235,226}
\definecolor{RdPu-5-B}{RGB}{254,235,226}
\definecolor{RdPu-5-2}{RGB}{251,180,185}
\definecolor{RdPu-5-E}{RGB}{251,180,185}
\definecolor{RdPu-5-3}{RGB}{247,104,161}
\definecolor{RdPu-5-G}{RGB}{247,104,161}
\definecolor{RdPu-5-4}{RGB}{197,27,138}
\definecolor{RdPu-5-I}{RGB}{197,27,138}
\definecolor{RdPu-5-5}{RGB}{122,1,119}
\definecolor{RdPu-5-K}{RGB}{122,1,119}
\definecolor{RdPu-6-1}{RGB}{254,235,226}
\definecolor{RdPu-6-B}{RGB}{254,235,226}
\definecolor{RdPu-6-2}{RGB}{252,197,192}
\definecolor{RdPu-6-D}{RGB}{252,197,192}
\definecolor{RdPu-6-3}{RGB}{250,159,181}
\definecolor{RdPu-6-F}{RGB}{250,159,181}
\definecolor{RdPu-6-4}{RGB}{247,104,161}
\definecolor{RdPu-6-G}{RGB}{247,104,161}
\definecolor{RdPu-6-5}{RGB}{197,27,138}
\definecolor{RdPu-6-I}{RGB}{197,27,138}
\definecolor{RdPu-6-6}{RGB}{122,1,119}
\definecolor{RdPu-6-K}{RGB}{122,1,119}
\definecolor{RdPu-7-1}{RGB}{254,235,226}
\definecolor{RdPu-7-B}{RGB}{254,235,226}
\definecolor{RdPu-7-2}{RGB}{252,197,192}
\definecolor{RdPu-7-D}{RGB}{252,197,192}
\definecolor{RdPu-7-3}{RGB}{250,159,181}
\definecolor{RdPu-7-F}{RGB}{250,159,181}
\definecolor{RdPu-7-4}{RGB}{247,104,161}
\definecolor{RdPu-7-G}{RGB}{247,104,161}
\definecolor{RdPu-7-5}{RGB}{221,52,151}
\definecolor{RdPu-7-H}{RGB}{221,52,151}
\definecolor{RdPu-7-6}{RGB}{174,1,126}
\definecolor{RdPu-7-J}{RGB}{174,1,126}
\definecolor{RdPu-7-7}{RGB}{122,1,119}
\definecolor{RdPu-7-L}{RGB}{122,1,119}
\definecolor{RdPu-8-1}{RGB}{255,247,243}
\definecolor{RdPu-8-A}{RGB}{255,247,243}
\definecolor{RdPu-8-2}{RGB}{253,224,221}
\definecolor{RdPu-8-C}{RGB}{253,224,221}
\definecolor{RdPu-8-3}{RGB}{252,197,192}
\definecolor{RdPu-8-D}{RGB}{252,197,192}
\definecolor{RdPu-8-4}{RGB}{250,159,181}
\definecolor{RdPu-8-F}{RGB}{250,159,181}
\definecolor{RdPu-8-5}{RGB}{247,104,161}
\definecolor{RdPu-8-G}{RGB}{247,104,161}
\definecolor{RdPu-8-6}{RGB}{221,52,151}
\definecolor{RdPu-8-H}{RGB}{221,52,151}
\definecolor{RdPu-8-7}{RGB}{174,1,126}
\definecolor{RdPu-8-J}{RGB}{174,1,126}
\definecolor{RdPu-8-8}{RGB}{122,1,119}
\definecolor{RdPu-8-L}{RGB}{122,1,119}
\definecolor{RdPu-9-1}{RGB}{255,247,243}
\definecolor{RdPu-9-A}{RGB}{255,247,243}
\definecolor{RdPu-9-2}{RGB}{253,224,221}
\definecolor{RdPu-9-C}{RGB}{253,224,221}
\definecolor{RdPu-9-3}{RGB}{252,197,192}
\definecolor{RdPu-9-D}{RGB}{252,197,192}
\definecolor{RdPu-9-4}{RGB}{250,159,181}
\definecolor{RdPu-9-F}{RGB}{250,159,181}
\definecolor{RdPu-9-5}{RGB}{247,104,161}
\definecolor{RdPu-9-G}{RGB}{247,104,161}
\definecolor{RdPu-9-6}{RGB}{221,52,151}
\definecolor{RdPu-9-H}{RGB}{221,52,151}
\definecolor{RdPu-9-7}{RGB}{174,1,126}
\definecolor{RdPu-9-J}{RGB}{174,1,126}
\definecolor{RdPu-9-8}{RGB}{122,1,119}
\definecolor{RdPu-9-K}{RGB}{122,1,119}
\definecolor{RdPu-9-9}{RGB}{73,0,106}
\definecolor{RdPu-9-M}{RGB}{73,0,106}
\definecolor{PuRd-3-1}{RGB}{231,225,239}
\definecolor{PuRd-3-C}{RGB}{231,225,239}
\definecolor{PuRd-3-2}{RGB}{201,148,199}
\definecolor{PuRd-3-F}{RGB}{201,148,199}
\definecolor{PuRd-3-3}{RGB}{221,28,119}
\definecolor{PuRd-3-I}{RGB}{221,28,119}
\definecolor{PuRd-4-1}{RGB}{241,238,246}
\definecolor{PuRd-4-B}{RGB}{241,238,246}
\definecolor{PuRd-4-2}{RGB}{215,181,216}
\definecolor{PuRd-4-E}{RGB}{215,181,216}
\definecolor{PuRd-4-3}{RGB}{223,101,176}
\definecolor{PuRd-4-G}{RGB}{223,101,176}
\definecolor{PuRd-4-4}{RGB}{206,18,86}
\definecolor{PuRd-4-J}{RGB}{206,18,86}
\definecolor{PuRd-5-1}{RGB}{241,238,246}
\definecolor{PuRd-5-B}{RGB}{241,238,246}
\definecolor{PuRd-5-2}{RGB}{215,181,216}
\definecolor{PuRd-5-E}{RGB}{215,181,216}
\definecolor{PuRd-5-3}{RGB}{223,101,176}
\definecolor{PuRd-5-G}{RGB}{223,101,176}
\definecolor{PuRd-5-4}{RGB}{221,28,119}
\definecolor{PuRd-5-I}{RGB}{221,28,119}
\definecolor{PuRd-5-5}{RGB}{152,0,67}
\definecolor{PuRd-5-K}{RGB}{152,0,67}
\definecolor{PuRd-6-1}{RGB}{241,238,246}
\definecolor{PuRd-6-B}{RGB}{241,238,246}
\definecolor{PuRd-6-2}{RGB}{212,185,218}
\definecolor{PuRd-6-D}{RGB}{212,185,218}
\definecolor{PuRd-6-3}{RGB}{201,148,199}
\definecolor{PuRd-6-F}{RGB}{201,148,199}
\definecolor{PuRd-6-4}{RGB}{223,101,176}
\definecolor{PuRd-6-G}{RGB}{223,101,176}
\definecolor{PuRd-6-5}{RGB}{221,28,119}
\definecolor{PuRd-6-I}{RGB}{221,28,119}
\definecolor{PuRd-6-6}{RGB}{152,0,67}
\definecolor{PuRd-6-K}{RGB}{152,0,67}
\definecolor{PuRd-7-1}{RGB}{241,238,246}
\definecolor{PuRd-7-B}{RGB}{241,238,246}
\definecolor{PuRd-7-2}{RGB}{212,185,218}
\definecolor{PuRd-7-D}{RGB}{212,185,218}
\definecolor{PuRd-7-3}{RGB}{201,148,199}
\definecolor{PuRd-7-F}{RGB}{201,148,199}
\definecolor{PuRd-7-4}{RGB}{223,101,176}
\definecolor{PuRd-7-G}{RGB}{223,101,176}
\definecolor{PuRd-7-5}{RGB}{231,41,138}
\definecolor{PuRd-7-H}{RGB}{231,41,138}
\definecolor{PuRd-7-6}{RGB}{206,18,86}
\definecolor{PuRd-7-J}{RGB}{206,18,86}
\definecolor{PuRd-7-7}{RGB}{145,0,63}
\definecolor{PuRd-7-L}{RGB}{145,0,63}
\definecolor{PuRd-8-1}{RGB}{247,244,249}
\definecolor{PuRd-8-A}{RGB}{247,244,249}
\definecolor{PuRd-8-2}{RGB}{231,225,239}
\definecolor{PuRd-8-C}{RGB}{231,225,239}
\definecolor{PuRd-8-3}{RGB}{212,185,218}
\definecolor{PuRd-8-D}{RGB}{212,185,218}
\definecolor{PuRd-8-4}{RGB}{201,148,199}
\definecolor{PuRd-8-F}{RGB}{201,148,199}
\definecolor{PuRd-8-5}{RGB}{223,101,176}
\definecolor{PuRd-8-G}{RGB}{223,101,176}
\definecolor{PuRd-8-6}{RGB}{231,41,138}
\definecolor{PuRd-8-H}{RGB}{231,41,138}
\definecolor{PuRd-8-7}{RGB}{206,18,86}
\definecolor{PuRd-8-J}{RGB}{206,18,86}
\definecolor{PuRd-8-8}{RGB}{145,0,63}
\definecolor{PuRd-8-L}{RGB}{145,0,63}
\definecolor{PuRd-9-1}{RGB}{247,244,249}
\definecolor{PuRd-9-A}{RGB}{247,244,249}
\definecolor{PuRd-9-2}{RGB}{231,225,239}
\definecolor{PuRd-9-C}{RGB}{231,225,239}
\definecolor{PuRd-9-3}{RGB}{212,185,218}
\definecolor{PuRd-9-D}{RGB}{212,185,218}
\definecolor{PuRd-9-4}{RGB}{201,148,199}
\definecolor{PuRd-9-F}{RGB}{201,148,199}
\definecolor{PuRd-9-5}{RGB}{223,101,176}
\definecolor{PuRd-9-G}{RGB}{223,101,176}
\definecolor{PuRd-9-6}{RGB}{231,41,138}
\definecolor{PuRd-9-H}{RGB}{231,41,138}
\definecolor{PuRd-9-7}{RGB}{206,18,86}
\definecolor{PuRd-9-J}{RGB}{206,18,86}
\definecolor{PuRd-9-8}{RGB}{152,0,67}
\definecolor{PuRd-9-K}{RGB}{152,0,67}
\definecolor{PuRd-9-9}{RGB}{103,0,31}
\definecolor{PuRd-9-M}{RGB}{103,0,31}
\definecolor{OrRd-3-1}{RGB}{254,232,200}
\definecolor{OrRd-3-C}{RGB}{254,232,200}
\definecolor{OrRd-3-2}{RGB}{253,187,132}
\definecolor{OrRd-3-F}{RGB}{253,187,132}
\definecolor{OrRd-3-3}{RGB}{227,74,51}
\definecolor{OrRd-3-I}{RGB}{227,74,51}
\definecolor{OrRd-4-1}{RGB}{254,240,217}
\definecolor{OrRd-4-B}{RGB}{254,240,217}
\definecolor{OrRd-4-2}{RGB}{253,204,138}
\definecolor{OrRd-4-E}{RGB}{253,204,138}
\definecolor{OrRd-4-3}{RGB}{252,141,89}
\definecolor{OrRd-4-G}{RGB}{252,141,89}
\definecolor{OrRd-4-4}{RGB}{215,48,31}
\definecolor{OrRd-4-J}{RGB}{215,48,31}
\definecolor{OrRd-5-1}{RGB}{254,240,217}
\definecolor{OrRd-5-B}{RGB}{254,240,217}
\definecolor{OrRd-5-2}{RGB}{253,204,138}
\definecolor{OrRd-5-E}{RGB}{253,204,138}
\definecolor{OrRd-5-3}{RGB}{252,141,89}
\definecolor{OrRd-5-G}{RGB}{252,141,89}
\definecolor{OrRd-5-4}{RGB}{227,74,51}
\definecolor{OrRd-5-I}{RGB}{227,74,51}
\definecolor{OrRd-5-5}{RGB}{179,0,0}
\definecolor{OrRd-5-K}{RGB}{179,0,0}
\definecolor{OrRd-6-1}{RGB}{254,240,217}
\definecolor{OrRd-6-B}{RGB}{254,240,217}
\definecolor{OrRd-6-2}{RGB}{253,212,158}
\definecolor{OrRd-6-D}{RGB}{253,212,158}
\definecolor{OrRd-6-3}{RGB}{253,187,132}
\definecolor{OrRd-6-F}{RGB}{253,187,132}
\definecolor{OrRd-6-4}{RGB}{252,141,89}
\definecolor{OrRd-6-G}{RGB}{252,141,89}
\definecolor{OrRd-6-5}{RGB}{227,74,51}
\definecolor{OrRd-6-I}{RGB}{227,74,51}
\definecolor{OrRd-6-6}{RGB}{179,0,0}
\definecolor{OrRd-6-K}{RGB}{179,0,0}
\definecolor{OrRd-7-1}{RGB}{254,240,217}
\definecolor{OrRd-7-B}{RGB}{254,240,217}
\definecolor{OrRd-7-2}{RGB}{253,212,158}
\definecolor{OrRd-7-D}{RGB}{253,212,158}
\definecolor{OrRd-7-3}{RGB}{253,187,132}
\definecolor{OrRd-7-F}{RGB}{253,187,132}
\definecolor{OrRd-7-4}{RGB}{252,141,89}
\definecolor{OrRd-7-G}{RGB}{252,141,89}
\definecolor{OrRd-7-5}{RGB}{239,101,72}
\definecolor{OrRd-7-H}{RGB}{239,101,72}
\definecolor{OrRd-7-6}{RGB}{215,48,31}
\definecolor{OrRd-7-J}{RGB}{215,48,31}
\definecolor{OrRd-7-7}{RGB}{153,0,0}
\definecolor{OrRd-7-L}{RGB}{153,0,0}
\definecolor{OrRd-8-1}{RGB}{255,247,236}
\definecolor{OrRd-8-A}{RGB}{255,247,236}
\definecolor{OrRd-8-2}{RGB}{254,232,200}
\definecolor{OrRd-8-C}{RGB}{254,232,200}
\definecolor{OrRd-8-3}{RGB}{253,212,158}
\definecolor{OrRd-8-D}{RGB}{253,212,158}
\definecolor{OrRd-8-4}{RGB}{253,187,132}
\definecolor{OrRd-8-F}{RGB}{253,187,132}
\definecolor{OrRd-8-5}{RGB}{252,141,89}
\definecolor{OrRd-8-G}{RGB}{252,141,89}
\definecolor{OrRd-8-6}{RGB}{239,101,72}
\definecolor{OrRd-8-H}{RGB}{239,101,72}
\definecolor{OrRd-8-7}{RGB}{215,48,31}
\definecolor{OrRd-8-J}{RGB}{215,48,31}
\definecolor{OrRd-8-8}{RGB}{153,0,0}
\definecolor{OrRd-8-L}{RGB}{153,0,0}
\definecolor{OrRd-9-1}{RGB}{255,247,236}
\definecolor{OrRd-9-A}{RGB}{255,247,236}
\definecolor{OrRd-9-2}{RGB}{254,232,200}
\definecolor{OrRd-9-C}{RGB}{254,232,200}
\definecolor{OrRd-9-3}{RGB}{253,212,158}
\definecolor{OrRd-9-D}{RGB}{253,212,158}
\definecolor{OrRd-9-4}{RGB}{253,187,132}
\definecolor{OrRd-9-F}{RGB}{253,187,132}
\definecolor{OrRd-9-5}{RGB}{252,141,89}
\definecolor{OrRd-9-G}{RGB}{252,141,89}
\definecolor{OrRd-9-6}{RGB}{239,101,72}
\definecolor{OrRd-9-H}{RGB}{239,101,72}
\definecolor{OrRd-9-7}{RGB}{215,48,31}
\definecolor{OrRd-9-J}{RGB}{215,48,31}
\definecolor{OrRd-9-8}{RGB}{179,0,0}
\definecolor{OrRd-9-K}{RGB}{179,0,0}
\definecolor{OrRd-9-9}{RGB}{127,0,0}
\definecolor{OrRd-9-M}{RGB}{127,0,0}
\definecolor{YlOrRd-3-1}{RGB}{255,237,160}
\definecolor{YlOrRd-3-C}{RGB}{255,237,160}
\definecolor{YlOrRd-3-2}{RGB}{254,178,76}
\definecolor{YlOrRd-3-F}{RGB}{254,178,76}
\definecolor{YlOrRd-3-3}{RGB}{240,59,32}
\definecolor{YlOrRd-3-I}{RGB}{240,59,32}
\definecolor{YlOrRd-4-1}{RGB}{255,255,178}
\definecolor{YlOrRd-4-B}{RGB}{255,255,178}
\definecolor{YlOrRd-4-2}{RGB}{254,204,92}
\definecolor{YlOrRd-4-E}{RGB}{254,204,92}
\definecolor{YlOrRd-4-3}{RGB}{253,141,60}
\definecolor{YlOrRd-4-G}{RGB}{253,141,60}
\definecolor{YlOrRd-4-4}{RGB}{227,26,28}
\definecolor{YlOrRd-4-J}{RGB}{227,26,28}
\definecolor{YlOrRd-5-1}{RGB}{255,255,178}
\definecolor{YlOrRd-5-B}{RGB}{255,255,178}
\definecolor{YlOrRd-5-2}{RGB}{254,204,92}
\definecolor{YlOrRd-5-E}{RGB}{254,204,92}
\definecolor{YlOrRd-5-3}{RGB}{253,141,60}
\definecolor{YlOrRd-5-G}{RGB}{253,141,60}
\definecolor{YlOrRd-5-4}{RGB}{240,59,32}
\definecolor{YlOrRd-5-I}{RGB}{240,59,32}
\definecolor{YlOrRd-5-5}{RGB}{189,0,38}
\definecolor{YlOrRd-5-K}{RGB}{189,0,38}
\definecolor{YlOrRd-6-1}{RGB}{255,255,178}
\definecolor{YlOrRd-6-B}{RGB}{255,255,178}
\definecolor{YlOrRd-6-2}{RGB}{254,217,118}
\definecolor{YlOrRd-6-D}{RGB}{254,217,118}
\definecolor{YlOrRd-6-3}{RGB}{254,178,76}
\definecolor{YlOrRd-6-F}{RGB}{254,178,76}
\definecolor{YlOrRd-6-4}{RGB}{253,141,60}
\definecolor{YlOrRd-6-G}{RGB}{253,141,60}
\definecolor{YlOrRd-6-5}{RGB}{240,59,32}
\definecolor{YlOrRd-6-I}{RGB}{240,59,32}
\definecolor{YlOrRd-6-6}{RGB}{189,0,38}
\definecolor{YlOrRd-6-K}{RGB}{189,0,38}
\definecolor{YlOrRd-7-1}{RGB}{255,255,178}
\definecolor{YlOrRd-7-B}{RGB}{255,255,178}
\definecolor{YlOrRd-7-2}{RGB}{254,217,118}
\definecolor{YlOrRd-7-D}{RGB}{254,217,118}
\definecolor{YlOrRd-7-3}{RGB}{254,178,76}
\definecolor{YlOrRd-7-F}{RGB}{254,178,76}
\definecolor{YlOrRd-7-4}{RGB}{253,141,60}
\definecolor{YlOrRd-7-G}{RGB}{253,141,60}
\definecolor{YlOrRd-7-5}{RGB}{252,78,42}
\definecolor{YlOrRd-7-H}{RGB}{252,78,42}
\definecolor{YlOrRd-7-6}{RGB}{227,26,28}
\definecolor{YlOrRd-7-J}{RGB}{227,26,28}
\definecolor{YlOrRd-7-7}{RGB}{177,0,38}
\definecolor{YlOrRd-7-L}{RGB}{177,0,38}
\definecolor{YlOrRd-8-1}{RGB}{255,255,204}
\definecolor{YlOrRd-8-A}{RGB}{255,255,204}
\definecolor{YlOrRd-8-2}{RGB}{255,237,160}
\definecolor{YlOrRd-8-C}{RGB}{255,237,160}
\definecolor{YlOrRd-8-3}{RGB}{254,217,118}
\definecolor{YlOrRd-8-D}{RGB}{254,217,118}
\definecolor{YlOrRd-8-4}{RGB}{254,178,76}
\definecolor{YlOrRd-8-F}{RGB}{254,178,76}
\definecolor{YlOrRd-8-5}{RGB}{253,141,60}
\definecolor{YlOrRd-8-G}{RGB}{253,141,60}
\definecolor{YlOrRd-8-6}{RGB}{252,78,42}
\definecolor{YlOrRd-8-H}{RGB}{252,78,42}
\definecolor{YlOrRd-8-7}{RGB}{227,26,28}
\definecolor{YlOrRd-8-J}{RGB}{227,26,28}
\definecolor{YlOrRd-8-8}{RGB}{177,0,38}
\definecolor{YlOrRd-8-L}{RGB}{177,0,38}
\definecolor{YlOrRd-9-1}{RGB}{255,255,204}
\definecolor{YlOrRd-9-A}{RGB}{255,255,204}
\definecolor{YlOrRd-9-2}{RGB}{255,237,160}
\definecolor{YlOrRd-9-C}{RGB}{255,237,160}
\definecolor{YlOrRd-9-3}{RGB}{254,217,118}
\definecolor{YlOrRd-9-D}{RGB}{254,217,118}
\definecolor{YlOrRd-9-4}{RGB}{254,178,76}
\definecolor{YlOrRd-9-F}{RGB}{254,178,76}
\definecolor{YlOrRd-9-5}{RGB}{253,141,60}
\definecolor{YlOrRd-9-G}{RGB}{253,141,60}
\definecolor{YlOrRd-9-6}{RGB}{252,78,42}
\definecolor{YlOrRd-9-H}{RGB}{252,78,42}
\definecolor{YlOrRd-9-7}{RGB}{227,26,28}
\definecolor{YlOrRd-9-J}{RGB}{227,26,28}
\definecolor{YlOrRd-9-8}{RGB}{189,0,38}
\definecolor{YlOrRd-9-K}{RGB}{189,0,38}
\definecolor{YlOrRd-9-9}{RGB}{128,0,38}
\definecolor{YlOrRd-9-M}{RGB}{128,0,38}
\definecolor{YlOrBr-3-1}{RGB}{255,247,188}
\definecolor{YlOrBr-3-C}{RGB}{255,247,188}
\definecolor{YlOrBr-3-2}{RGB}{254,196,79}
\definecolor{YlOrBr-3-F}{RGB}{254,196,79}
\definecolor{YlOrBr-3-3}{RGB}{217,95,14}
\definecolor{YlOrBr-3-I}{RGB}{217,95,14}
\definecolor{YlOrBr-4-1}{RGB}{255,255,212}
\definecolor{YlOrBr-4-B}{RGB}{255,255,212}
\definecolor{YlOrBr-4-2}{RGB}{254,217,142}
\definecolor{YlOrBr-4-E}{RGB}{254,217,142}
\definecolor{YlOrBr-4-3}{RGB}{254,153,41}
\definecolor{YlOrBr-4-G}{RGB}{254,153,41}
\definecolor{YlOrBr-4-4}{RGB}{204,76,2}
\definecolor{YlOrBr-4-J}{RGB}{204,76,2}
\definecolor{YlOrBr-5-1}{RGB}{255,255,212}
\definecolor{YlOrBr-5-B}{RGB}{255,255,212}
\definecolor{YlOrBr-5-2}{RGB}{254,217,142}
\definecolor{YlOrBr-5-E}{RGB}{254,217,142}
\definecolor{YlOrBr-5-3}{RGB}{254,153,41}
\definecolor{YlOrBr-5-G}{RGB}{254,153,41}
\definecolor{YlOrBr-5-4}{RGB}{217,95,14}
\definecolor{YlOrBr-5-I}{RGB}{217,95,14}
\definecolor{YlOrBr-5-5}{RGB}{153,52,4}
\definecolor{YlOrBr-5-K}{RGB}{153,52,4}
\definecolor{YlOrBr-6-1}{RGB}{255,255,212}
\definecolor{YlOrBr-6-B}{RGB}{255,255,212}
\definecolor{YlOrBr-6-2}{RGB}{254,227,145}
\definecolor{YlOrBr-6-D}{RGB}{254,227,145}
\definecolor{YlOrBr-6-3}{RGB}{254,196,79}
\definecolor{YlOrBr-6-F}{RGB}{254,196,79}
\definecolor{YlOrBr-6-4}{RGB}{254,153,41}
\definecolor{YlOrBr-6-G}{RGB}{254,153,41}
\definecolor{YlOrBr-6-5}{RGB}{217,95,14}
\definecolor{YlOrBr-6-I}{RGB}{217,95,14}
\definecolor{YlOrBr-6-6}{RGB}{153,52,4}
\definecolor{YlOrBr-6-K}{RGB}{153,52,4}
\definecolor{YlOrBr-7-1}{RGB}{255,255,212}
\definecolor{YlOrBr-7-B}{RGB}{255,255,212}
\definecolor{YlOrBr-7-2}{RGB}{254,227,145}
\definecolor{YlOrBr-7-D}{RGB}{254,227,145}
\definecolor{YlOrBr-7-3}{RGB}{254,196,79}
\definecolor{YlOrBr-7-F}{RGB}{254,196,79}
\definecolor{YlOrBr-7-4}{RGB}{254,153,41}
\definecolor{YlOrBr-7-G}{RGB}{254,153,41}
\definecolor{YlOrBr-7-5}{RGB}{236,112,20}
\definecolor{YlOrBr-7-H}{RGB}{236,112,20}
\definecolor{YlOrBr-7-6}{RGB}{204,76,2}
\definecolor{YlOrBr-7-J}{RGB}{204,76,2}
\definecolor{YlOrBr-7-7}{RGB}{140,45,4}
\definecolor{YlOrBr-7-L}{RGB}{140,45,4}
\definecolor{YlOrBr-8-1}{RGB}{255,255,229}
\definecolor{YlOrBr-8-A}{RGB}{255,255,229}
\definecolor{YlOrBr-8-2}{RGB}{255,247,188}
\definecolor{YlOrBr-8-C}{RGB}{255,247,188}
\definecolor{YlOrBr-8-3}{RGB}{254,227,145}
\definecolor{YlOrBr-8-D}{RGB}{254,227,145}
\definecolor{YlOrBr-8-4}{RGB}{254,196,79}
\definecolor{YlOrBr-8-F}{RGB}{254,196,79}
\definecolor{YlOrBr-8-5}{RGB}{254,153,41}
\definecolor{YlOrBr-8-G}{RGB}{254,153,41}
\definecolor{YlOrBr-8-6}{RGB}{236,112,20}
\definecolor{YlOrBr-8-H}{RGB}{236,112,20}
\definecolor{YlOrBr-8-7}{RGB}{204,76,2}
\definecolor{YlOrBr-8-J}{RGB}{204,76,2}
\definecolor{YlOrBr-8-8}{RGB}{140,45,4}
\definecolor{YlOrBr-8-L}{RGB}{140,45,4}
\definecolor{YlOrBr-9-1}{RGB}{255,255,229}
\definecolor{YlOrBr-9-A}{RGB}{255,255,229}
\definecolor{YlOrBr-9-2}{RGB}{255,247,188}
\definecolor{YlOrBr-9-C}{RGB}{255,247,188}
\definecolor{YlOrBr-9-3}{RGB}{254,227,145}
\definecolor{YlOrBr-9-D}{RGB}{254,227,145}
\definecolor{YlOrBr-9-4}{RGB}{254,196,79}
\definecolor{YlOrBr-9-F}{RGB}{254,196,79}
\definecolor{YlOrBr-9-5}{RGB}{254,153,41}
\definecolor{YlOrBr-9-G}{RGB}{254,153,41}
\definecolor{YlOrBr-9-6}{RGB}{236,112,20}
\definecolor{YlOrBr-9-H}{RGB}{236,112,20}
\definecolor{YlOrBr-9-7}{RGB}{204,76,2}
\definecolor{YlOrBr-9-J}{RGB}{204,76,2}
\definecolor{YlOrBr-9-8}{RGB}{153,52,4}
\definecolor{YlOrBr-9-K}{RGB}{153,52,4}
\definecolor{YlOrBr-9-9}{RGB}{102,37,6}
\definecolor{YlOrBr-9-M}{RGB}{102,37,6}
\definecolor{Purples-3-1}{RGB}{239,237,245}
\definecolor{Purples-3-C}{RGB}{239,237,245}
\definecolor{Purples-3-2}{RGB}{188,189,220}
\definecolor{Purples-3-F}{RGB}{188,189,220}
\definecolor{Purples-3-3}{RGB}{117,107,177}
\definecolor{Purples-3-I}{RGB}{117,107,177}
\definecolor{Purples-4-1}{RGB}{242,240,247}
\definecolor{Purples-4-B}{RGB}{242,240,247}
\definecolor{Purples-4-2}{RGB}{203,201,226}
\definecolor{Purples-4-E}{RGB}{203,201,226}
\definecolor{Purples-4-3}{RGB}{158,154,200}
\definecolor{Purples-4-G}{RGB}{158,154,200}
\definecolor{Purples-4-4}{RGB}{106,81,163}
\definecolor{Purples-4-J}{RGB}{106,81,163}
\definecolor{Purples-5-1}{RGB}{242,240,247}
\definecolor{Purples-5-B}{RGB}{242,240,247}
\definecolor{Purples-5-2}{RGB}{203,201,226}
\definecolor{Purples-5-E}{RGB}{203,201,226}
\definecolor{Purples-5-3}{RGB}{158,154,200}
\definecolor{Purples-5-G}{RGB}{158,154,200}
\definecolor{Purples-5-4}{RGB}{117,107,177}
\definecolor{Purples-5-I}{RGB}{117,107,177}
\definecolor{Purples-5-5}{RGB}{84,39,143}
\definecolor{Purples-5-K}{RGB}{84,39,143}
\definecolor{Purples-6-1}{RGB}{242,240,247}
\definecolor{Purples-6-B}{RGB}{242,240,247}
\definecolor{Purples-6-2}{RGB}{218,218,235}
\definecolor{Purples-6-D}{RGB}{218,218,235}
\definecolor{Purples-6-3}{RGB}{188,189,220}
\definecolor{Purples-6-F}{RGB}{188,189,220}
\definecolor{Purples-6-4}{RGB}{158,154,200}
\definecolor{Purples-6-G}{RGB}{158,154,200}
\definecolor{Purples-6-5}{RGB}{117,107,177}
\definecolor{Purples-6-I}{RGB}{117,107,177}
\definecolor{Purples-6-6}{RGB}{84,39,143}
\definecolor{Purples-6-K}{RGB}{84,39,143}
\definecolor{Purples-7-1}{RGB}{242,240,247}
\definecolor{Purples-7-B}{RGB}{242,240,247}
\definecolor{Purples-7-2}{RGB}{218,218,235}
\definecolor{Purples-7-D}{RGB}{218,218,235}
\definecolor{Purples-7-3}{RGB}{188,189,220}
\definecolor{Purples-7-F}{RGB}{188,189,220}
\definecolor{Purples-7-4}{RGB}{158,154,200}
\definecolor{Purples-7-G}{RGB}{158,154,200}
\definecolor{Purples-7-5}{RGB}{128,125,186}
\definecolor{Purples-7-H}{RGB}{128,125,186}
\definecolor{Purples-7-6}{RGB}{106,81,163}
\definecolor{Purples-7-J}{RGB}{106,81,163}
\definecolor{Purples-7-7}{RGB}{74,20,134}
\definecolor{Purples-7-L}{RGB}{74,20,134}
\definecolor{Purples-8-1}{RGB}{252,251,253}
\definecolor{Purples-8-A}{RGB}{252,251,253}
\definecolor{Purples-8-2}{RGB}{239,237,245}
\definecolor{Purples-8-C}{RGB}{239,237,245}
\definecolor{Purples-8-3}{RGB}{218,218,235}
\definecolor{Purples-8-D}{RGB}{218,218,235}
\definecolor{Purples-8-4}{RGB}{188,189,220}
\definecolor{Purples-8-F}{RGB}{188,189,220}
\definecolor{Purples-8-5}{RGB}{158,154,200}
\definecolor{Purples-8-G}{RGB}{158,154,200}
\definecolor{Purples-8-6}{RGB}{128,125,186}
\definecolor{Purples-8-H}{RGB}{128,125,186}
\definecolor{Purples-8-7}{RGB}{106,81,163}
\definecolor{Purples-8-J}{RGB}{106,81,163}
\definecolor{Purples-8-8}{RGB}{74,20,134}
\definecolor{Purples-8-L}{RGB}{74,20,134}
\definecolor{Purples-9-1}{RGB}{252,251,253}
\definecolor{Purples-9-A}{RGB}{252,251,253}
\definecolor{Purples-9-2}{RGB}{239,237,245}
\definecolor{Purples-9-C}{RGB}{239,237,245}
\definecolor{Purples-9-3}{RGB}{218,218,235}
\definecolor{Purples-9-D}{RGB}{218,218,235}
\definecolor{Purples-9-4}{RGB}{188,189,220}
\definecolor{Purples-9-F}{RGB}{188,189,220}
\definecolor{Purples-9-5}{RGB}{158,154,200}
\definecolor{Purples-9-G}{RGB}{158,154,200}
\definecolor{Purples-9-6}{RGB}{128,125,186}
\definecolor{Purples-9-H}{RGB}{128,125,186}
\definecolor{Purples-9-7}{RGB}{106,81,163}
\definecolor{Purples-9-J}{RGB}{106,81,163}
\definecolor{Purples-9-8}{RGB}{84,39,143}
\definecolor{Purples-9-K}{RGB}{84,39,143}
\definecolor{Purples-9-9}{RGB}{63,0,125}
\definecolor{Purples-9-M}{RGB}{63,0,125}
\definecolor{Blues-3-1}{RGB}{222,235,247}
\definecolor{Blues-3-C}{RGB}{222,235,247}
\definecolor{Blues-3-2}{RGB}{158,202,225}
\definecolor{Blues-3-F}{RGB}{158,202,225}
\definecolor{Blues-3-3}{RGB}{49,130,189}
\definecolor{Blues-3-I}{RGB}{49,130,189}
\definecolor{Blues-4-1}{RGB}{239,243,255}
\definecolor{Blues-4-B}{RGB}{239,243,255}
\definecolor{Blues-4-2}{RGB}{189,215,231}
\definecolor{Blues-4-E}{RGB}{189,215,231}
\definecolor{Blues-4-3}{RGB}{107,174,214}
\definecolor{Blues-4-G}{RGB}{107,174,214}
\definecolor{Blues-4-4}{RGB}{33,113,181}
\definecolor{Blues-4-J}{RGB}{33,113,181}
\definecolor{Blues-5-1}{RGB}{239,243,255}
\definecolor{Blues-5-B}{RGB}{239,243,255}
\definecolor{Blues-5-2}{RGB}{189,215,231}
\definecolor{Blues-5-E}{RGB}{189,215,231}
\definecolor{Blues-5-3}{RGB}{107,174,214}
\definecolor{Blues-5-G}{RGB}{107,174,214}
\definecolor{Blues-5-4}{RGB}{49,130,189}
\definecolor{Blues-5-I}{RGB}{49,130,189}
\definecolor{Blues-5-5}{RGB}{8,81,156}
\definecolor{Blues-5-K}{RGB}{8,81,156}
\definecolor{Blues-6-1}{RGB}{239,243,255}
\definecolor{Blues-6-B}{RGB}{239,243,255}
\definecolor{Blues-6-2}{RGB}{198,219,239}
\definecolor{Blues-6-D}{RGB}{198,219,239}
\definecolor{Blues-6-3}{RGB}{158,202,225}
\definecolor{Blues-6-F}{RGB}{158,202,225}
\definecolor{Blues-6-4}{RGB}{107,174,214}
\definecolor{Blues-6-G}{RGB}{107,174,214}
\definecolor{Blues-6-5}{RGB}{49,130,189}
\definecolor{Blues-6-I}{RGB}{49,130,189}
\definecolor{Blues-6-6}{RGB}{8,81,156}
\definecolor{Blues-6-K}{RGB}{8,81,156}
\definecolor{Blues-7-1}{RGB}{239,243,255}
\definecolor{Blues-7-B}{RGB}{239,243,255}
\definecolor{Blues-7-2}{RGB}{198,219,239}
\definecolor{Blues-7-D}{RGB}{198,219,239}
\definecolor{Blues-7-3}{RGB}{158,202,225}
\definecolor{Blues-7-F}{RGB}{158,202,225}
\definecolor{Blues-7-4}{RGB}{107,174,214}
\definecolor{Blues-7-G}{RGB}{107,174,214}
\definecolor{Blues-7-5}{RGB}{66,146,198}
\definecolor{Blues-7-H}{RGB}{66,146,198}
\definecolor{Blues-7-6}{RGB}{33,113,181}
\definecolor{Blues-7-J}{RGB}{33,113,181}
\definecolor{Blues-7-7}{RGB}{8,69,148}
\definecolor{Blues-7-L}{RGB}{8,69,148}
\definecolor{Blues-8-1}{RGB}{247,251,255}
\definecolor{Blues-8-A}{RGB}{247,251,255}
\definecolor{Blues-8-2}{RGB}{222,235,247}
\definecolor{Blues-8-C}{RGB}{222,235,247}
\definecolor{Blues-8-3}{RGB}{198,219,239}
\definecolor{Blues-8-D}{RGB}{198,219,239}
\definecolor{Blues-8-4}{RGB}{158,202,225}
\definecolor{Blues-8-F}{RGB}{158,202,225}
\definecolor{Blues-8-5}{RGB}{107,174,214}
\definecolor{Blues-8-G}{RGB}{107,174,214}
\definecolor{Blues-8-6}{RGB}{66,146,198}
\definecolor{Blues-8-H}{RGB}{66,146,198}
\definecolor{Blues-8-7}{RGB}{33,113,181}
\definecolor{Blues-8-J}{RGB}{33,113,181}
\definecolor{Blues-8-8}{RGB}{8,69,148}
\definecolor{Blues-8-L}{RGB}{8,69,148}
\definecolor{Blues-9-1}{RGB}{247,251,255}
\definecolor{Blues-9-A}{RGB}{247,251,255}
\definecolor{Blues-9-2}{RGB}{222,235,247}
\definecolor{Blues-9-C}{RGB}{222,235,247}
\definecolor{Blues-9-3}{RGB}{198,219,239}
\definecolor{Blues-9-D}{RGB}{198,219,239}
\definecolor{Blues-9-4}{RGB}{158,202,225}
\definecolor{Blues-9-F}{RGB}{158,202,225}
\definecolor{Blues-9-5}{RGB}{107,174,214}
\definecolor{Blues-9-G}{RGB}{107,174,214}
\definecolor{Blues-9-6}{RGB}{66,146,198}
\definecolor{Blues-9-H}{RGB}{66,146,198}
\definecolor{Blues-9-7}{RGB}{33,113,181}
\definecolor{Blues-9-J}{RGB}{33,113,181}
\definecolor{Blues-9-8}{RGB}{8,81,156}
\definecolor{Blues-9-K}{RGB}{8,81,156}
\definecolor{Blues-9-9}{RGB}{8,48,107}
\definecolor{Blues-9-M}{RGB}{8,48,107}
\definecolor{Greens-3-1}{RGB}{229,245,224}
\definecolor{Greens-3-C}{RGB}{229,245,224}
\definecolor{Greens-3-2}{RGB}{161,217,155}
\definecolor{Greens-3-F}{RGB}{161,217,155}
\definecolor{Greens-3-3}{RGB}{49,163,84}
\definecolor{Greens-3-I}{RGB}{49,163,84}
\definecolor{Greens-4-1}{RGB}{237,248,233}
\definecolor{Greens-4-B}{RGB}{237,248,233}
\definecolor{Greens-4-2}{RGB}{186,228,179}
\definecolor{Greens-4-E}{RGB}{186,228,179}
\definecolor{Greens-4-3}{RGB}{116,196,118}
\definecolor{Greens-4-G}{RGB}{116,196,118}
\definecolor{Greens-4-4}{RGB}{35,139,69}
\definecolor{Greens-4-J}{RGB}{35,139,69}
\definecolor{Greens-5-1}{RGB}{237,248,233}
\definecolor{Greens-5-B}{RGB}{237,248,233}
\definecolor{Greens-5-2}{RGB}{186,228,179}
\definecolor{Greens-5-E}{RGB}{186,228,179}
\definecolor{Greens-5-3}{RGB}{116,196,118}
\definecolor{Greens-5-G}{RGB}{116,196,118}
\definecolor{Greens-5-4}{RGB}{49,163,84}
\definecolor{Greens-5-I}{RGB}{49,163,84}
\definecolor{Greens-5-5}{RGB}{0,109,44}
\definecolor{Greens-5-K}{RGB}{0,109,44}
\definecolor{Greens-6-1}{RGB}{237,248,233}
\definecolor{Greens-6-B}{RGB}{237,248,233}
\definecolor{Greens-6-2}{RGB}{199,233,192}
\definecolor{Greens-6-D}{RGB}{199,233,192}
\definecolor{Greens-6-3}{RGB}{161,217,155}
\definecolor{Greens-6-F}{RGB}{161,217,155}
\definecolor{Greens-6-4}{RGB}{116,196,118}
\definecolor{Greens-6-G}{RGB}{116,196,118}
\definecolor{Greens-6-5}{RGB}{49,163,84}
\definecolor{Greens-6-I}{RGB}{49,163,84}
\definecolor{Greens-6-6}{RGB}{0,109,44}
\definecolor{Greens-6-K}{RGB}{0,109,44}
\definecolor{Greens-7-1}{RGB}{237,248,233}
\definecolor{Greens-7-B}{RGB}{237,248,233}
\definecolor{Greens-7-2}{RGB}{199,233,192}
\definecolor{Greens-7-D}{RGB}{199,233,192}
\definecolor{Greens-7-3}{RGB}{161,217,155}
\definecolor{Greens-7-F}{RGB}{161,217,155}
\definecolor{Greens-7-4}{RGB}{116,196,118}
\definecolor{Greens-7-G}{RGB}{116,196,118}
\definecolor{Greens-7-5}{RGB}{65,171,93}
\definecolor{Greens-7-H}{RGB}{65,171,93}
\definecolor{Greens-7-6}{RGB}{35,139,69}
\definecolor{Greens-7-J}{RGB}{35,139,69}
\definecolor{Greens-7-7}{RGB}{0,90,50}
\definecolor{Greens-7-L}{RGB}{0,90,50}
\definecolor{Greens-8-1}{RGB}{247,252,245}
\definecolor{Greens-8-A}{RGB}{247,252,245}
\definecolor{Greens-8-2}{RGB}{229,245,224}
\definecolor{Greens-8-C}{RGB}{229,245,224}
\definecolor{Greens-8-3}{RGB}{199,233,192}
\definecolor{Greens-8-D}{RGB}{199,233,192}
\definecolor{Greens-8-4}{RGB}{161,217,155}
\definecolor{Greens-8-F}{RGB}{161,217,155}
\definecolor{Greens-8-5}{RGB}{116,196,118}
\definecolor{Greens-8-G}{RGB}{116,196,118}
\definecolor{Greens-8-6}{RGB}{65,171,93}
\definecolor{Greens-8-H}{RGB}{65,171,93}
\definecolor{Greens-8-7}{RGB}{35,139,69}
\definecolor{Greens-8-J}{RGB}{35,139,69}
\definecolor{Greens-8-8}{RGB}{0,90,50}
\definecolor{Greens-8-L}{RGB}{0,90,50}
\definecolor{Greens-9-1}{RGB}{247,252,245}
\definecolor{Greens-9-A}{RGB}{247,252,245}
\definecolor{Greens-9-2}{RGB}{229,245,224}
\definecolor{Greens-9-C}{RGB}{229,245,224}
\definecolor{Greens-9-3}{RGB}{199,233,192}
\definecolor{Greens-9-D}{RGB}{199,233,192}
\definecolor{Greens-9-4}{RGB}{161,217,155}
\definecolor{Greens-9-F}{RGB}{161,217,155}
\definecolor{Greens-9-5}{RGB}{116,196,118}
\definecolor{Greens-9-G}{RGB}{116,196,118}
\definecolor{Greens-9-6}{RGB}{65,171,93}
\definecolor{Greens-9-H}{RGB}{65,171,93}
\definecolor{Greens-9-7}{RGB}{35,139,69}
\definecolor{Greens-9-J}{RGB}{35,139,69}
\definecolor{Greens-9-8}{RGB}{0,109,44}
\definecolor{Greens-9-K}{RGB}{0,109,44}
\definecolor{Greens-9-9}{RGB}{0,68,27}
\definecolor{Greens-9-M}{RGB}{0,68,27}
\definecolor{Oranges-3-1}{RGB}{254,230,206}
\definecolor{Oranges-3-C}{RGB}{254,230,206}
\definecolor{Oranges-3-2}{RGB}{253,174,107}
\definecolor{Oranges-3-F}{RGB}{253,174,107}
\definecolor{Oranges-3-3}{RGB}{230,85,13}
\definecolor{Oranges-3-I}{RGB}{230,85,13}
\definecolor{Oranges-4-1}{RGB}{254,237,222}
\definecolor{Oranges-4-B}{RGB}{254,237,222}
\definecolor{Oranges-4-2}{RGB}{253,190,133}
\definecolor{Oranges-4-E}{RGB}{253,190,133}
\definecolor{Oranges-4-3}{RGB}{253,141,60}
\definecolor{Oranges-4-G}{RGB}{253,141,60}
\definecolor{Oranges-4-4}{RGB}{217,71,1}
\definecolor{Oranges-4-J}{RGB}{217,71,1}
\definecolor{Oranges-5-1}{RGB}{254,237,222}
\definecolor{Oranges-5-B}{RGB}{254,237,222}
\definecolor{Oranges-5-2}{RGB}{253,190,133}
\definecolor{Oranges-5-E}{RGB}{253,190,133}
\definecolor{Oranges-5-3}{RGB}{253,141,60}
\definecolor{Oranges-5-G}{RGB}{253,141,60}
\definecolor{Oranges-5-4}{RGB}{230,85,13}
\definecolor{Oranges-5-I}{RGB}{230,85,13}
\definecolor{Oranges-5-5}{RGB}{166,54,3}
\definecolor{Oranges-5-K}{RGB}{166,54,3}
\definecolor{Oranges-6-1}{RGB}{254,237,222}
\definecolor{Oranges-6-B}{RGB}{254,237,222}
\definecolor{Oranges-6-2}{RGB}{253,208,162}
\definecolor{Oranges-6-D}{RGB}{253,208,162}
\definecolor{Oranges-6-3}{RGB}{253,174,107}
\definecolor{Oranges-6-F}{RGB}{253,174,107}
\definecolor{Oranges-6-4}{RGB}{253,141,60}
\definecolor{Oranges-6-G}{RGB}{253,141,60}
\definecolor{Oranges-6-5}{RGB}{230,85,13}
\definecolor{Oranges-6-I}{RGB}{230,85,13}
\definecolor{Oranges-6-6}{RGB}{166,54,3}
\definecolor{Oranges-6-K}{RGB}{166,54,3}
\definecolor{Oranges-7-1}{RGB}{254,237,222}
\definecolor{Oranges-7-B}{RGB}{254,237,222}
\definecolor{Oranges-7-2}{RGB}{253,208,162}
\definecolor{Oranges-7-D}{RGB}{253,208,162}
\definecolor{Oranges-7-3}{RGB}{253,174,107}
\definecolor{Oranges-7-F}{RGB}{253,174,107}
\definecolor{Oranges-7-4}{RGB}{253,141,60}
\definecolor{Oranges-7-G}{RGB}{253,141,60}
\definecolor{Oranges-7-5}{RGB}{241,105,19}
\definecolor{Oranges-7-H}{RGB}{241,105,19}
\definecolor{Oranges-7-6}{RGB}{217,72,1}
\definecolor{Oranges-7-J}{RGB}{217,72,1}
\definecolor{Oranges-7-7}{RGB}{140,45,4}
\definecolor{Oranges-7-L}{RGB}{140,45,4}
\definecolor{Oranges-8-1}{RGB}{255,245,235}
\definecolor{Oranges-8-A}{RGB}{255,245,235}
\definecolor{Oranges-8-2}{RGB}{254,230,206}
\definecolor{Oranges-8-C}{RGB}{254,230,206}
\definecolor{Oranges-8-3}{RGB}{253,208,162}
\definecolor{Oranges-8-D}{RGB}{253,208,162}
\definecolor{Oranges-8-4}{RGB}{253,174,107}
\definecolor{Oranges-8-F}{RGB}{253,174,107}
\definecolor{Oranges-8-5}{RGB}{253,141,60}
\definecolor{Oranges-8-G}{RGB}{253,141,60}
\definecolor{Oranges-8-6}{RGB}{241,105,19}
\definecolor{Oranges-8-H}{RGB}{241,105,19}
\definecolor{Oranges-8-7}{RGB}{217,72,1}
\definecolor{Oranges-8-J}{RGB}{217,72,1}
\definecolor{Oranges-8-8}{RGB}{140,45,4}
\definecolor{Oranges-8-L}{RGB}{140,45,4}
\definecolor{Oranges-9-1}{RGB}{255,245,235}
\definecolor{Oranges-9-A}{RGB}{255,245,235}
\definecolor{Oranges-9-2}{RGB}{254,230,206}
\definecolor{Oranges-9-C}{RGB}{254,230,206}
\definecolor{Oranges-9-3}{RGB}{253,208,162}
\definecolor{Oranges-9-D}{RGB}{253,208,162}
\definecolor{Oranges-9-4}{RGB}{253,174,107}
\definecolor{Oranges-9-F}{RGB}{253,174,107}
\definecolor{Oranges-9-5}{RGB}{253,141,60}
\definecolor{Oranges-9-G}{RGB}{253,141,60}
\definecolor{Oranges-9-6}{RGB}{241,105,19}
\definecolor{Oranges-9-H}{RGB}{241,105,19}
\definecolor{Oranges-9-7}{RGB}{217,72,1}
\definecolor{Oranges-9-J}{RGB}{217,72,1}
\definecolor{Oranges-9-8}{RGB}{166,54,3}
\definecolor{Oranges-9-K}{RGB}{166,54,3}
\definecolor{Oranges-9-9}{RGB}{127,39,4}
\definecolor{Oranges-9-M}{RGB}{127,39,4}
\definecolor{Reds-3-1}{RGB}{254,224,210}
\definecolor{Reds-3-C}{RGB}{254,224,210}
\definecolor{Reds-3-2}{RGB}{252,146,114}
\definecolor{Reds-3-F}{RGB}{252,146,114}
\definecolor{Reds-3-3}{RGB}{222,45,38}
\definecolor{Reds-3-I}{RGB}{222,45,38}
\definecolor{Reds-4-1}{RGB}{254,229,217}
\definecolor{Reds-4-B}{RGB}{254,229,217}
\definecolor{Reds-4-2}{RGB}{252,174,145}
\definecolor{Reds-4-E}{RGB}{252,174,145}
\definecolor{Reds-4-3}{RGB}{251,106,74}
\definecolor{Reds-4-G}{RGB}{251,106,74}
\definecolor{Reds-4-4}{RGB}{203,24,29}
\definecolor{Reds-4-J}{RGB}{203,24,29}
\definecolor{Reds-5-1}{RGB}{254,229,217}
\definecolor{Reds-5-B}{RGB}{254,229,217}
\definecolor{Reds-5-2}{RGB}{252,174,145}
\definecolor{Reds-5-E}{RGB}{252,174,145}
\definecolor{Reds-5-3}{RGB}{251,106,74}
\definecolor{Reds-5-G}{RGB}{251,106,74}
\definecolor{Reds-5-4}{RGB}{222,45,38}
\definecolor{Reds-5-I}{RGB}{222,45,38}
\definecolor{Reds-5-5}{RGB}{165,15,21}
\definecolor{Reds-5-K}{RGB}{165,15,21}
\definecolor{Reds-6-1}{RGB}{254,229,217}
\definecolor{Reds-6-B}{RGB}{254,229,217}
\definecolor{Reds-6-2}{RGB}{252,187,161}
\definecolor{Reds-6-D}{RGB}{252,187,161}
\definecolor{Reds-6-3}{RGB}{252,146,114}
\definecolor{Reds-6-F}{RGB}{252,146,114}
\definecolor{Reds-6-4}{RGB}{251,106,74}
\definecolor{Reds-6-G}{RGB}{251,106,74}
\definecolor{Reds-6-5}{RGB}{222,45,38}
\definecolor{Reds-6-I}{RGB}{222,45,38}
\definecolor{Reds-6-6}{RGB}{165,15,21}
\definecolor{Reds-6-K}{RGB}{165,15,21}
\definecolor{Reds-7-1}{RGB}{254,229,217}
\definecolor{Reds-7-B}{RGB}{254,229,217}
\definecolor{Reds-7-2}{RGB}{252,187,161}
\definecolor{Reds-7-D}{RGB}{252,187,161}
\definecolor{Reds-7-3}{RGB}{252,146,114}
\definecolor{Reds-7-F}{RGB}{252,146,114}
\definecolor{Reds-7-4}{RGB}{251,106,74}
\definecolor{Reds-7-G}{RGB}{251,106,74}
\definecolor{Reds-7-5}{RGB}{239,59,44}
\definecolor{Reds-7-H}{RGB}{239,59,44}
\definecolor{Reds-7-6}{RGB}{203,24,29}
\definecolor{Reds-7-J}{RGB}{203,24,29}
\definecolor{Reds-7-7}{RGB}{153,0,13}
\definecolor{Reds-7-L}{RGB}{153,0,13}
\definecolor{Reds-8-1}{RGB}{255,245,240}
\definecolor{Reds-8-A}{RGB}{255,245,240}
\definecolor{Reds-8-2}{RGB}{254,224,210}
\definecolor{Reds-8-C}{RGB}{254,224,210}
\definecolor{Reds-8-3}{RGB}{252,187,161}
\definecolor{Reds-8-D}{RGB}{252,187,161}
\definecolor{Reds-8-4}{RGB}{252,146,114}
\definecolor{Reds-8-F}{RGB}{252,146,114}
\definecolor{Reds-8-5}{RGB}{251,106,74}
\definecolor{Reds-8-G}{RGB}{251,106,74}
\definecolor{Reds-8-6}{RGB}{239,59,44}
\definecolor{Reds-8-H}{RGB}{239,59,44}
\definecolor{Reds-8-7}{RGB}{203,24,29}
\definecolor{Reds-8-J}{RGB}{203,24,29}
\definecolor{Reds-8-8}{RGB}{153,0,13}
\definecolor{Reds-8-L}{RGB}{153,0,13}
\definecolor{Reds-9-1}{RGB}{255,245,240}
\definecolor{Reds-9-A}{RGB}{255,245,240}
\definecolor{Reds-9-2}{RGB}{254,224,210}
\definecolor{Reds-9-C}{RGB}{254,224,210}
\definecolor{Reds-9-3}{RGB}{252,187,161}
\definecolor{Reds-9-D}{RGB}{252,187,161}
\definecolor{Reds-9-4}{RGB}{252,146,114}
\definecolor{Reds-9-F}{RGB}{252,146,114}
\definecolor{Reds-9-5}{RGB}{251,106,74}
\definecolor{Reds-9-G}{RGB}{251,106,74}
\definecolor{Reds-9-6}{RGB}{239,59,44}
\definecolor{Reds-9-H}{RGB}{239,59,44}
\definecolor{Reds-9-7}{RGB}{203,24,29}
\definecolor{Reds-9-J}{RGB}{203,24,29}
\definecolor{Reds-9-8}{RGB}{165,15,21}
\definecolor{Reds-9-K}{RGB}{165,15,21}
\definecolor{Reds-9-9}{RGB}{103,0,13}
\definecolor{Reds-9-M}{RGB}{103,0,13}
\definecolor{Greys-3-1}{RGB}{240,240,240}
\definecolor{Greys-3-C}{RGB}{240,240,240}
\definecolor{Greys-3-2}{RGB}{189,189,189}
\definecolor{Greys-3-F}{RGB}{189,189,189}
\definecolor{Greys-3-3}{RGB}{99,99,99}
\definecolor{Greys-3-I}{RGB}{99,99,99}
\definecolor{Greys-4-1}{RGB}{247,247,247}
\definecolor{Greys-4-B}{RGB}{247,247,247}
\definecolor{Greys-4-2}{RGB}{204,204,204}
\definecolor{Greys-4-E}{RGB}{204,204,204}
\definecolor{Greys-4-3}{RGB}{150,150,150}
\definecolor{Greys-4-G}{RGB}{150,150,150}
\definecolor{Greys-4-4}{RGB}{82,82,82}
\definecolor{Greys-4-J}{RGB}{82,82,82}
\definecolor{Greys-5-1}{RGB}{247,247,247}
\definecolor{Greys-5-B}{RGB}{247,247,247}
\definecolor{Greys-5-2}{RGB}{204,204,204}
\definecolor{Greys-5-E}{RGB}{204,204,204}
\definecolor{Greys-5-3}{RGB}{150,150,150}
\definecolor{Greys-5-G}{RGB}{150,150,150}
\definecolor{Greys-5-4}{RGB}{99,99,99}
\definecolor{Greys-5-I}{RGB}{99,99,99}
\definecolor{Greys-5-5}{RGB}{37,37,37}
\definecolor{Greys-5-K}{RGB}{37,37,37}
\definecolor{Greys-6-1}{RGB}{247,247,247}
\definecolor{Greys-6-B}{RGB}{247,247,247}
\definecolor{Greys-6-2}{RGB}{217,217,217}
\definecolor{Greys-6-D}{RGB}{217,217,217}
\definecolor{Greys-6-3}{RGB}{189,189,189}
\definecolor{Greys-6-F}{RGB}{189,189,189}
\definecolor{Greys-6-4}{RGB}{150,150,150}
\definecolor{Greys-6-G}{RGB}{150,150,150}
\definecolor{Greys-6-5}{RGB}{99,99,99}
\definecolor{Greys-6-I}{RGB}{99,99,99}
\definecolor{Greys-6-6}{RGB}{37,37,37}
\definecolor{Greys-6-K}{RGB}{37,37,37}
\definecolor{Greys-7-1}{RGB}{247,247,247}
\definecolor{Greys-7-B}{RGB}{247,247,247}
\definecolor{Greys-7-2}{RGB}{217,217,217}
\definecolor{Greys-7-D}{RGB}{217,217,217}
\definecolor{Greys-7-3}{RGB}{189,189,189}
\definecolor{Greys-7-F}{RGB}{189,189,189}
\definecolor{Greys-7-4}{RGB}{150,150,150}
\definecolor{Greys-7-G}{RGB}{150,150,150}
\definecolor{Greys-7-5}{RGB}{115,115,115}
\definecolor{Greys-7-H}{RGB}{115,115,115}
\definecolor{Greys-7-6}{RGB}{82,82,82}
\definecolor{Greys-7-J}{RGB}{82,82,82}
\definecolor{Greys-7-7}{RGB}{37,37,37}
\definecolor{Greys-7-L}{RGB}{37,37,37}
\definecolor{Greys-8-1}{RGB}{255,255,255}
\definecolor{Greys-8-A}{RGB}{255,255,255}
\definecolor{Greys-8-2}{RGB}{240,240,240}
\definecolor{Greys-8-C}{RGB}{240,240,240}
\definecolor{Greys-8-3}{RGB}{217,217,217}
\definecolor{Greys-8-D}{RGB}{217,217,217}
\definecolor{Greys-8-4}{RGB}{189,189,189}
\definecolor{Greys-8-F}{RGB}{189,189,189}
\definecolor{Greys-8-5}{RGB}{150,150,150}
\definecolor{Greys-8-G}{RGB}{150,150,150}
\definecolor{Greys-8-6}{RGB}{115,115,115}
\definecolor{Greys-8-H}{RGB}{115,115,115}
\definecolor{Greys-8-7}{RGB}{82,82,82}
\definecolor{Greys-8-J}{RGB}{82,82,82}
\definecolor{Greys-8-8}{RGB}{37,37,37}
\definecolor{Greys-8-L}{RGB}{37,37,37}
\definecolor{Greys-9-1}{RGB}{255,255,255}
\definecolor{Greys-9-A}{RGB}{255,255,255}
\definecolor{Greys-9-2}{RGB}{240,240,240}
\definecolor{Greys-9-C}{RGB}{240,240,240}
\definecolor{Greys-9-3}{RGB}{217,217,217}
\definecolor{Greys-9-D}{RGB}{217,217,217}
\definecolor{Greys-9-4}{RGB}{189,189,189}
\definecolor{Greys-9-F}{RGB}{189,189,189}
\definecolor{Greys-9-5}{RGB}{150,150,150}
\definecolor{Greys-9-G}{RGB}{150,150,150}
\definecolor{Greys-9-6}{RGB}{115,115,115}
\definecolor{Greys-9-H}{RGB}{115,115,115}
\definecolor{Greys-9-7}{RGB}{82,82,82}
\definecolor{Greys-9-J}{RGB}{82,82,82}
\definecolor{Greys-9-8}{RGB}{37,37,37}
\definecolor{Greys-9-K}{RGB}{37,37,37}
\definecolor{Greys-9-9}{RGB}{0,0,0}
\definecolor{Greys-9-M}{RGB}{0,0,0}
\definecolor{PuOr-3-1}{RGB}{241,163,64}
\definecolor{PuOr-3-E}{RGB}{241,163,64}
\definecolor{PuOr-3-2}{RGB}{247,247,247}
\definecolor{PuOr-3-H}{RGB}{247,247,247}
\definecolor{PuOr-3-3}{RGB}{153,142,195}
\definecolor{PuOr-3-K}{RGB}{153,142,195}
\definecolor{PuOr-4-1}{RGB}{230,97,1}
\definecolor{PuOr-4-C}{RGB}{230,97,1}
\definecolor{PuOr-4-2}{RGB}{253,184,99}
\definecolor{PuOr-4-F}{RGB}{253,184,99}
\definecolor{PuOr-4-3}{RGB}{178,171,210}
\definecolor{PuOr-4-J}{RGB}{178,171,210}
\definecolor{PuOr-4-4}{RGB}{94,60,153}
\definecolor{PuOr-4-M}{RGB}{94,60,153}
\definecolor{PuOr-5-1}{RGB}{230,97,1}
\definecolor{PuOr-5-C}{RGB}{230,97,1}
\definecolor{PuOr-5-2}{RGB}{253,184,99}
\definecolor{PuOr-5-F}{RGB}{253,184,99}
\definecolor{PuOr-5-3}{RGB}{247,247,247}
\definecolor{PuOr-5-H}{RGB}{247,247,247}
\definecolor{PuOr-5-4}{RGB}{178,171,210}
\definecolor{PuOr-5-J}{RGB}{178,171,210}
\definecolor{PuOr-5-5}{RGB}{94,60,153}
\definecolor{PuOr-5-M}{RGB}{94,60,153}
\definecolor{PuOr-6-1}{RGB}{179,88,6}
\definecolor{PuOr-6-B}{RGB}{179,88,6}
\definecolor{PuOr-6-2}{RGB}{241,163,64}
\definecolor{PuOr-6-E}{RGB}{241,163,64}
\definecolor{PuOr-6-3}{RGB}{254,224,182}
\definecolor{PuOr-6-G}{RGB}{254,224,182}
\definecolor{PuOr-6-4}{RGB}{216,218,235}
\definecolor{PuOr-6-I}{RGB}{216,218,235}
\definecolor{PuOr-6-5}{RGB}{153,142,195}
\definecolor{PuOr-6-K}{RGB}{153,142,195}
\definecolor{PuOr-6-6}{RGB}{84,39,136}
\definecolor{PuOr-6-N}{RGB}{84,39,136}
\definecolor{PuOr-7-1}{RGB}{179,88,6}
\definecolor{PuOr-7-B}{RGB}{179,88,6}
\definecolor{PuOr-7-2}{RGB}{241,163,64}
\definecolor{PuOr-7-E}{RGB}{241,163,64}
\definecolor{PuOr-7-3}{RGB}{254,224,182}
\definecolor{PuOr-7-G}{RGB}{254,224,182}
\definecolor{PuOr-7-4}{RGB}{247,247,247}
\definecolor{PuOr-7-H}{RGB}{247,247,247}
\definecolor{PuOr-7-5}{RGB}{216,218,235}
\definecolor{PuOr-7-I}{RGB}{216,218,235}
\definecolor{PuOr-7-6}{RGB}{153,142,195}
\definecolor{PuOr-7-K}{RGB}{153,142,195}
\definecolor{PuOr-7-7}{RGB}{84,39,136}
\definecolor{PuOr-7-N}{RGB}{84,39,136}
\definecolor{PuOr-8-1}{RGB}{179,88,6}
\definecolor{PuOr-8-B}{RGB}{179,88,6}
\definecolor{PuOr-8-2}{RGB}{224,130,20}
\definecolor{PuOr-8-D}{RGB}{224,130,20}
\definecolor{PuOr-8-3}{RGB}{253,184,99}
\definecolor{PuOr-8-F}{RGB}{253,184,99}
\definecolor{PuOr-8-4}{RGB}{254,224,182}
\definecolor{PuOr-8-G}{RGB}{254,224,182}
\definecolor{PuOr-8-5}{RGB}{216,218,235}
\definecolor{PuOr-8-I}{RGB}{216,218,235}
\definecolor{PuOr-8-6}{RGB}{178,171,210}
\definecolor{PuOr-8-J}{RGB}{178,171,210}
\definecolor{PuOr-8-7}{RGB}{128,115,172}
\definecolor{PuOr-8-L}{RGB}{128,115,172}
\definecolor{PuOr-8-8}{RGB}{84,39,136}
\definecolor{PuOr-8-N}{RGB}{84,39,136}
\definecolor{PuOr-9-1}{RGB}{179,88,6}
\definecolor{PuOr-9-B}{RGB}{179,88,6}
\definecolor{PuOr-9-2}{RGB}{224,130,20}
\definecolor{PuOr-9-D}{RGB}{224,130,20}
\definecolor{PuOr-9-3}{RGB}{253,184,99}
\definecolor{PuOr-9-F}{RGB}{253,184,99}
\definecolor{PuOr-9-4}{RGB}{254,224,182}
\definecolor{PuOr-9-G}{RGB}{254,224,182}
\definecolor{PuOr-9-5}{RGB}{247,247,247}
\definecolor{PuOr-9-H}{RGB}{247,247,247}
\definecolor{PuOr-9-6}{RGB}{216,218,235}
\definecolor{PuOr-9-I}{RGB}{216,218,235}
\definecolor{PuOr-9-7}{RGB}{178,171,210}
\definecolor{PuOr-9-J}{RGB}{178,171,210}
\definecolor{PuOr-9-8}{RGB}{128,115,172}
\definecolor{PuOr-9-L}{RGB}{128,115,172}
\definecolor{PuOr-9-9}{RGB}{84,39,136}
\definecolor{PuOr-9-N}{RGB}{84,39,136}
\definecolor{PuOr-10-1}{RGB}{127,59,8}
\definecolor{PuOr-10-A}{RGB}{127,59,8}
\definecolor{PuOr-10-2}{RGB}{179,88,6}
\definecolor{PuOr-10-B}{RGB}{179,88,6}
\definecolor{PuOr-10-3}{RGB}{224,130,20}
\definecolor{PuOr-10-D}{RGB}{224,130,20}
\definecolor{PuOr-10-4}{RGB}{253,184,99}
\definecolor{PuOr-10-F}{RGB}{253,184,99}
\definecolor{PuOr-10-5}{RGB}{254,224,182}
\definecolor{PuOr-10-G}{RGB}{254,224,182}
\definecolor{PuOr-10-6}{RGB}{216,218,235}
\definecolor{PuOr-10-I}{RGB}{216,218,235}
\definecolor{PuOr-10-7}{RGB}{178,171,210}
\definecolor{PuOr-10-J}{RGB}{178,171,210}
\definecolor{PuOr-10-8}{RGB}{128,115,172}
\definecolor{PuOr-10-L}{RGB}{128,115,172}
\definecolor{PuOr-10-9}{RGB}{84,39,136}
\definecolor{PuOr-10-N}{RGB}{84,39,136}
\definecolor{PuOr-10-10}{RGB}{45,0,75}
\definecolor{PuOr-10-O}{RGB}{45,0,75}
\definecolor{PuOr-11-1}{RGB}{127,59,8}
\definecolor{PuOr-11-A}{RGB}{127,59,8}
\definecolor{PuOr-11-2}{RGB}{179,88,6}
\definecolor{PuOr-11-B}{RGB}{179,88,6}
\definecolor{PuOr-11-3}{RGB}{224,130,20}
\definecolor{PuOr-11-D}{RGB}{224,130,20}
\definecolor{PuOr-11-4}{RGB}{253,184,99}
\definecolor{PuOr-11-F}{RGB}{253,184,99}
\definecolor{PuOr-11-5}{RGB}{254,224,182}
\definecolor{PuOr-11-G}{RGB}{254,224,182}
\definecolor{PuOr-11-6}{RGB}{247,247,247}
\definecolor{PuOr-11-H}{RGB}{247,247,247}
\definecolor{PuOr-11-7}{RGB}{216,218,235}
\definecolor{PuOr-11-I}{RGB}{216,218,235}
\definecolor{PuOr-11-8}{RGB}{178,171,210}
\definecolor{PuOr-11-J}{RGB}{178,171,210}
\definecolor{PuOr-11-9}{RGB}{128,115,172}
\definecolor{PuOr-11-L}{RGB}{128,115,172}
\definecolor{PuOr-11-10}{RGB}{84,39,136}
\definecolor{PuOr-11-N}{RGB}{84,39,136}
\definecolor{PuOr-11-11}{RGB}{45,0,75}
\definecolor{PuOr-11-O}{RGB}{45,0,75}
\definecolor{BrBG-3-1}{RGB}{216,179,101}
\definecolor{BrBG-3-E}{RGB}{216,179,101}
\definecolor{BrBG-3-2}{RGB}{245,245,245}
\definecolor{BrBG-3-H}{RGB}{245,245,245}
\definecolor{BrBG-3-3}{RGB}{90,180,172}
\definecolor{BrBG-3-K}{RGB}{90,180,172}
\definecolor{BrBG-4-1}{RGB}{166,97,26}
\definecolor{BrBG-4-C}{RGB}{166,97,26}
\definecolor{BrBG-4-2}{RGB}{223,194,125}
\definecolor{BrBG-4-F}{RGB}{223,194,125}
\definecolor{BrBG-4-3}{RGB}{128,205,193}
\definecolor{BrBG-4-J}{RGB}{128,205,193}
\definecolor{BrBG-4-4}{RGB}{1,133,113}
\definecolor{BrBG-4-M}{RGB}{1,133,113}
\definecolor{BrBG-5-1}{RGB}{166,97,26}
\definecolor{BrBG-5-C}{RGB}{166,97,26}
\definecolor{BrBG-5-2}{RGB}{223,194,125}
\definecolor{BrBG-5-F}{RGB}{223,194,125}
\definecolor{BrBG-5-3}{RGB}{245,245,245}
\definecolor{BrBG-5-H}{RGB}{245,245,245}
\definecolor{BrBG-5-4}{RGB}{128,205,193}
\definecolor{BrBG-5-J}{RGB}{128,205,193}
\definecolor{BrBG-5-5}{RGB}{1,133,113}
\definecolor{BrBG-5-M}{RGB}{1,133,113}
\definecolor{BrBG-6-1}{RGB}{140,81,10}
\definecolor{BrBG-6-B}{RGB}{140,81,10}
\definecolor{BrBG-6-2}{RGB}{216,179,101}
\definecolor{BrBG-6-E}{RGB}{216,179,101}
\definecolor{BrBG-6-3}{RGB}{246,232,195}
\definecolor{BrBG-6-G}{RGB}{246,232,195}
\definecolor{BrBG-6-4}{RGB}{199,234,229}
\definecolor{BrBG-6-I}{RGB}{199,234,229}
\definecolor{BrBG-6-5}{RGB}{90,180,172}
\definecolor{BrBG-6-K}{RGB}{90,180,172}
\definecolor{BrBG-6-6}{RGB}{1,102,94}
\definecolor{BrBG-6-N}{RGB}{1,102,94}
\definecolor{BrBG-7-1}{RGB}{140,81,10}
\definecolor{BrBG-7-B}{RGB}{140,81,10}
\definecolor{BrBG-7-2}{RGB}{216,179,101}
\definecolor{BrBG-7-E}{RGB}{216,179,101}
\definecolor{BrBG-7-3}{RGB}{246,232,195}
\definecolor{BrBG-7-G}{RGB}{246,232,195}
\definecolor{BrBG-7-4}{RGB}{245,245,245}
\definecolor{BrBG-7-H}{RGB}{245,245,245}
\definecolor{BrBG-7-5}{RGB}{199,234,229}
\definecolor{BrBG-7-I}{RGB}{199,234,229}
\definecolor{BrBG-7-6}{RGB}{90,180,172}
\definecolor{BrBG-7-K}{RGB}{90,180,172}
\definecolor{BrBG-7-7}{RGB}{1,102,94}
\definecolor{BrBG-7-N}{RGB}{1,102,94}
\definecolor{BrBG-8-1}{RGB}{140,81,10}
\definecolor{BrBG-8-B}{RGB}{140,81,10}
\definecolor{BrBG-8-2}{RGB}{191,129,45}
\definecolor{BrBG-8-D}{RGB}{191,129,45}
\definecolor{BrBG-8-3}{RGB}{223,194,125}
\definecolor{BrBG-8-F}{RGB}{223,194,125}
\definecolor{BrBG-8-4}{RGB}{246,232,195}
\definecolor{BrBG-8-G}{RGB}{246,232,195}
\definecolor{BrBG-8-5}{RGB}{199,234,229}
\definecolor{BrBG-8-I}{RGB}{199,234,229}
\definecolor{BrBG-8-6}{RGB}{128,205,193}
\definecolor{BrBG-8-J}{RGB}{128,205,193}
\definecolor{BrBG-8-7}{RGB}{53,151,143}
\definecolor{BrBG-8-L}{RGB}{53,151,143}
\definecolor{BrBG-8-8}{RGB}{1,102,94}
\definecolor{BrBG-8-N}{RGB}{1,102,94}
\definecolor{BrBG-9-1}{RGB}{140,81,10}
\definecolor{BrBG-9-B}{RGB}{140,81,10}
\definecolor{BrBG-9-2}{RGB}{191,129,45}
\definecolor{BrBG-9-D}{RGB}{191,129,45}
\definecolor{BrBG-9-3}{RGB}{223,194,125}
\definecolor{BrBG-9-F}{RGB}{223,194,125}
\definecolor{BrBG-9-4}{RGB}{246,232,195}
\definecolor{BrBG-9-G}{RGB}{246,232,195}
\definecolor{BrBG-9-5}{RGB}{245,245,245}
\definecolor{BrBG-9-H}{RGB}{245,245,245}
\definecolor{BrBG-9-6}{RGB}{199,234,229}
\definecolor{BrBG-9-I}{RGB}{199,234,229}
\definecolor{BrBG-9-7}{RGB}{128,205,193}
\definecolor{BrBG-9-J}{RGB}{128,205,193}
\definecolor{BrBG-9-8}{RGB}{53,151,143}
\definecolor{BrBG-9-L}{RGB}{53,151,143}
\definecolor{BrBG-9-9}{RGB}{1,102,94}
\definecolor{BrBG-9-N}{RGB}{1,102,94}
\definecolor{BrBG-10-1}{RGB}{84,48,5}
\definecolor{BrBG-10-A}{RGB}{84,48,5}
\definecolor{BrBG-10-2}{RGB}{140,81,10}
\definecolor{BrBG-10-B}{RGB}{140,81,10}
\definecolor{BrBG-10-3}{RGB}{191,129,45}
\definecolor{BrBG-10-D}{RGB}{191,129,45}
\definecolor{BrBG-10-4}{RGB}{223,194,125}
\definecolor{BrBG-10-F}{RGB}{223,194,125}
\definecolor{BrBG-10-5}{RGB}{246,232,195}
\definecolor{BrBG-10-G}{RGB}{246,232,195}
\definecolor{BrBG-10-6}{RGB}{199,234,229}
\definecolor{BrBG-10-I}{RGB}{199,234,229}
\definecolor{BrBG-10-7}{RGB}{128,205,193}
\definecolor{BrBG-10-J}{RGB}{128,205,193}
\definecolor{BrBG-10-8}{RGB}{53,151,143}
\definecolor{BrBG-10-L}{RGB}{53,151,143}
\definecolor{BrBG-10-9}{RGB}{1,102,94}
\definecolor{BrBG-10-N}{RGB}{1,102,94}
\definecolor{BrBG-10-10}{RGB}{0,60,48}
\definecolor{BrBG-10-O}{RGB}{0,60,48}
\definecolor{BrBG-11-1}{RGB}{84,48,5}
\definecolor{BrBG-11-A}{RGB}{84,48,5}
\definecolor{BrBG-11-2}{RGB}{140,81,10}
\definecolor{BrBG-11-B}{RGB}{140,81,10}
\definecolor{BrBG-11-3}{RGB}{191,129,45}
\definecolor{BrBG-11-D}{RGB}{191,129,45}
\definecolor{BrBG-11-4}{RGB}{223,194,125}
\definecolor{BrBG-11-F}{RGB}{223,194,125}
\definecolor{BrBG-11-5}{RGB}{246,232,195}
\definecolor{BrBG-11-G}{RGB}{246,232,195}
\definecolor{BrBG-11-6}{RGB}{245,245,245}
\definecolor{BrBG-11-H}{RGB}{245,245,245}
\definecolor{BrBG-11-7}{RGB}{199,234,229}
\definecolor{BrBG-11-I}{RGB}{199,234,229}
\definecolor{BrBG-11-8}{RGB}{128,205,193}
\definecolor{BrBG-11-J}{RGB}{128,205,193}
\definecolor{BrBG-11-9}{RGB}{53,151,143}
\definecolor{BrBG-11-L}{RGB}{53,151,143}
\definecolor{BrBG-11-10}{RGB}{1,102,94}
\definecolor{BrBG-11-N}{RGB}{1,102,94}
\definecolor{BrBG-11-11}{RGB}{0,60,48}
\definecolor{BrBG-11-O}{RGB}{0,60,48}
\definecolor{PRGn-3-1}{RGB}{175,141,195}
\definecolor{PRGn-3-E}{RGB}{175,141,195}
\definecolor{PRGn-3-2}{RGB}{247,247,247}
\definecolor{PRGn-3-H}{RGB}{247,247,247}
\definecolor{PRGn-3-3}{RGB}{127,191,123}
\definecolor{PRGn-3-K}{RGB}{127,191,123}
\definecolor{PRGn-4-1}{RGB}{123,50,148}
\definecolor{PRGn-4-C}{RGB}{123,50,148}
\definecolor{PRGn-4-2}{RGB}{194,165,207}
\definecolor{PRGn-4-F}{RGB}{194,165,207}
\definecolor{PRGn-4-3}{RGB}{166,219,160}
\definecolor{PRGn-4-J}{RGB}{166,219,160}
\definecolor{PRGn-4-4}{RGB}{0,136,55}
\definecolor{PRGn-4-M}{RGB}{0,136,55}
\definecolor{PRGn-5-1}{RGB}{123,50,148}
\definecolor{PRGn-5-C}{RGB}{123,50,148}
\definecolor{PRGn-5-2}{RGB}{194,165,207}
\definecolor{PRGn-5-F}{RGB}{194,165,207}
\definecolor{PRGn-5-3}{RGB}{247,247,247}
\definecolor{PRGn-5-H}{RGB}{247,247,247}
\definecolor{PRGn-5-4}{RGB}{166,219,160}
\definecolor{PRGn-5-J}{RGB}{166,219,160}
\definecolor{PRGn-5-5}{RGB}{0,136,55}
\definecolor{PRGn-5-M}{RGB}{0,136,55}
\definecolor{PRGn-6-1}{RGB}{118,42,131}
\definecolor{PRGn-6-B}{RGB}{118,42,131}
\definecolor{PRGn-6-2}{RGB}{175,141,195}
\definecolor{PRGn-6-E}{RGB}{175,141,195}
\definecolor{PRGn-6-3}{RGB}{231,212,232}
\definecolor{PRGn-6-G}{RGB}{231,212,232}
\definecolor{PRGn-6-4}{RGB}{217,240,211}
\definecolor{PRGn-6-I}{RGB}{217,240,211}
\definecolor{PRGn-6-5}{RGB}{127,191,123}
\definecolor{PRGn-6-K}{RGB}{127,191,123}
\definecolor{PRGn-6-6}{RGB}{27,120,55}
\definecolor{PRGn-6-N}{RGB}{27,120,55}
\definecolor{PRGn-7-1}{RGB}{118,42,131}
\definecolor{PRGn-7-B}{RGB}{118,42,131}
\definecolor{PRGn-7-2}{RGB}{175,141,195}
\definecolor{PRGn-7-E}{RGB}{175,141,195}
\definecolor{PRGn-7-3}{RGB}{231,212,232}
\definecolor{PRGn-7-G}{RGB}{231,212,232}
\definecolor{PRGn-7-4}{RGB}{247,247,247}
\definecolor{PRGn-7-H}{RGB}{247,247,247}
\definecolor{PRGn-7-5}{RGB}{217,240,211}
\definecolor{PRGn-7-I}{RGB}{217,240,211}
\definecolor{PRGn-7-6}{RGB}{127,191,123}
\definecolor{PRGn-7-K}{RGB}{127,191,123}
\definecolor{PRGn-7-7}{RGB}{27,120,55}
\definecolor{PRGn-7-N}{RGB}{27,120,55}
\definecolor{PRGn-8-1}{RGB}{118,42,131}
\definecolor{PRGn-8-B}{RGB}{118,42,131}
\definecolor{PRGn-8-2}{RGB}{153,112,171}
\definecolor{PRGn-8-D}{RGB}{153,112,171}
\definecolor{PRGn-8-3}{RGB}{194,165,207}
\definecolor{PRGn-8-F}{RGB}{194,165,207}
\definecolor{PRGn-8-4}{RGB}{231,212,232}
\definecolor{PRGn-8-G}{RGB}{231,212,232}
\definecolor{PRGn-8-5}{RGB}{217,240,211}
\definecolor{PRGn-8-I}{RGB}{217,240,211}
\definecolor{PRGn-8-6}{RGB}{166,219,160}
\definecolor{PRGn-8-J}{RGB}{166,219,160}
\definecolor{PRGn-8-7}{RGB}{90,174,97}
\definecolor{PRGn-8-L}{RGB}{90,174,97}
\definecolor{PRGn-8-8}{RGB}{27,120,55}
\definecolor{PRGn-8-N}{RGB}{27,120,55}
\definecolor{PRGn-9-1}{RGB}{118,42,131}
\definecolor{PRGn-9-B}{RGB}{118,42,131}
\definecolor{PRGn-9-2}{RGB}{153,112,171}
\definecolor{PRGn-9-D}{RGB}{153,112,171}
\definecolor{PRGn-9-3}{RGB}{194,165,207}
\definecolor{PRGn-9-F}{RGB}{194,165,207}
\definecolor{PRGn-9-4}{RGB}{231,212,232}
\definecolor{PRGn-9-G}{RGB}{231,212,232}
\definecolor{PRGn-9-5}{RGB}{247,247,247}
\definecolor{PRGn-9-H}{RGB}{247,247,247}
\definecolor{PRGn-9-6}{RGB}{217,240,211}
\definecolor{PRGn-9-I}{RGB}{217,240,211}
\definecolor{PRGn-9-7}{RGB}{166,219,160}
\definecolor{PRGn-9-J}{RGB}{166,219,160}
\definecolor{PRGn-9-8}{RGB}{90,174,97}
\definecolor{PRGn-9-L}{RGB}{90,174,97}
\definecolor{PRGn-9-9}{RGB}{27,120,55}
\definecolor{PRGn-9-N}{RGB}{27,120,55}
\definecolor{PRGn-10-1}{RGB}{64,0,75}
\definecolor{PRGn-10-A}{RGB}{64,0,75}
\definecolor{PRGn-10-2}{RGB}{118,42,131}
\definecolor{PRGn-10-B}{RGB}{118,42,131}
\definecolor{PRGn-10-3}{RGB}{153,112,171}
\definecolor{PRGn-10-D}{RGB}{153,112,171}
\definecolor{PRGn-10-4}{RGB}{194,165,207}
\definecolor{PRGn-10-F}{RGB}{194,165,207}
\definecolor{PRGn-10-5}{RGB}{231,212,232}
\definecolor{PRGn-10-G}{RGB}{231,212,232}
\definecolor{PRGn-10-6}{RGB}{217,240,211}
\definecolor{PRGn-10-I}{RGB}{217,240,211}
\definecolor{PRGn-10-7}{RGB}{166,219,160}
\definecolor{PRGn-10-J}{RGB}{166,219,160}
\definecolor{PRGn-10-8}{RGB}{90,174,97}
\definecolor{PRGn-10-L}{RGB}{90,174,97}
\definecolor{PRGn-10-9}{RGB}{27,120,55}
\definecolor{PRGn-10-N}{RGB}{27,120,55}
\definecolor{PRGn-10-10}{RGB}{0,68,27}
\definecolor{PRGn-10-O}{RGB}{0,68,27}
\definecolor{PRGn-11-1}{RGB}{64,0,75}
\definecolor{PRGn-11-A}{RGB}{64,0,75}
\definecolor{PRGn-11-2}{RGB}{118,42,131}
\definecolor{PRGn-11-B}{RGB}{118,42,131}
\definecolor{PRGn-11-3}{RGB}{153,112,171}
\definecolor{PRGn-11-D}{RGB}{153,112,171}
\definecolor{PRGn-11-4}{RGB}{194,165,207}
\definecolor{PRGn-11-F}{RGB}{194,165,207}
\definecolor{PRGn-11-5}{RGB}{231,212,232}
\definecolor{PRGn-11-G}{RGB}{231,212,232}
\definecolor{PRGn-11-6}{RGB}{247,247,247}
\definecolor{PRGn-11-H}{RGB}{247,247,247}
\definecolor{PRGn-11-7}{RGB}{217,240,211}
\definecolor{PRGn-11-I}{RGB}{217,240,211}
\definecolor{PRGn-11-8}{RGB}{166,219,160}
\definecolor{PRGn-11-J}{RGB}{166,219,160}
\definecolor{PRGn-11-9}{RGB}{90,174,97}
\definecolor{PRGn-11-L}{RGB}{90,174,97}
\definecolor{PRGn-11-10}{RGB}{27,120,55}
\definecolor{PRGn-11-N}{RGB}{27,120,55}
\definecolor{PRGn-11-11}{RGB}{0,68,27}
\definecolor{PRGn-11-O}{RGB}{0,68,27}
\definecolor{PiYG-3-1}{RGB}{233,163,201}
\definecolor{PiYG-3-E}{RGB}{233,163,201}
\definecolor{PiYG-3-2}{RGB}{247,247,247}
\definecolor{PiYG-3-H}{RGB}{247,247,247}
\definecolor{PiYG-3-3}{RGB}{161,215,106}
\definecolor{PiYG-3-K}{RGB}{161,215,106}
\definecolor{PiYG-4-1}{RGB}{208,28,139}
\definecolor{PiYG-4-C}{RGB}{208,28,139}
\definecolor{PiYG-4-2}{RGB}{241,182,218}
\definecolor{PiYG-4-F}{RGB}{241,182,218}
\definecolor{PiYG-4-3}{RGB}{184,225,134}
\definecolor{PiYG-4-J}{RGB}{184,225,134}
\definecolor{PiYG-4-4}{RGB}{77,172,38}
\definecolor{PiYG-4-M}{RGB}{77,172,38}
\definecolor{PiYG-5-1}{RGB}{208,28,139}
\definecolor{PiYG-5-C}{RGB}{208,28,139}
\definecolor{PiYG-5-2}{RGB}{241,182,218}
\definecolor{PiYG-5-F}{RGB}{241,182,218}
\definecolor{PiYG-5-3}{RGB}{247,247,247}
\definecolor{PiYG-5-H}{RGB}{247,247,247}
\definecolor{PiYG-5-4}{RGB}{184,225,134}
\definecolor{PiYG-5-J}{RGB}{184,225,134}
\definecolor{PiYG-5-5}{RGB}{77,172,38}
\definecolor{PiYG-5-M}{RGB}{77,172,38}
\definecolor{PiYG-6-1}{RGB}{197,27,125}
\definecolor{PiYG-6-B}{RGB}{197,27,125}
\definecolor{PiYG-6-2}{RGB}{233,163,201}
\definecolor{PiYG-6-E}{RGB}{233,163,201}
\definecolor{PiYG-6-3}{RGB}{253,224,239}
\definecolor{PiYG-6-G}{RGB}{253,224,239}
\definecolor{PiYG-6-4}{RGB}{230,245,208}
\definecolor{PiYG-6-I}{RGB}{230,245,208}
\definecolor{PiYG-6-5}{RGB}{161,215,106}
\definecolor{PiYG-6-K}{RGB}{161,215,106}
\definecolor{PiYG-6-6}{RGB}{77,146,33}
\definecolor{PiYG-6-N}{RGB}{77,146,33}
\definecolor{PiYG-7-1}{RGB}{197,27,125}
\definecolor{PiYG-7-B}{RGB}{197,27,125}
\definecolor{PiYG-7-2}{RGB}{233,163,201}
\definecolor{PiYG-7-E}{RGB}{233,163,201}
\definecolor{PiYG-7-3}{RGB}{253,224,239}
\definecolor{PiYG-7-G}{RGB}{253,224,239}
\definecolor{PiYG-7-4}{RGB}{247,247,247}
\definecolor{PiYG-7-H}{RGB}{247,247,247}
\definecolor{PiYG-7-5}{RGB}{230,245,208}
\definecolor{PiYG-7-I}{RGB}{230,245,208}
\definecolor{PiYG-7-6}{RGB}{161,215,106}
\definecolor{PiYG-7-K}{RGB}{161,215,106}
\definecolor{PiYG-7-7}{RGB}{77,146,33}
\definecolor{PiYG-7-N}{RGB}{77,146,33}
\definecolor{PiYG-8-1}{RGB}{197,27,125}
\definecolor{PiYG-8-B}{RGB}{197,27,125}
\definecolor{PiYG-8-2}{RGB}{222,119,174}
\definecolor{PiYG-8-D}{RGB}{222,119,174}
\definecolor{PiYG-8-3}{RGB}{241,182,218}
\definecolor{PiYG-8-F}{RGB}{241,182,218}
\definecolor{PiYG-8-4}{RGB}{253,224,239}
\definecolor{PiYG-8-G}{RGB}{253,224,239}
\definecolor{PiYG-8-5}{RGB}{230,245,208}
\definecolor{PiYG-8-I}{RGB}{230,245,208}
\definecolor{PiYG-8-6}{RGB}{184,225,134}
\definecolor{PiYG-8-J}{RGB}{184,225,134}
\definecolor{PiYG-8-7}{RGB}{127,188,65}
\definecolor{PiYG-8-L}{RGB}{127,188,65}
\definecolor{PiYG-8-8}{RGB}{77,146,33}
\definecolor{PiYG-8-N}{RGB}{77,146,33}
\definecolor{PiYG-9-1}{RGB}{197,27,125}
\definecolor{PiYG-9-B}{RGB}{197,27,125}
\definecolor{PiYG-9-2}{RGB}{222,119,174}
\definecolor{PiYG-9-D}{RGB}{222,119,174}
\definecolor{PiYG-9-3}{RGB}{241,182,218}
\definecolor{PiYG-9-F}{RGB}{241,182,218}
\definecolor{PiYG-9-4}{RGB}{253,224,239}
\definecolor{PiYG-9-G}{RGB}{253,224,239}
\definecolor{PiYG-9-5}{RGB}{247,247,247}
\definecolor{PiYG-9-H}{RGB}{247,247,247}
\definecolor{PiYG-9-6}{RGB}{230,245,208}
\definecolor{PiYG-9-I}{RGB}{230,245,208}
\definecolor{PiYG-9-7}{RGB}{184,225,134}
\definecolor{PiYG-9-J}{RGB}{184,225,134}
\definecolor{PiYG-9-8}{RGB}{127,188,65}
\definecolor{PiYG-9-L}{RGB}{127,188,65}
\definecolor{PiYG-9-9}{RGB}{77,146,33}
\definecolor{PiYG-9-N}{RGB}{77,146,33}
\definecolor{PiYG-10-1}{RGB}{142,1,82}
\definecolor{PiYG-10-A}{RGB}{142,1,82}
\definecolor{PiYG-10-2}{RGB}{197,27,125}
\definecolor{PiYG-10-B}{RGB}{197,27,125}
\definecolor{PiYG-10-3}{RGB}{222,119,174}
\definecolor{PiYG-10-D}{RGB}{222,119,174}
\definecolor{PiYG-10-4}{RGB}{241,182,218}
\definecolor{PiYG-10-F}{RGB}{241,182,218}
\definecolor{PiYG-10-5}{RGB}{253,224,239}
\definecolor{PiYG-10-G}{RGB}{253,224,239}
\definecolor{PiYG-10-6}{RGB}{230,245,208}
\definecolor{PiYG-10-I}{RGB}{230,245,208}
\definecolor{PiYG-10-7}{RGB}{184,225,134}
\definecolor{PiYG-10-J}{RGB}{184,225,134}
\definecolor{PiYG-10-8}{RGB}{127,188,65}
\definecolor{PiYG-10-L}{RGB}{127,188,65}
\definecolor{PiYG-10-9}{RGB}{77,146,33}
\definecolor{PiYG-10-N}{RGB}{77,146,33}
\definecolor{PiYG-10-10}{RGB}{39,100,25}
\definecolor{PiYG-10-O}{RGB}{39,100,25}
\definecolor{PiYG-11-1}{RGB}{142,1,82}
\definecolor{PiYG-11-A}{RGB}{142,1,82}
\definecolor{PiYG-11-2}{RGB}{197,27,125}
\definecolor{PiYG-11-B}{RGB}{197,27,125}
\definecolor{PiYG-11-3}{RGB}{222,119,174}
\definecolor{PiYG-11-D}{RGB}{222,119,174}
\definecolor{PiYG-11-4}{RGB}{241,182,218}
\definecolor{PiYG-11-F}{RGB}{241,182,218}
\definecolor{PiYG-11-5}{RGB}{253,224,239}
\definecolor{PiYG-11-G}{RGB}{253,224,239}
\definecolor{PiYG-11-6}{RGB}{247,247,247}
\definecolor{PiYG-11-H}{RGB}{247,247,247}
\definecolor{PiYG-11-7}{RGB}{230,245,208}
\definecolor{PiYG-11-I}{RGB}{230,245,208}
\definecolor{PiYG-11-8}{RGB}{184,225,134}
\definecolor{PiYG-11-J}{RGB}{184,225,134}
\definecolor{PiYG-11-9}{RGB}{127,188,65}
\definecolor{PiYG-11-L}{RGB}{127,188,65}
\definecolor{PiYG-11-10}{RGB}{77,146,33}
\definecolor{PiYG-11-N}{RGB}{77,146,33}
\definecolor{PiYG-11-11}{RGB}{39,100,25}
\definecolor{PiYG-11-O}{RGB}{39,100,25}
\definecolor{RdBu-3-1}{RGB}{239,138,98}
\definecolor{RdBu-3-E}{RGB}{239,138,98}
\definecolor{RdBu-3-2}{RGB}{247,247,247}
\definecolor{RdBu-3-H}{RGB}{247,247,247}
\definecolor{RdBu-3-3}{RGB}{103,169,207}
\definecolor{RdBu-3-K}{RGB}{103,169,207}
\definecolor{RdBu-4-1}{RGB}{202,0,32}
\definecolor{RdBu-4-C}{RGB}{202,0,32}
\definecolor{RdBu-4-2}{RGB}{244,165,130}
\definecolor{RdBu-4-F}{RGB}{244,165,130}
\definecolor{RdBu-4-3}{RGB}{146,197,222}
\definecolor{RdBu-4-J}{RGB}{146,197,222}
\definecolor{RdBu-4-4}{RGB}{5,113,176}
\definecolor{RdBu-4-M}{RGB}{5,113,176}
\definecolor{RdBu-5-1}{RGB}{202,0,32}
\definecolor{RdBu-5-C}{RGB}{202,0,32}
\definecolor{RdBu-5-2}{RGB}{244,165,130}
\definecolor{RdBu-5-F}{RGB}{244,165,130}
\definecolor{RdBu-5-3}{RGB}{247,247,247}
\definecolor{RdBu-5-H}{RGB}{247,247,247}
\definecolor{RdBu-5-4}{RGB}{146,197,222}
\definecolor{RdBu-5-J}{RGB}{146,197,222}
\definecolor{RdBu-5-5}{RGB}{5,113,176}
\definecolor{RdBu-5-M}{RGB}{5,113,176}
\definecolor{RdBu-6-1}{RGB}{178,24,43}
\definecolor{RdBu-6-B}{RGB}{178,24,43}
\definecolor{RdBu-6-2}{RGB}{239,138,98}
\definecolor{RdBu-6-E}{RGB}{239,138,98}
\definecolor{RdBu-6-3}{RGB}{253,219,199}
\definecolor{RdBu-6-G}{RGB}{253,219,199}
\definecolor{RdBu-6-4}{RGB}{209,229,240}
\definecolor{RdBu-6-I}{RGB}{209,229,240}
\definecolor{RdBu-6-5}{RGB}{103,169,207}
\definecolor{RdBu-6-K}{RGB}{103,169,207}
\definecolor{RdBu-6-6}{RGB}{33,102,172}
\definecolor{RdBu-6-N}{RGB}{33,102,172}
\definecolor{RdBu-7-1}{RGB}{178,24,43}
\definecolor{RdBu-7-B}{RGB}{178,24,43}
\definecolor{RdBu-7-2}{RGB}{239,138,98}
\definecolor{RdBu-7-E}{RGB}{239,138,98}
\definecolor{RdBu-7-3}{RGB}{253,219,199}
\definecolor{RdBu-7-G}{RGB}{253,219,199}
\definecolor{RdBu-7-4}{RGB}{247,247,247}
\definecolor{RdBu-7-H}{RGB}{247,247,247}
\definecolor{RdBu-7-5}{RGB}{209,229,240}
\definecolor{RdBu-7-I}{RGB}{209,229,240}
\definecolor{RdBu-7-6}{RGB}{103,169,207}
\definecolor{RdBu-7-K}{RGB}{103,169,207}
\definecolor{RdBu-7-7}{RGB}{33,102,172}
\definecolor{RdBu-7-N}{RGB}{33,102,172}
\definecolor{RdBu-8-1}{RGB}{178,24,43}
\definecolor{RdBu-8-B}{RGB}{178,24,43}
\definecolor{RdBu-8-2}{RGB}{214,96,77}
\definecolor{RdBu-8-D}{RGB}{214,96,77}
\definecolor{RdBu-8-3}{RGB}{244,165,130}
\definecolor{RdBu-8-F}{RGB}{244,165,130}
\definecolor{RdBu-8-4}{RGB}{253,219,199}
\definecolor{RdBu-8-G}{RGB}{253,219,199}
\definecolor{RdBu-8-5}{RGB}{209,229,240}
\definecolor{RdBu-8-I}{RGB}{209,229,240}
\definecolor{RdBu-8-6}{RGB}{146,197,222}
\definecolor{RdBu-8-J}{RGB}{146,197,222}
\definecolor{RdBu-8-7}{RGB}{67,147,195}
\definecolor{RdBu-8-L}{RGB}{67,147,195}
\definecolor{RdBu-8-8}{RGB}{33,102,172}
\definecolor{RdBu-8-N}{RGB}{33,102,172}
\definecolor{RdBu-9-1}{RGB}{178,24,43}
\definecolor{RdBu-9-B}{RGB}{178,24,43}
\definecolor{RdBu-9-2}{RGB}{214,96,77}
\definecolor{RdBu-9-D}{RGB}{214,96,77}
\definecolor{RdBu-9-3}{RGB}{244,165,130}
\definecolor{RdBu-9-F}{RGB}{244,165,130}
\definecolor{RdBu-9-4}{RGB}{253,219,199}
\definecolor{RdBu-9-G}{RGB}{253,219,199}
\definecolor{RdBu-9-5}{RGB}{247,247,247}
\definecolor{RdBu-9-H}{RGB}{247,247,247}
\definecolor{RdBu-9-6}{RGB}{209,229,240}
\definecolor{RdBu-9-I}{RGB}{209,229,240}
\definecolor{RdBu-9-7}{RGB}{146,197,222}
\definecolor{RdBu-9-J}{RGB}{146,197,222}
\definecolor{RdBu-9-8}{RGB}{67,147,195}
\definecolor{RdBu-9-L}{RGB}{67,147,195}
\definecolor{RdBu-9-9}{RGB}{33,102,172}
\definecolor{RdBu-9-N}{RGB}{33,102,172}
\definecolor{RdBu-10-1}{RGB}{103,0,31}
\definecolor{RdBu-10-A}{RGB}{103,0,31}
\definecolor{RdBu-10-2}{RGB}{178,24,43}
\definecolor{RdBu-10-B}{RGB}{178,24,43}
\definecolor{RdBu-10-3}{RGB}{214,96,77}
\definecolor{RdBu-10-D}{RGB}{214,96,77}
\definecolor{RdBu-10-4}{RGB}{244,165,130}
\definecolor{RdBu-10-F}{RGB}{244,165,130}
\definecolor{RdBu-10-5}{RGB}{253,219,199}
\definecolor{RdBu-10-G}{RGB}{253,219,199}
\definecolor{RdBu-10-6}{RGB}{209,229,240}
\definecolor{RdBu-10-I}{RGB}{209,229,240}
\definecolor{RdBu-10-7}{RGB}{146,197,222}
\definecolor{RdBu-10-J}{RGB}{146,197,222}
\definecolor{RdBu-10-8}{RGB}{67,147,195}
\definecolor{RdBu-10-L}{RGB}{67,147,195}
\definecolor{RdBu-10-9}{RGB}{33,102,172}
\definecolor{RdBu-10-N}{RGB}{33,102,172}
\definecolor{RdBu-10-10}{RGB}{5,48,97}
\definecolor{RdBu-10-O}{RGB}{5,48,97}
\definecolor{RdBu-11-1}{RGB}{103,0,31}
\definecolor{RdBu-11-A}{RGB}{103,0,31}
\definecolor{RdBu-11-2}{RGB}{178,24,43}
\definecolor{RdBu-11-B}{RGB}{178,24,43}
\definecolor{RdBu-11-3}{RGB}{214,96,77}
\definecolor{RdBu-11-D}{RGB}{214,96,77}
\definecolor{RdBu-11-4}{RGB}{244,165,130}
\definecolor{RdBu-11-F}{RGB}{244,165,130}
\definecolor{RdBu-11-5}{RGB}{253,219,199}
\definecolor{RdBu-11-G}{RGB}{253,219,199}
\definecolor{RdBu-11-6}{RGB}{247,247,247}
\definecolor{RdBu-11-H}{RGB}{247,247,247}
\definecolor{RdBu-11-7}{RGB}{209,229,240}
\definecolor{RdBu-11-I}{RGB}{209,229,240}
\definecolor{RdBu-11-8}{RGB}{146,197,222}
\definecolor{RdBu-11-J}{RGB}{146,197,222}
\definecolor{RdBu-11-9}{RGB}{67,147,195}
\definecolor{RdBu-11-L}{RGB}{67,147,195}
\definecolor{RdBu-11-10}{RGB}{33,102,172}
\definecolor{RdBu-11-N}{RGB}{33,102,172}
\definecolor{RdBu-11-11}{RGB}{5,48,97}
\definecolor{RdBu-11-O}{RGB}{5,48,97}
\definecolor{RdGy-3-1}{RGB}{239,138,98}
\definecolor{RdGy-3-E}{RGB}{239,138,98}
\definecolor{RdGy-3-2}{RGB}{255,255,255}
\definecolor{RdGy-3-H}{RGB}{255,255,255}
\definecolor{RdGy-3-3}{RGB}{153,153,153}
\definecolor{RdGy-3-K}{RGB}{153,153,153}
\definecolor{RdGy-4-1}{RGB}{202,0,32}
\definecolor{RdGy-4-C}{RGB}{202,0,32}
\definecolor{RdGy-4-2}{RGB}{244,165,130}
\definecolor{RdGy-4-F}{RGB}{244,165,130}
\definecolor{RdGy-4-3}{RGB}{186,186,186}
\definecolor{RdGy-4-J}{RGB}{186,186,186}
\definecolor{RdGy-4-4}{RGB}{64,64,64}
\definecolor{RdGy-4-M}{RGB}{64,64,64}
\definecolor{RdGy-5-1}{RGB}{202,0,32}
\definecolor{RdGy-5-C}{RGB}{202,0,32}
\definecolor{RdGy-5-2}{RGB}{244,165,130}
\definecolor{RdGy-5-F}{RGB}{244,165,130}
\definecolor{RdGy-5-3}{RGB}{255,255,255}
\definecolor{RdGy-5-H}{RGB}{255,255,255}
\definecolor{RdGy-5-4}{RGB}{186,186,186}
\definecolor{RdGy-5-J}{RGB}{186,186,186}
\definecolor{RdGy-5-5}{RGB}{64,64,64}
\definecolor{RdGy-5-M}{RGB}{64,64,64}
\definecolor{RdGy-6-1}{RGB}{178,24,43}
\definecolor{RdGy-6-B}{RGB}{178,24,43}
\definecolor{RdGy-6-2}{RGB}{239,138,98}
\definecolor{RdGy-6-E}{RGB}{239,138,98}
\definecolor{RdGy-6-3}{RGB}{253,219,199}
\definecolor{RdGy-6-G}{RGB}{253,219,199}
\definecolor{RdGy-6-4}{RGB}{224,224,224}
\definecolor{RdGy-6-I}{RGB}{224,224,224}
\definecolor{RdGy-6-5}{RGB}{153,153,153}
\definecolor{RdGy-6-K}{RGB}{153,153,153}
\definecolor{RdGy-6-6}{RGB}{77,77,77}
\definecolor{RdGy-6-N}{RGB}{77,77,77}
\definecolor{RdGy-7-1}{RGB}{178,24,43}
\definecolor{RdGy-7-B}{RGB}{178,24,43}
\definecolor{RdGy-7-2}{RGB}{239,138,98}
\definecolor{RdGy-7-E}{RGB}{239,138,98}
\definecolor{RdGy-7-3}{RGB}{253,219,199}
\definecolor{RdGy-7-G}{RGB}{253,219,199}
\definecolor{RdGy-7-4}{RGB}{255,255,255}
\definecolor{RdGy-7-H}{RGB}{255,255,255}
\definecolor{RdGy-7-5}{RGB}{224,224,224}
\definecolor{RdGy-7-I}{RGB}{224,224,224}
\definecolor{RdGy-7-6}{RGB}{153,153,153}
\definecolor{RdGy-7-K}{RGB}{153,153,153}
\definecolor{RdGy-7-7}{RGB}{77,77,77}
\definecolor{RdGy-7-N}{RGB}{77,77,77}
\definecolor{RdGy-8-1}{RGB}{178,24,43}
\definecolor{RdGy-8-B}{RGB}{178,24,43}
\definecolor{RdGy-8-2}{RGB}{214,96,77}
\definecolor{RdGy-8-D}{RGB}{214,96,77}
\definecolor{RdGy-8-3}{RGB}{244,165,130}
\definecolor{RdGy-8-F}{RGB}{244,165,130}
\definecolor{RdGy-8-4}{RGB}{253,219,199}
\definecolor{RdGy-8-G}{RGB}{253,219,199}
\definecolor{RdGy-8-5}{RGB}{224,224,224}
\definecolor{RdGy-8-I}{RGB}{224,224,224}
\definecolor{RdGy-8-6}{RGB}{186,186,186}
\definecolor{RdGy-8-J}{RGB}{186,186,186}
\definecolor{RdGy-8-7}{RGB}{135,135,135}
\definecolor{RdGy-8-L}{RGB}{135,135,135}
\definecolor{RdGy-8-8}{RGB}{77,77,77}
\definecolor{RdGy-8-N}{RGB}{77,77,77}
\definecolor{RdGy-9-1}{RGB}{178,24,43}
\definecolor{RdGy-9-B}{RGB}{178,24,43}
\definecolor{RdGy-9-2}{RGB}{214,96,77}
\definecolor{RdGy-9-D}{RGB}{214,96,77}
\definecolor{RdGy-9-3}{RGB}{244,165,130}
\definecolor{RdGy-9-F}{RGB}{244,165,130}
\definecolor{RdGy-9-4}{RGB}{253,219,199}
\definecolor{RdGy-9-G}{RGB}{253,219,199}
\definecolor{RdGy-9-5}{RGB}{255,255,255}
\definecolor{RdGy-9-H}{RGB}{255,255,255}
\definecolor{RdGy-9-6}{RGB}{224,224,224}
\definecolor{RdGy-9-I}{RGB}{224,224,224}
\definecolor{RdGy-9-7}{RGB}{186,186,186}
\definecolor{RdGy-9-J}{RGB}{186,186,186}
\definecolor{RdGy-9-8}{RGB}{135,135,135}
\definecolor{RdGy-9-L}{RGB}{135,135,135}
\definecolor{RdGy-9-9}{RGB}{77,77,77}
\definecolor{RdGy-9-N}{RGB}{77,77,77}
\definecolor{RdGy-10-1}{RGB}{103,0,31}
\definecolor{RdGy-10-A}{RGB}{103,0,31}
\definecolor{RdGy-10-2}{RGB}{178,24,43}
\definecolor{RdGy-10-B}{RGB}{178,24,43}
\definecolor{RdGy-10-3}{RGB}{214,96,77}
\definecolor{RdGy-10-D}{RGB}{214,96,77}
\definecolor{RdGy-10-4}{RGB}{244,165,130}
\definecolor{RdGy-10-F}{RGB}{244,165,130}
\definecolor{RdGy-10-5}{RGB}{253,219,199}
\definecolor{RdGy-10-G}{RGB}{253,219,199}
\definecolor{RdGy-10-6}{RGB}{224,224,224}
\definecolor{RdGy-10-I}{RGB}{224,224,224}
\definecolor{RdGy-10-7}{RGB}{186,186,186}
\definecolor{RdGy-10-J}{RGB}{186,186,186}
\definecolor{RdGy-10-8}{RGB}{135,135,135}
\definecolor{RdGy-10-L}{RGB}{135,135,135}
\definecolor{RdGy-10-9}{RGB}{77,77,77}
\definecolor{RdGy-10-N}{RGB}{77,77,77}
\definecolor{RdGy-10-10}{RGB}{26,26,26}
\definecolor{RdGy-10-O}{RGB}{26,26,26}
\definecolor{RdGy-11-1}{RGB}{103,0,31}
\definecolor{RdGy-11-A}{RGB}{103,0,31}
\definecolor{RdGy-11-2}{RGB}{178,24,43}
\definecolor{RdGy-11-B}{RGB}{178,24,43}
\definecolor{RdGy-11-3}{RGB}{214,96,77}
\definecolor{RdGy-11-D}{RGB}{214,96,77}
\definecolor{RdGy-11-4}{RGB}{244,165,130}
\definecolor{RdGy-11-F}{RGB}{244,165,130}
\definecolor{RdGy-11-5}{RGB}{253,219,199}
\definecolor{RdGy-11-G}{RGB}{253,219,199}
\definecolor{RdGy-11-6}{RGB}{255,255,255}
\definecolor{RdGy-11-H}{RGB}{255,255,255}
\definecolor{RdGy-11-7}{RGB}{224,224,224}
\definecolor{RdGy-11-I}{RGB}{224,224,224}
\definecolor{RdGy-11-8}{RGB}{186,186,186}
\definecolor{RdGy-11-J}{RGB}{186,186,186}
\definecolor{RdGy-11-9}{RGB}{135,135,135}
\definecolor{RdGy-11-L}{RGB}{135,135,135}
\definecolor{RdGy-11-10}{RGB}{77,77,77}
\definecolor{RdGy-11-N}{RGB}{77,77,77}
\definecolor{RdGy-11-11}{RGB}{26,26,26}
\definecolor{RdGy-11-O}{RGB}{26,26,26}
\definecolor{RdYlBu-3-1}{RGB}{252,141,89}
\definecolor{RdYlBu-3-E}{RGB}{252,141,89}
\definecolor{RdYlBu-3-2}{RGB}{255,255,191}
\definecolor{RdYlBu-3-H}{RGB}{255,255,191}
\definecolor{RdYlBu-3-3}{RGB}{145,191,219}
\definecolor{RdYlBu-3-K}{RGB}{145,191,219}
\definecolor{RdYlBu-4-1}{RGB}{215,25,28}
\definecolor{RdYlBu-4-C}{RGB}{215,25,28}
\definecolor{RdYlBu-4-2}{RGB}{253,174,97}
\definecolor{RdYlBu-4-F}{RGB}{253,174,97}
\definecolor{RdYlBu-4-3}{RGB}{171,217,233}
\definecolor{RdYlBu-4-J}{RGB}{171,217,233}
\definecolor{RdYlBu-4-4}{RGB}{44,123,182}
\definecolor{RdYlBu-4-M}{RGB}{44,123,182}
\definecolor{RdYlBu-5-1}{RGB}{215,25,28}
\definecolor{RdYlBu-5-C}{RGB}{215,25,28}
\definecolor{RdYlBu-5-2}{RGB}{253,174,97}
\definecolor{RdYlBu-5-F}{RGB}{253,174,97}
\definecolor{RdYlBu-5-3}{RGB}{255,255,191}
\definecolor{RdYlBu-5-H}{RGB}{255,255,191}
\definecolor{RdYlBu-5-4}{RGB}{171,217,233}
\definecolor{RdYlBu-5-J}{RGB}{171,217,233}
\definecolor{RdYlBu-5-5}{RGB}{44,123,182}
\definecolor{RdYlBu-5-M}{RGB}{44,123,182}
\definecolor{RdYlBu-6-1}{RGB}{215,48,39}
\definecolor{RdYlBu-6-B}{RGB}{215,48,39}
\definecolor{RdYlBu-6-2}{RGB}{252,141,89}
\definecolor{RdYlBu-6-E}{RGB}{252,141,89}
\definecolor{RdYlBu-6-3}{RGB}{254,224,144}
\definecolor{RdYlBu-6-G}{RGB}{254,224,144}
\definecolor{RdYlBu-6-4}{RGB}{224,243,248}
\definecolor{RdYlBu-6-I}{RGB}{224,243,248}
\definecolor{RdYlBu-6-5}{RGB}{145,191,219}
\definecolor{RdYlBu-6-K}{RGB}{145,191,219}
\definecolor{RdYlBu-6-6}{RGB}{69,117,180}
\definecolor{RdYlBu-6-N}{RGB}{69,117,180}
\definecolor{RdYlBu-7-1}{RGB}{215,48,39}
\definecolor{RdYlBu-7-B}{RGB}{215,48,39}
\definecolor{RdYlBu-7-2}{RGB}{252,141,89}
\definecolor{RdYlBu-7-E}{RGB}{252,141,89}
\definecolor{RdYlBu-7-3}{RGB}{254,224,144}
\definecolor{RdYlBu-7-G}{RGB}{254,224,144}
\definecolor{RdYlBu-7-4}{RGB}{255,255,191}
\definecolor{RdYlBu-7-H}{RGB}{255,255,191}
\definecolor{RdYlBu-7-5}{RGB}{224,243,248}
\definecolor{RdYlBu-7-I}{RGB}{224,243,248}
\definecolor{RdYlBu-7-6}{RGB}{145,191,219}
\definecolor{RdYlBu-7-K}{RGB}{145,191,219}
\definecolor{RdYlBu-7-7}{RGB}{69,117,180}
\definecolor{RdYlBu-7-N}{RGB}{69,117,180}
\definecolor{RdYlBu-8-1}{RGB}{215,48,39}
\definecolor{RdYlBu-8-B}{RGB}{215,48,39}
\definecolor{RdYlBu-8-2}{RGB}{244,109,67}
\definecolor{RdYlBu-8-D}{RGB}{244,109,67}
\definecolor{RdYlBu-8-3}{RGB}{253,174,97}
\definecolor{RdYlBu-8-F}{RGB}{253,174,97}
\definecolor{RdYlBu-8-4}{RGB}{254,224,144}
\definecolor{RdYlBu-8-G}{RGB}{254,224,144}
\definecolor{RdYlBu-8-5}{RGB}{224,243,248}
\definecolor{RdYlBu-8-I}{RGB}{224,243,248}
\definecolor{RdYlBu-8-6}{RGB}{171,217,233}
\definecolor{RdYlBu-8-J}{RGB}{171,217,233}
\definecolor{RdYlBu-8-7}{RGB}{116,173,209}
\definecolor{RdYlBu-8-L}{RGB}{116,173,209}
\definecolor{RdYlBu-8-8}{RGB}{69,117,180}
\definecolor{RdYlBu-8-N}{RGB}{69,117,180}
\definecolor{RdYlBu-9-1}{RGB}{215,48,39}
\definecolor{RdYlBu-9-B}{RGB}{215,48,39}
\definecolor{RdYlBu-9-2}{RGB}{244,109,67}
\definecolor{RdYlBu-9-D}{RGB}{244,109,67}
\definecolor{RdYlBu-9-3}{RGB}{253,174,97}
\definecolor{RdYlBu-9-F}{RGB}{253,174,97}
\definecolor{RdYlBu-9-4}{RGB}{254,224,144}
\definecolor{RdYlBu-9-G}{RGB}{254,224,144}
\definecolor{RdYlBu-9-5}{RGB}{255,255,191}
\definecolor{RdYlBu-9-H}{RGB}{255,255,191}
\definecolor{RdYlBu-9-6}{RGB}{224,243,248}
\definecolor{RdYlBu-9-I}{RGB}{224,243,248}
\definecolor{RdYlBu-9-7}{RGB}{171,217,233}
\definecolor{RdYlBu-9-J}{RGB}{171,217,233}
\definecolor{RdYlBu-9-8}{RGB}{116,173,209}
\definecolor{RdYlBu-9-L}{RGB}{116,173,209}
\definecolor{RdYlBu-9-9}{RGB}{69,117,180}
\definecolor{RdYlBu-9-N}{RGB}{69,117,180}
\definecolor{RdYlBu-10-1}{RGB}{165,0,38}
\definecolor{RdYlBu-10-A}{RGB}{165,0,38}
\definecolor{RdYlBu-10-2}{RGB}{215,48,39}
\definecolor{RdYlBu-10-B}{RGB}{215,48,39}
\definecolor{RdYlBu-10-3}{RGB}{244,109,67}
\definecolor{RdYlBu-10-D}{RGB}{244,109,67}
\definecolor{RdYlBu-10-4}{RGB}{253,174,97}
\definecolor{RdYlBu-10-F}{RGB}{253,174,97}
\definecolor{RdYlBu-10-5}{RGB}{254,224,144}
\definecolor{RdYlBu-10-G}{RGB}{254,224,144}
\definecolor{RdYlBu-10-6}{RGB}{224,243,248}
\definecolor{RdYlBu-10-I}{RGB}{224,243,248}
\definecolor{RdYlBu-10-7}{RGB}{171,217,233}
\definecolor{RdYlBu-10-J}{RGB}{171,217,233}
\definecolor{RdYlBu-10-8}{RGB}{116,173,209}
\definecolor{RdYlBu-10-L}{RGB}{116,173,209}
\definecolor{RdYlBu-10-9}{RGB}{69,117,180}
\definecolor{RdYlBu-10-N}{RGB}{69,117,180}
\definecolor{RdYlBu-10-10}{RGB}{49,54,149}
\definecolor{RdYlBu-10-O}{RGB}{49,54,149}
\definecolor{RdYlBu-11-1}{RGB}{165,0,38}
\definecolor{RdYlBu-11-A}{RGB}{165,0,38}
\definecolor{RdYlBu-11-2}{RGB}{215,48,39}
\definecolor{RdYlBu-11-B}{RGB}{215,48,39}
\definecolor{RdYlBu-11-3}{RGB}{244,109,67}
\definecolor{RdYlBu-11-D}{RGB}{244,109,67}
\definecolor{RdYlBu-11-4}{RGB}{253,174,97}
\definecolor{RdYlBu-11-F}{RGB}{253,174,97}
\definecolor{RdYlBu-11-5}{RGB}{254,224,144}
\definecolor{RdYlBu-11-G}{RGB}{254,224,144}
\definecolor{RdYlBu-11-6}{RGB}{255,255,191}
\definecolor{RdYlBu-11-H}{RGB}{255,255,191}
\definecolor{RdYlBu-11-7}{RGB}{224,243,248}
\definecolor{RdYlBu-11-I}{RGB}{224,243,248}
\definecolor{RdYlBu-11-8}{RGB}{171,217,233}
\definecolor{RdYlBu-11-J}{RGB}{171,217,233}
\definecolor{RdYlBu-11-9}{RGB}{116,173,209}
\definecolor{RdYlBu-11-L}{RGB}{116,173,209}
\definecolor{RdYlBu-11-10}{RGB}{69,117,180}
\definecolor{RdYlBu-11-N}{RGB}{69,117,180}
\definecolor{RdYlBu-11-11}{RGB}{49,54,149}
\definecolor{RdYlBu-11-O}{RGB}{49,54,149}
\definecolor{Spectral-3-1}{RGB}{252,141,89}
\definecolor{Spectral-3-E}{RGB}{252,141,89}
\definecolor{Spectral-3-2}{RGB}{255,255,191}
\definecolor{Spectral-3-H}{RGB}{255,255,191}
\definecolor{Spectral-3-3}{RGB}{153,213,148}
\definecolor{Spectral-3-K}{RGB}{153,213,148}
\definecolor{Spectral-4-1}{RGB}{215,25,28}
\definecolor{Spectral-4-C}{RGB}{215,25,28}
\definecolor{Spectral-4-2}{RGB}{253,174,97}
\definecolor{Spectral-4-F}{RGB}{253,174,97}
\definecolor{Spectral-4-3}{RGB}{171,221,164}
\definecolor{Spectral-4-J}{RGB}{171,221,164}
\definecolor{Spectral-4-4}{RGB}{43,131,186}
\definecolor{Spectral-4-M}{RGB}{43,131,186}
\definecolor{Spectral-5-1}{RGB}{215,25,28}
\definecolor{Spectral-5-C}{RGB}{215,25,28}
\definecolor{Spectral-5-2}{RGB}{253,174,97}
\definecolor{Spectral-5-F}{RGB}{253,174,97}
\definecolor{Spectral-5-3}{RGB}{255,255,191}
\definecolor{Spectral-5-H}{RGB}{255,255,191}
\definecolor{Spectral-5-4}{RGB}{171,221,164}
\definecolor{Spectral-5-J}{RGB}{171,221,164}
\definecolor{Spectral-5-5}{RGB}{43,131,186}
\definecolor{Spectral-5-M}{RGB}{43,131,186}
\definecolor{Spectral-6-1}{RGB}{213,62,79}
\definecolor{Spectral-6-B}{RGB}{213,62,79}
\definecolor{Spectral-6-2}{RGB}{252,141,89}
\definecolor{Spectral-6-E}{RGB}{252,141,89}
\definecolor{Spectral-6-3}{RGB}{254,224,139}
\definecolor{Spectral-6-G}{RGB}{254,224,139}
\definecolor{Spectral-6-4}{RGB}{230,245,152}
\definecolor{Spectral-6-I}{RGB}{230,245,152}
\definecolor{Spectral-6-5}{RGB}{153,213,148}
\definecolor{Spectral-6-K}{RGB}{153,213,148}
\definecolor{Spectral-6-6}{RGB}{50,136,189}
\definecolor{Spectral-6-N}{RGB}{50,136,189}
\definecolor{Spectral-7-1}{RGB}{213,62,79}
\definecolor{Spectral-7-B}{RGB}{213,62,79}
\definecolor{Spectral-7-2}{RGB}{252,141,89}
\definecolor{Spectral-7-E}{RGB}{252,141,89}
\definecolor{Spectral-7-3}{RGB}{254,224,139}
\definecolor{Spectral-7-G}{RGB}{254,224,139}
\definecolor{Spectral-7-4}{RGB}{255,255,191}
\definecolor{Spectral-7-H}{RGB}{255,255,191}
\definecolor{Spectral-7-5}{RGB}{230,245,152}
\definecolor{Spectral-7-I}{RGB}{230,245,152}
\definecolor{Spectral-7-6}{RGB}{153,213,148}
\definecolor{Spectral-7-K}{RGB}{153,213,148}
\definecolor{Spectral-7-7}{RGB}{50,136,189}
\definecolor{Spectral-7-N}{RGB}{50,136,189}
\definecolor{Spectral-8-1}{RGB}{213,62,79}
\definecolor{Spectral-8-B}{RGB}{213,62,79}
\definecolor{Spectral-8-2}{RGB}{244,109,67}
\definecolor{Spectral-8-D}{RGB}{244,109,67}
\definecolor{Spectral-8-3}{RGB}{253,174,97}
\definecolor{Spectral-8-F}{RGB}{253,174,97}
\definecolor{Spectral-8-4}{RGB}{254,224,139}
\definecolor{Spectral-8-G}{RGB}{254,224,139}
\definecolor{Spectral-8-5}{RGB}{230,245,152}
\definecolor{Spectral-8-I}{RGB}{230,245,152}
\definecolor{Spectral-8-6}{RGB}{171,221,164}
\definecolor{Spectral-8-J}{RGB}{171,221,164}
\definecolor{Spectral-8-7}{RGB}{102,194,165}
\definecolor{Spectral-8-L}{RGB}{102,194,165}
\definecolor{Spectral-8-8}{RGB}{50,136,189}
\definecolor{Spectral-8-N}{RGB}{50,136,189}
\definecolor{Spectral-9-1}{RGB}{213,62,79}
\definecolor{Spectral-9-B}{RGB}{213,62,79}
\definecolor{Spectral-9-2}{RGB}{244,109,67}
\definecolor{Spectral-9-D}{RGB}{244,109,67}
\definecolor{Spectral-9-3}{RGB}{253,174,97}
\definecolor{Spectral-9-F}{RGB}{253,174,97}
\definecolor{Spectral-9-4}{RGB}{254,224,139}
\definecolor{Spectral-9-G}{RGB}{254,224,139}
\definecolor{Spectral-9-5}{RGB}{255,255,191}
\definecolor{Spectral-9-H}{RGB}{255,255,191}
\definecolor{Spectral-9-6}{RGB}{230,245,152}
\definecolor{Spectral-9-I}{RGB}{230,245,152}
\definecolor{Spectral-9-7}{RGB}{171,221,164}
\definecolor{Spectral-9-J}{RGB}{171,221,164}
\definecolor{Spectral-9-8}{RGB}{102,194,165}
\definecolor{Spectral-9-L}{RGB}{102,194,165}
\definecolor{Spectral-9-9}{RGB}{50,136,189}
\definecolor{Spectral-9-N}{RGB}{50,136,189}
\definecolor{Spectral-10-1}{RGB}{158,1,66}
\definecolor{Spectral-10-A}{RGB}{158,1,66}
\definecolor{Spectral-10-2}{RGB}{213,62,79}
\definecolor{Spectral-10-B}{RGB}{213,62,79}
\definecolor{Spectral-10-3}{RGB}{244,109,67}
\definecolor{Spectral-10-D}{RGB}{244,109,67}
\definecolor{Spectral-10-4}{RGB}{253,174,97}
\definecolor{Spectral-10-F}{RGB}{253,174,97}
\definecolor{Spectral-10-5}{RGB}{254,224,139}
\definecolor{Spectral-10-G}{RGB}{254,224,139}
\definecolor{Spectral-10-6}{RGB}{230,245,152}
\definecolor{Spectral-10-I}{RGB}{230,245,152}
\definecolor{Spectral-10-7}{RGB}{171,221,164}
\definecolor{Spectral-10-J}{RGB}{171,221,164}
\definecolor{Spectral-10-8}{RGB}{102,194,165}
\definecolor{Spectral-10-L}{RGB}{102,194,165}
\definecolor{Spectral-10-9}{RGB}{50,136,189}
\definecolor{Spectral-10-N}{RGB}{50,136,189}
\definecolor{Spectral-10-10}{RGB}{94,79,162}
\definecolor{Spectral-10-O}{RGB}{94,79,162}
\definecolor{Spectral-11-1}{RGB}{158,1,66}
\definecolor{Spectral-11-A}{RGB}{158,1,66}
\definecolor{Spectral-11-2}{RGB}{213,62,79}
\definecolor{Spectral-11-B}{RGB}{213,62,79}
\definecolor{Spectral-11-3}{RGB}{244,109,67}
\definecolor{Spectral-11-D}{RGB}{244,109,67}
\definecolor{Spectral-11-4}{RGB}{253,174,97}
\definecolor{Spectral-11-F}{RGB}{253,174,97}
\definecolor{Spectral-11-5}{RGB}{254,224,139}
\definecolor{Spectral-11-G}{RGB}{254,224,139}
\definecolor{Spectral-11-6}{RGB}{255,255,191}
\definecolor{Spectral-11-H}{RGB}{255,255,191}
\definecolor{Spectral-11-7}{RGB}{230,245,152}
\definecolor{Spectral-11-I}{RGB}{230,245,152}
\definecolor{Spectral-11-8}{RGB}{171,221,164}
\definecolor{Spectral-11-J}{RGB}{171,221,164}
\definecolor{Spectral-11-9}{RGB}{102,194,165}
\definecolor{Spectral-11-L}{RGB}{102,194,165}
\definecolor{Spectral-11-10}{RGB}{50,136,189}
\definecolor{Spectral-11-N}{RGB}{50,136,189}
\definecolor{Spectral-11-11}{RGB}{94,79,162}
\definecolor{Spectral-11-O}{RGB}{94,79,162}
\definecolor{RdYlGn-3-1}{RGB}{252,141,89}
\definecolor{RdYlGn-3-E}{RGB}{252,141,89}
\definecolor{RdYlGn-3-2}{RGB}{255,255,191}
\definecolor{RdYlGn-3-H}{RGB}{255,255,191}
\definecolor{RdYlGn-3-3}{RGB}{145,207,96}
\definecolor{RdYlGn-3-K}{RGB}{145,207,96}
\definecolor{RdYlGn-4-1}{RGB}{215,25,28}
\definecolor{RdYlGn-4-C}{RGB}{215,25,28}
\definecolor{RdYlGn-4-2}{RGB}{253,174,97}
\definecolor{RdYlGn-4-F}{RGB}{253,174,97}
\definecolor{RdYlGn-4-3}{RGB}{166,217,106}
\definecolor{RdYlGn-4-J}{RGB}{166,217,106}
\definecolor{RdYlGn-4-4}{RGB}{26,150,65}
\definecolor{RdYlGn-4-M}{RGB}{26,150,65}
\definecolor{RdYlGn-5-1}{RGB}{215,25,28}
\definecolor{RdYlGn-5-C}{RGB}{215,25,28}
\definecolor{RdYlGn-5-2}{RGB}{253,174,97}
\definecolor{RdYlGn-5-F}{RGB}{253,174,97}
\definecolor{RdYlGn-5-3}{RGB}{255,255,191}
\definecolor{RdYlGn-5-H}{RGB}{255,255,191}
\definecolor{RdYlGn-5-4}{RGB}{166,217,106}
\definecolor{RdYlGn-5-J}{RGB}{166,217,106}
\definecolor{RdYlGn-5-5}{RGB}{26,150,65}
\definecolor{RdYlGn-5-M}{RGB}{26,150,65}
\definecolor{RdYlGn-6-1}{RGB}{215,48,39}
\definecolor{RdYlGn-6-B}{RGB}{215,48,39}
\definecolor{RdYlGn-6-2}{RGB}{252,141,89}
\definecolor{RdYlGn-6-E}{RGB}{252,141,89}
\definecolor{RdYlGn-6-3}{RGB}{254,224,139}
\definecolor{RdYlGn-6-G}{RGB}{254,224,139}
\definecolor{RdYlGn-6-4}{RGB}{217,239,139}
\definecolor{RdYlGn-6-I}{RGB}{217,239,139}
\definecolor{RdYlGn-6-5}{RGB}{145,207,96}
\definecolor{RdYlGn-6-K}{RGB}{145,207,96}
\definecolor{RdYlGn-6-6}{RGB}{26,152,80}
\definecolor{RdYlGn-6-N}{RGB}{26,152,80}
\definecolor{RdYlGn-7-1}{RGB}{215,48,39}
\definecolor{RdYlGn-7-B}{RGB}{215,48,39}
\definecolor{RdYlGn-7-2}{RGB}{252,141,89}
\definecolor{RdYlGn-7-E}{RGB}{252,141,89}
\definecolor{RdYlGn-7-3}{RGB}{254,224,139}
\definecolor{RdYlGn-7-G}{RGB}{254,224,139}
\definecolor{RdYlGn-7-4}{RGB}{255,255,191}
\definecolor{RdYlGn-7-H}{RGB}{255,255,191}
\definecolor{RdYlGn-7-5}{RGB}{217,239,139}
\definecolor{RdYlGn-7-I}{RGB}{217,239,139}
\definecolor{RdYlGn-7-6}{RGB}{145,207,96}
\definecolor{RdYlGn-7-K}{RGB}{145,207,96}
\definecolor{RdYlGn-7-7}{RGB}{26,152,80}
\definecolor{RdYlGn-7-N}{RGB}{26,152,80}
\definecolor{RdYlGn-8-1}{RGB}{215,48,39}
\definecolor{RdYlGn-8-B}{RGB}{215,48,39}
\definecolor{RdYlGn-8-2}{RGB}{244,109,67}
\definecolor{RdYlGn-8-D}{RGB}{244,109,67}
\definecolor{RdYlGn-8-3}{RGB}{253,174,97}
\definecolor{RdYlGn-8-F}{RGB}{253,174,97}
\definecolor{RdYlGn-8-4}{RGB}{254,224,139}
\definecolor{RdYlGn-8-G}{RGB}{254,224,139}
\definecolor{RdYlGn-8-5}{RGB}{217,239,139}
\definecolor{RdYlGn-8-I}{RGB}{217,239,139}
\definecolor{RdYlGn-8-6}{RGB}{166,217,106}
\definecolor{RdYlGn-8-J}{RGB}{166,217,106}
\definecolor{RdYlGn-8-7}{RGB}{102,189,99}
\definecolor{RdYlGn-8-L}{RGB}{102,189,99}
\definecolor{RdYlGn-8-8}{RGB}{26,152,80}
\definecolor{RdYlGn-8-N}{RGB}{26,152,80}
\definecolor{RdYlGn-9-1}{RGB}{215,48,39}
\definecolor{RdYlGn-9-B}{RGB}{215,48,39}
\definecolor{RdYlGn-9-2}{RGB}{244,109,67}
\definecolor{RdYlGn-9-D}{RGB}{244,109,67}
\definecolor{RdYlGn-9-3}{RGB}{253,174,97}
\definecolor{RdYlGn-9-F}{RGB}{253,174,97}
\definecolor{RdYlGn-9-4}{RGB}{254,224,139}
\definecolor{RdYlGn-9-G}{RGB}{254,224,139}
\definecolor{RdYlGn-9-5}{RGB}{255,255,191}
\definecolor{RdYlGn-9-H}{RGB}{255,255,191}
\definecolor{RdYlGn-9-6}{RGB}{217,239,139}
\definecolor{RdYlGn-9-I}{RGB}{217,239,139}
\definecolor{RdYlGn-9-7}{RGB}{166,217,106}
\definecolor{RdYlGn-9-J}{RGB}{166,217,106}
\definecolor{RdYlGn-9-8}{RGB}{102,189,99}
\definecolor{RdYlGn-9-L}{RGB}{102,189,99}
\definecolor{RdYlGn-9-9}{RGB}{26,152,80}
\definecolor{RdYlGn-9-N}{RGB}{26,152,80}
\definecolor{RdYlGn-10-1}{RGB}{165,0,38}
\definecolor{RdYlGn-10-A}{RGB}{165,0,38}
\definecolor{RdYlGn-10-2}{RGB}{215,48,39}
\definecolor{RdYlGn-10-B}{RGB}{215,48,39}
\definecolor{RdYlGn-10-3}{RGB}{244,109,67}
\definecolor{RdYlGn-10-D}{RGB}{244,109,67}
\definecolor{RdYlGn-10-4}{RGB}{253,174,97}
\definecolor{RdYlGn-10-F}{RGB}{253,174,97}
\definecolor{RdYlGn-10-5}{RGB}{254,224,139}
\definecolor{RdYlGn-10-G}{RGB}{254,224,139}
\definecolor{RdYlGn-10-6}{RGB}{217,239,139}
\definecolor{RdYlGn-10-I}{RGB}{217,239,139}
\definecolor{RdYlGn-10-7}{RGB}{166,217,106}
\definecolor{RdYlGn-10-J}{RGB}{166,217,106}
\definecolor{RdYlGn-10-8}{RGB}{102,189,99}
\definecolor{RdYlGn-10-L}{RGB}{102,189,99}
\definecolor{RdYlGn-10-9}{RGB}{26,152,80}
\definecolor{RdYlGn-10-N}{RGB}{26,152,80}
\definecolor{RdYlGn-10-10}{RGB}{0,104,55}
\definecolor{RdYlGn-10-O}{RGB}{0,104,55}
\definecolor{RdYlGn-11-1}{RGB}{165,0,38}
\definecolor{RdYlGn-11-A}{RGB}{165,0,38}
\definecolor{RdYlGn-11-2}{RGB}{215,48,39}
\definecolor{RdYlGn-11-B}{RGB}{215,48,39}
\definecolor{RdYlGn-11-3}{RGB}{244,109,67}
\definecolor{RdYlGn-11-D}{RGB}{244,109,67}
\definecolor{RdYlGn-11-4}{RGB}{253,174,97}
\definecolor{RdYlGn-11-F}{RGB}{253,174,97}
\definecolor{RdYlGn-11-5}{RGB}{254,224,139}
\definecolor{RdYlGn-11-G}{RGB}{254,224,139}
\definecolor{RdYlGn-11-6}{RGB}{255,255,191}
\definecolor{RdYlGn-11-H}{RGB}{255,255,191}
\definecolor{RdYlGn-11-7}{RGB}{217,239,139}
\definecolor{RdYlGn-11-I}{RGB}{217,239,139}
\definecolor{RdYlGn-11-8}{RGB}{166,217,106}
\definecolor{RdYlGn-11-J}{RGB}{166,217,106}
\definecolor{RdYlGn-11-9}{RGB}{102,189,99}
\definecolor{RdYlGn-11-L}{RGB}{102,189,99}
\definecolor{RdYlGn-11-10}{RGB}{26,152,80}
\definecolor{RdYlGn-11-N}{RGB}{26,152,80}
\definecolor{RdYlGn-11-11}{RGB}{0,104,55}
\definecolor{RdYlGn-11-O}{RGB}{0,104,55}
\definecolor{Set3-3-1}{RGB}{141,211,199}
\definecolor{Set3-3-A}{RGB}{141,211,199}
\definecolor{Set3-3-2}{RGB}{255,255,179}
\definecolor{Set3-3-B}{RGB}{255,255,179}
\definecolor{Set3-3-3}{RGB}{190,186,218}
\definecolor{Set3-3-C}{RGB}{190,186,218}
\definecolor{Set3-4-1}{RGB}{141,211,199}
\definecolor{Set3-4-A}{RGB}{141,211,199}
\definecolor{Set3-4-2}{RGB}{255,255,179}
\definecolor{Set3-4-B}{RGB}{255,255,179}
\definecolor{Set3-4-3}{RGB}{190,186,218}
\definecolor{Set3-4-C}{RGB}{190,186,218}
\definecolor{Set3-4-4}{RGB}{251,128,114}
\definecolor{Set3-4-D}{RGB}{251,128,114}
\definecolor{Set3-5-1}{RGB}{141,211,199}
\definecolor{Set3-5-A}{RGB}{141,211,199}
\definecolor{Set3-5-2}{RGB}{255,255,179}
\definecolor{Set3-5-B}{RGB}{255,255,179}
\definecolor{Set3-5-3}{RGB}{190,186,218}
\definecolor{Set3-5-C}{RGB}{190,186,218}
\definecolor{Set3-5-4}{RGB}{251,128,114}
\definecolor{Set3-5-D}{RGB}{251,128,114}
\definecolor{Set3-5-5}{RGB}{128,177,211}
\definecolor{Set3-5-E}{RGB}{128,177,211}
\definecolor{Set3-6-1}{RGB}{141,211,199}
\definecolor{Set3-6-A}{RGB}{141,211,199}
\definecolor{Set3-6-2}{RGB}{255,255,179}
\definecolor{Set3-6-B}{RGB}{255,255,179}
\definecolor{Set3-6-3}{RGB}{190,186,218}
\definecolor{Set3-6-C}{RGB}{190,186,218}
\definecolor{Set3-6-4}{RGB}{251,128,114}
\definecolor{Set3-6-D}{RGB}{251,128,114}
\definecolor{Set3-6-5}{RGB}{128,177,211}
\definecolor{Set3-6-E}{RGB}{128,177,211}
\definecolor{Set3-6-6}{RGB}{253,180,98}
\definecolor{Set3-6-F}{RGB}{253,180,98}
\definecolor{Set3-7-1}{RGB}{141,211,199}
\definecolor{Set3-7-A}{RGB}{141,211,199}
\definecolor{Set3-7-2}{RGB}{255,255,179}
\definecolor{Set3-7-B}{RGB}{255,255,179}
\definecolor{Set3-7-3}{RGB}{190,186,218}
\definecolor{Set3-7-C}{RGB}{190,186,218}
\definecolor{Set3-7-4}{RGB}{251,128,114}
\definecolor{Set3-7-D}{RGB}{251,128,114}
\definecolor{Set3-7-5}{RGB}{128,177,211}
\definecolor{Set3-7-E}{RGB}{128,177,211}
\definecolor{Set3-7-6}{RGB}{253,180,98}
\definecolor{Set3-7-F}{RGB}{253,180,98}
\definecolor{Set3-7-7}{RGB}{179,222,105}
\definecolor{Set3-7-G}{RGB}{179,222,105}
\definecolor{Set3-8-1}{RGB}{141,211,199}
\definecolor{Set3-8-A}{RGB}{141,211,199}
\definecolor{Set3-8-2}{RGB}{255,255,179}
\definecolor{Set3-8-B}{RGB}{255,255,179}
\definecolor{Set3-8-3}{RGB}{190,186,218}
\definecolor{Set3-8-C}{RGB}{190,186,218}
\definecolor{Set3-8-4}{RGB}{251,128,114}
\definecolor{Set3-8-D}{RGB}{251,128,114}
\definecolor{Set3-8-5}{RGB}{128,177,211}
\definecolor{Set3-8-E}{RGB}{128,177,211}
\definecolor{Set3-8-6}{RGB}{253,180,98}
\definecolor{Set3-8-F}{RGB}{253,180,98}
\definecolor{Set3-8-7}{RGB}{179,222,105}
\definecolor{Set3-8-G}{RGB}{179,222,105}
\definecolor{Set3-8-8}{RGB}{252,205,229}
\definecolor{Set3-8-H}{RGB}{252,205,229}
\definecolor{Set3-9-1}{RGB}{141,211,199}
\definecolor{Set3-9-A}{RGB}{141,211,199}
\definecolor{Set3-9-2}{RGB}{255,255,179}
\definecolor{Set3-9-B}{RGB}{255,255,179}
\definecolor{Set3-9-3}{RGB}{190,186,218}
\definecolor{Set3-9-C}{RGB}{190,186,218}
\definecolor{Set3-9-4}{RGB}{251,128,114}
\definecolor{Set3-9-D}{RGB}{251,128,114}
\definecolor{Set3-9-5}{RGB}{128,177,211}
\definecolor{Set3-9-E}{RGB}{128,177,211}
\definecolor{Set3-9-6}{RGB}{253,180,98}
\definecolor{Set3-9-F}{RGB}{253,180,98}
\definecolor{Set3-9-7}{RGB}{179,222,105}
\definecolor{Set3-9-G}{RGB}{179,222,105}
\definecolor{Set3-9-8}{RGB}{252,205,229}
\definecolor{Set3-9-H}{RGB}{252,205,229}
\definecolor{Set3-9-9}{RGB}{217,217,217}
\definecolor{Set3-9-I}{RGB}{217,217,217}
\definecolor{Set3-10-1}{RGB}{141,211,199}
\definecolor{Set3-10-A}{RGB}{141,211,199}
\definecolor{Set3-10-2}{RGB}{255,255,179}
\definecolor{Set3-10-B}{RGB}{255,255,179}
\definecolor{Set3-10-3}{RGB}{190,186,218}
\definecolor{Set3-10-C}{RGB}{190,186,218}
\definecolor{Set3-10-4}{RGB}{251,128,114}
\definecolor{Set3-10-D}{RGB}{251,128,114}
\definecolor{Set3-10-5}{RGB}{128,177,211}
\definecolor{Set3-10-E}{RGB}{128,177,211}
\definecolor{Set3-10-6}{RGB}{253,180,98}
\definecolor{Set3-10-F}{RGB}{253,180,98}
\definecolor{Set3-10-7}{RGB}{179,222,105}
\definecolor{Set3-10-G}{RGB}{179,222,105}
\definecolor{Set3-10-8}{RGB}{252,205,229}
\definecolor{Set3-10-H}{RGB}{252,205,229}
\definecolor{Set3-10-9}{RGB}{217,217,217}
\definecolor{Set3-10-I}{RGB}{217,217,217}
\definecolor{Set3-10-10}{RGB}{188,128,189}
\definecolor{Set3-10-J}{RGB}{188,128,189}
\definecolor{Set3-11-1}{RGB}{141,211,199}
\definecolor{Set3-11-A}{RGB}{141,211,199}
\definecolor{Set3-11-2}{RGB}{255,255,179}
\definecolor{Set3-11-B}{RGB}{255,255,179}
\definecolor{Set3-11-3}{RGB}{190,186,218}
\definecolor{Set3-11-C}{RGB}{190,186,218}
\definecolor{Set3-11-4}{RGB}{251,128,114}
\definecolor{Set3-11-D}{RGB}{251,128,114}
\definecolor{Set3-11-5}{RGB}{128,177,211}
\definecolor{Set3-11-E}{RGB}{128,177,211}
\definecolor{Set3-11-6}{RGB}{253,180,98}
\definecolor{Set3-11-F}{RGB}{253,180,98}
\definecolor{Set3-11-7}{RGB}{179,222,105}
\definecolor{Set3-11-G}{RGB}{179,222,105}
\definecolor{Set3-11-8}{RGB}{252,205,229}
\definecolor{Set3-11-H}{RGB}{252,205,229}
\definecolor{Set3-11-9}{RGB}{217,217,217}
\definecolor{Set3-11-I}{RGB}{217,217,217}
\definecolor{Set3-11-10}{RGB}{188,128,189}
\definecolor{Set3-11-J}{RGB}{188,128,189}
\definecolor{Set3-11-11}{RGB}{204,235,197}
\definecolor{Set3-11-K}{RGB}{204,235,197}
\definecolor{Set3-12-1}{RGB}{141,211,199}
\definecolor{Set3-12-A}{RGB}{141,211,199}
\definecolor{Set3-12-2}{RGB}{255,255,179}
\definecolor{Set3-12-B}{RGB}{255,255,179}
\definecolor{Set3-12-3}{RGB}{190,186,218}
\definecolor{Set3-12-C}{RGB}{190,186,218}
\definecolor{Set3-12-4}{RGB}{251,128,114}
\definecolor{Set3-12-D}{RGB}{251,128,114}
\definecolor{Set3-12-5}{RGB}{128,177,211}
\definecolor{Set3-12-E}{RGB}{128,177,211}
\definecolor{Set3-12-6}{RGB}{253,180,98}
\definecolor{Set3-12-F}{RGB}{253,180,98}
\definecolor{Set3-12-7}{RGB}{179,222,105}
\definecolor{Set3-12-G}{RGB}{179,222,105}
\definecolor{Set3-12-8}{RGB}{252,205,229}
\definecolor{Set3-12-H}{RGB}{252,205,229}
\definecolor{Set3-12-9}{RGB}{217,217,217}
\definecolor{Set3-12-I}{RGB}{217,217,217}
\definecolor{Set3-12-10}{RGB}{188,128,189}
\definecolor{Set3-12-J}{RGB}{188,128,189}
\definecolor{Set3-12-11}{RGB}{204,235,197}
\definecolor{Set3-12-K}{RGB}{204,235,197}
\definecolor{Set3-12-12}{RGB}{255,237,111}
\definecolor{Set3-12-L}{RGB}{255,237,111}
\definecolor{Pastel1-3-1}{RGB}{251,180,174}
\definecolor{Pastel1-3-A}{RGB}{251,180,174}
\definecolor{Pastel1-3-2}{RGB}{179,205,227}
\definecolor{Pastel1-3-B}{RGB}{179,205,227}
\definecolor{Pastel1-3-3}{RGB}{204,235,197}
\definecolor{Pastel1-3-C}{RGB}{204,235,197}
\definecolor{Pastel1-4-1}{RGB}{251,180,174}
\definecolor{Pastel1-4-A}{RGB}{251,180,174}
\definecolor{Pastel1-4-2}{RGB}{179,205,227}
\definecolor{Pastel1-4-B}{RGB}{179,205,227}
\definecolor{Pastel1-4-3}{RGB}{204,235,197}
\definecolor{Pastel1-4-C}{RGB}{204,235,197}
\definecolor{Pastel1-4-4}{RGB}{222,203,228}
\definecolor{Pastel1-4-D}{RGB}{222,203,228}
\definecolor{Pastel1-5-1}{RGB}{251,180,174}
\definecolor{Pastel1-5-A}{RGB}{251,180,174}
\definecolor{Pastel1-5-2}{RGB}{179,205,227}
\definecolor{Pastel1-5-B}{RGB}{179,205,227}
\definecolor{Pastel1-5-3}{RGB}{204,235,197}
\definecolor{Pastel1-5-C}{RGB}{204,235,197}
\definecolor{Pastel1-5-4}{RGB}{222,203,228}
\definecolor{Pastel1-5-D}{RGB}{222,203,228}
\definecolor{Pastel1-5-5}{RGB}{254,217,166}
\definecolor{Pastel1-5-E}{RGB}{254,217,166}
\definecolor{Pastel1-6-1}{RGB}{251,180,174}
\definecolor{Pastel1-6-A}{RGB}{251,180,174}
\definecolor{Pastel1-6-2}{RGB}{179,205,227}
\definecolor{Pastel1-6-B}{RGB}{179,205,227}
\definecolor{Pastel1-6-3}{RGB}{204,235,197}
\definecolor{Pastel1-6-C}{RGB}{204,235,197}
\definecolor{Pastel1-6-4}{RGB}{222,203,228}
\definecolor{Pastel1-6-D}{RGB}{222,203,228}
\definecolor{Pastel1-6-5}{RGB}{254,217,166}
\definecolor{Pastel1-6-E}{RGB}{254,217,166}
\definecolor{Pastel1-6-6}{RGB}{255,255,204}
\definecolor{Pastel1-6-F}{RGB}{255,255,204}
\definecolor{Pastel1-7-1}{RGB}{251,180,174}
\definecolor{Pastel1-7-A}{RGB}{251,180,174}
\definecolor{Pastel1-7-2}{RGB}{179,205,227}
\definecolor{Pastel1-7-B}{RGB}{179,205,227}
\definecolor{Pastel1-7-3}{RGB}{204,235,197}
\definecolor{Pastel1-7-C}{RGB}{204,235,197}
\definecolor{Pastel1-7-4}{RGB}{222,203,228}
\definecolor{Pastel1-7-D}{RGB}{222,203,228}
\definecolor{Pastel1-7-5}{RGB}{254,217,166}
\definecolor{Pastel1-7-E}{RGB}{254,217,166}
\definecolor{Pastel1-7-6}{RGB}{255,255,204}
\definecolor{Pastel1-7-F}{RGB}{255,255,204}
\definecolor{Pastel1-7-7}{RGB}{229,216,189}
\definecolor{Pastel1-7-G}{RGB}{229,216,189}
\definecolor{Pastel1-8-1}{RGB}{251,180,174}
\definecolor{Pastel1-8-A}{RGB}{251,180,174}
\definecolor{Pastel1-8-2}{RGB}{179,205,227}
\definecolor{Pastel1-8-B}{RGB}{179,205,227}
\definecolor{Pastel1-8-3}{RGB}{204,235,197}
\definecolor{Pastel1-8-C}{RGB}{204,235,197}
\definecolor{Pastel1-8-4}{RGB}{222,203,228}
\definecolor{Pastel1-8-D}{RGB}{222,203,228}
\definecolor{Pastel1-8-5}{RGB}{254,217,166}
\definecolor{Pastel1-8-E}{RGB}{254,217,166}
\definecolor{Pastel1-8-6}{RGB}{255,255,204}
\definecolor{Pastel1-8-F}{RGB}{255,255,204}
\definecolor{Pastel1-8-7}{RGB}{229,216,189}
\definecolor{Pastel1-8-G}{RGB}{229,216,189}
\definecolor{Pastel1-8-8}{RGB}{253,218,236}
\definecolor{Pastel1-8-H}{RGB}{253,218,236}
\definecolor{Pastel1-9-1}{RGB}{251,180,174}
\definecolor{Pastel1-9-A}{RGB}{251,180,174}
\definecolor{Pastel1-9-2}{RGB}{179,205,227}
\definecolor{Pastel1-9-B}{RGB}{179,205,227}
\definecolor{Pastel1-9-3}{RGB}{204,235,197}
\definecolor{Pastel1-9-C}{RGB}{204,235,197}
\definecolor{Pastel1-9-4}{RGB}{222,203,228}
\definecolor{Pastel1-9-D}{RGB}{222,203,228}
\definecolor{Pastel1-9-5}{RGB}{254,217,166}
\definecolor{Pastel1-9-E}{RGB}{254,217,166}
\definecolor{Pastel1-9-6}{RGB}{255,255,204}
\definecolor{Pastel1-9-F}{RGB}{255,255,204}
\definecolor{Pastel1-9-7}{RGB}{229,216,189}
\definecolor{Pastel1-9-G}{RGB}{229,216,189}
\definecolor{Pastel1-9-8}{RGB}{253,218,236}
\definecolor{Pastel1-9-H}{RGB}{253,218,236}
\definecolor{Pastel1-9-9}{RGB}{242,242,242}
\definecolor{Pastel1-9-I}{RGB}{242,242,242}
\definecolor{Set1-3-1}{RGB}{228,26,28}
\definecolor{Set1-3-A}{RGB}{228,26,28}
\definecolor{Set1-3-2}{RGB}{55,126,184}
\definecolor{Set1-3-B}{RGB}{55,126,184}
\definecolor{Set1-3-3}{RGB}{77,175,74}
\definecolor{Set1-3-C}{RGB}{77,175,74}
\definecolor{Set1-4-1}{RGB}{228,26,28}
\definecolor{Set1-4-A}{RGB}{228,26,28}
\definecolor{Set1-4-2}{RGB}{55,126,184}
\definecolor{Set1-4-B}{RGB}{55,126,184}
\definecolor{Set1-4-3}{RGB}{77,175,74}
\definecolor{Set1-4-C}{RGB}{77,175,74}
\definecolor{Set1-4-4}{RGB}{152,78,163}
\definecolor{Set1-4-D}{RGB}{152,78,163}
\definecolor{Set1-5-1}{RGB}{228,26,28}
\definecolor{Set1-5-A}{RGB}{228,26,28}
\definecolor{Set1-5-2}{RGB}{55,126,184}
\definecolor{Set1-5-B}{RGB}{55,126,184}
\definecolor{Set1-5-3}{RGB}{77,175,74}
\definecolor{Set1-5-C}{RGB}{77,175,74}
\definecolor{Set1-5-4}{RGB}{152,78,163}
\definecolor{Set1-5-D}{RGB}{152,78,163}
\definecolor{Set1-5-5}{RGB}{255,127,0}
\definecolor{Set1-5-E}{RGB}{255,127,0}
\definecolor{Set1-6-1}{RGB}{228,26,28}
\definecolor{Set1-6-A}{RGB}{228,26,28}
\definecolor{Set1-6-2}{RGB}{55,126,184}
\definecolor{Set1-6-B}{RGB}{55,126,184}
\definecolor{Set1-6-3}{RGB}{77,175,74}
\definecolor{Set1-6-C}{RGB}{77,175,74}
\definecolor{Set1-6-4}{RGB}{152,78,163}
\definecolor{Set1-6-D}{RGB}{152,78,163}
\definecolor{Set1-6-5}{RGB}{255,127,0}
\definecolor{Set1-6-E}{RGB}{255,127,0}
\definecolor{Set1-6-6}{RGB}{255,255,51}
\definecolor{Set1-6-F}{RGB}{255,255,51}
\definecolor{Set1-7-1}{RGB}{228,26,28}
\definecolor{Set1-7-A}{RGB}{228,26,28}
\definecolor{Set1-7-2}{RGB}{55,126,184}
\definecolor{Set1-7-B}{RGB}{55,126,184}
\definecolor{Set1-7-3}{RGB}{77,175,74}
\definecolor{Set1-7-C}{RGB}{77,175,74}
\definecolor{Set1-7-4}{RGB}{152,78,163}
\definecolor{Set1-7-D}{RGB}{152,78,163}
\definecolor{Set1-7-5}{RGB}{255,127,0}
\definecolor{Set1-7-E}{RGB}{255,127,0}
\definecolor{Set1-7-6}{RGB}{255,255,51}
\definecolor{Set1-7-F}{RGB}{255,255,51}
\definecolor{Set1-7-7}{RGB}{166,86,40}
\definecolor{Set1-7-G}{RGB}{166,86,40}
\definecolor{Set1-8-1}{RGB}{228,26,28}
\definecolor{Set1-8-A}{RGB}{228,26,28}
\definecolor{Set1-8-2}{RGB}{55,126,184}
\definecolor{Set1-8-B}{RGB}{55,126,184}
\definecolor{Set1-8-3}{RGB}{77,175,74}
\definecolor{Set1-8-C}{RGB}{77,175,74}
\definecolor{Set1-8-4}{RGB}{152,78,163}
\definecolor{Set1-8-D}{RGB}{152,78,163}
\definecolor{Set1-8-5}{RGB}{255,127,0}
\definecolor{Set1-8-E}{RGB}{255,127,0}
\definecolor{Set1-8-6}{RGB}{255,255,51}
\definecolor{Set1-8-F}{RGB}{255,255,51}
\definecolor{Set1-8-7}{RGB}{166,86,40}
\definecolor{Set1-8-G}{RGB}{166,86,40}
\definecolor{Set1-8-8}{RGB}{247,129,191}
\definecolor{Set1-8-H}{RGB}{247,129,191}
\definecolor{Set1-9-1}{RGB}{228,26,28}
\definecolor{Set1-9-A}{RGB}{228,26,28}
\definecolor{Set1-9-2}{RGB}{55,126,184}
\definecolor{Set1-9-B}{RGB}{55,126,184}
\definecolor{Set1-9-3}{RGB}{77,175,74}
\definecolor{Set1-9-C}{RGB}{77,175,74}
\definecolor{Set1-9-4}{RGB}{152,78,163}
\definecolor{Set1-9-D}{RGB}{152,78,163}
\definecolor{Set1-9-5}{RGB}{255,127,0}
\definecolor{Set1-9-E}{RGB}{255,127,0}
\definecolor{Set1-9-6}{RGB}{255,255,51}
\definecolor{Set1-9-F}{RGB}{255,255,51}
\definecolor{Set1-9-7}{RGB}{166,86,40}
\definecolor{Set1-9-G}{RGB}{166,86,40}
\definecolor{Set1-9-8}{RGB}{247,129,191}
\definecolor{Set1-9-H}{RGB}{247,129,191}
\definecolor{Set1-9-9}{RGB}{153,153,153}
\definecolor{Set1-9-I}{RGB}{153,153,153}
\definecolor{Pastel2-3-1}{RGB}{179,226,205}
\definecolor{Pastel2-3-A}{RGB}{179,226,205}
\definecolor{Pastel2-3-2}{RGB}{253,205,172}
\definecolor{Pastel2-3-B}{RGB}{253,205,172}
\definecolor{Pastel2-3-3}{RGB}{203,213,232}
\definecolor{Pastel2-3-C}{RGB}{203,213,232}
\definecolor{Pastel2-4-1}{RGB}{179,226,205}
\definecolor{Pastel2-4-A}{RGB}{179,226,205}
\definecolor{Pastel2-4-2}{RGB}{253,205,172}
\definecolor{Pastel2-4-B}{RGB}{253,205,172}
\definecolor{Pastel2-4-3}{RGB}{203,213,232}
\definecolor{Pastel2-4-C}{RGB}{203,213,232}
\definecolor{Pastel2-4-4}{RGB}{244,202,228}
\definecolor{Pastel2-4-D}{RGB}{244,202,228}
\definecolor{Pastel2-5-1}{RGB}{179,226,205}
\definecolor{Pastel2-5-A}{RGB}{179,226,205}
\definecolor{Pastel2-5-2}{RGB}{253,205,172}
\definecolor{Pastel2-5-B}{RGB}{253,205,172}
\definecolor{Pastel2-5-3}{RGB}{203,213,232}
\definecolor{Pastel2-5-C}{RGB}{203,213,232}
\definecolor{Pastel2-5-4}{RGB}{244,202,228}
\definecolor{Pastel2-5-D}{RGB}{244,202,228}
\definecolor{Pastel2-5-5}{RGB}{230,245,201}
\definecolor{Pastel2-5-E}{RGB}{230,245,201}
\definecolor{Pastel2-6-1}{RGB}{179,226,205}
\definecolor{Pastel2-6-A}{RGB}{179,226,205}
\definecolor{Pastel2-6-2}{RGB}{253,205,172}
\definecolor{Pastel2-6-B}{RGB}{253,205,172}
\definecolor{Pastel2-6-3}{RGB}{203,213,232}
\definecolor{Pastel2-6-C}{RGB}{203,213,232}
\definecolor{Pastel2-6-4}{RGB}{244,202,228}
\definecolor{Pastel2-6-D}{RGB}{244,202,228}
\definecolor{Pastel2-6-5}{RGB}{230,245,201}
\definecolor{Pastel2-6-E}{RGB}{230,245,201}
\definecolor{Pastel2-6-6}{RGB}{255,242,174}
\definecolor{Pastel2-6-F}{RGB}{255,242,174}
\definecolor{Pastel2-7-1}{RGB}{179,226,205}
\definecolor{Pastel2-7-A}{RGB}{179,226,205}
\definecolor{Pastel2-7-2}{RGB}{253,205,172}
\definecolor{Pastel2-7-B}{RGB}{253,205,172}
\definecolor{Pastel2-7-3}{RGB}{203,213,232}
\definecolor{Pastel2-7-C}{RGB}{203,213,232}
\definecolor{Pastel2-7-4}{RGB}{244,202,228}
\definecolor{Pastel2-7-D}{RGB}{244,202,228}
\definecolor{Pastel2-7-5}{RGB}{230,245,201}
\definecolor{Pastel2-7-E}{RGB}{230,245,201}
\definecolor{Pastel2-7-6}{RGB}{255,242,174}
\definecolor{Pastel2-7-F}{RGB}{255,242,174}
\definecolor{Pastel2-7-7}{RGB}{241,226,204}
\definecolor{Pastel2-7-G}{RGB}{241,226,204}
\definecolor{Pastel2-8-1}{RGB}{179,226,205}
\definecolor{Pastel2-8-A}{RGB}{179,226,205}
\definecolor{Pastel2-8-2}{RGB}{253,205,172}
\definecolor{Pastel2-8-B}{RGB}{253,205,172}
\definecolor{Pastel2-8-3}{RGB}{203,213,232}
\definecolor{Pastel2-8-C}{RGB}{203,213,232}
\definecolor{Pastel2-8-4}{RGB}{244,202,228}
\definecolor{Pastel2-8-D}{RGB}{244,202,228}
\definecolor{Pastel2-8-5}{RGB}{230,245,201}
\definecolor{Pastel2-8-E}{RGB}{230,245,201}
\definecolor{Pastel2-8-6}{RGB}{255,242,174}
\definecolor{Pastel2-8-F}{RGB}{255,242,174}
\definecolor{Pastel2-8-7}{RGB}{241,226,204}
\definecolor{Pastel2-8-G}{RGB}{241,226,204}
\definecolor{Pastel2-8-8}{RGB}{204,204,204}
\definecolor{Pastel2-8-H}{RGB}{204,204,204}
\definecolor{Set2-3-1}{RGB}{102,194,165}
\definecolor{Set2-3-A}{RGB}{102,194,165}
\definecolor{Set2-3-2}{RGB}{252,141,98}
\definecolor{Set2-3-B}{RGB}{252,141,98}
\definecolor{Set2-3-3}{RGB}{141,160,203}
\definecolor{Set2-3-C}{RGB}{141,160,203}
\definecolor{Set2-4-1}{RGB}{102,194,165}
\definecolor{Set2-4-A}{RGB}{102,194,165}
\definecolor{Set2-4-2}{RGB}{252,141,98}
\definecolor{Set2-4-B}{RGB}{252,141,98}
\definecolor{Set2-4-3}{RGB}{141,160,203}
\definecolor{Set2-4-C}{RGB}{141,160,203}
\definecolor{Set2-4-4}{RGB}{231,138,195}
\definecolor{Set2-4-D}{RGB}{231,138,195}
\definecolor{Set2-5-1}{RGB}{102,194,165}
\definecolor{Set2-5-A}{RGB}{102,194,165}
\definecolor{Set2-5-2}{RGB}{252,141,98}
\definecolor{Set2-5-B}{RGB}{252,141,98}
\definecolor{Set2-5-3}{RGB}{141,160,203}
\definecolor{Set2-5-C}{RGB}{141,160,203}
\definecolor{Set2-5-4}{RGB}{231,138,195}
\definecolor{Set2-5-D}{RGB}{231,138,195}
\definecolor{Set2-5-5}{RGB}{166,216,84}
\definecolor{Set2-5-E}{RGB}{166,216,84}
\definecolor{Set2-6-1}{RGB}{102,194,165}
\definecolor{Set2-6-A}{RGB}{102,194,165}
\definecolor{Set2-6-2}{RGB}{252,141,98}
\definecolor{Set2-6-B}{RGB}{252,141,98}
\definecolor{Set2-6-3}{RGB}{141,160,203}
\definecolor{Set2-6-C}{RGB}{141,160,203}
\definecolor{Set2-6-4}{RGB}{231,138,195}
\definecolor{Set2-6-D}{RGB}{231,138,195}
\definecolor{Set2-6-5}{RGB}{166,216,84}
\definecolor{Set2-6-E}{RGB}{166,216,84}
\definecolor{Set2-6-6}{RGB}{255,217,47}
\definecolor{Set2-6-F}{RGB}{255,217,47}
\definecolor{Set2-7-1}{RGB}{102,194,165}
\definecolor{Set2-7-A}{RGB}{102,194,165}
\definecolor{Set2-7-2}{RGB}{252,141,98}
\definecolor{Set2-7-B}{RGB}{252,141,98}
\definecolor{Set2-7-3}{RGB}{141,160,203}
\definecolor{Set2-7-C}{RGB}{141,160,203}
\definecolor{Set2-7-4}{RGB}{231,138,195}
\definecolor{Set2-7-D}{RGB}{231,138,195}
\definecolor{Set2-7-5}{RGB}{166,216,84}
\definecolor{Set2-7-E}{RGB}{166,216,84}
\definecolor{Set2-7-6}{RGB}{255,217,47}
\definecolor{Set2-7-F}{RGB}{255,217,47}
\definecolor{Set2-7-7}{RGB}{229,196,148}
\definecolor{Set2-7-G}{RGB}{229,196,148}
\definecolor{Set2-8-1}{RGB}{102,194,165}
\definecolor{Set2-8-A}{RGB}{102,194,165}
\definecolor{Set2-8-2}{RGB}{252,141,98}
\definecolor{Set2-8-B}{RGB}{252,141,98}
\definecolor{Set2-8-3}{RGB}{141,160,203}
\definecolor{Set2-8-C}{RGB}{141,160,203}
\definecolor{Set2-8-4}{RGB}{231,138,195}
\definecolor{Set2-8-D}{RGB}{231,138,195}
\definecolor{Set2-8-5}{RGB}{166,216,84}
\definecolor{Set2-8-E}{RGB}{166,216,84}
\definecolor{Set2-8-6}{RGB}{255,217,47}
\definecolor{Set2-8-F}{RGB}{255,217,47}
\definecolor{Set2-8-7}{RGB}{229,196,148}
\definecolor{Set2-8-G}{RGB}{229,196,148}
\definecolor{Set2-8-8}{RGB}{179,179,179}
\definecolor{Set2-8-H}{RGB}{179,179,179}
\definecolor{Dark2-3-1}{RGB}{27,158,119}
\definecolor{Dark2-3-A}{RGB}{27,158,119}
\definecolor{Dark2-3-2}{RGB}{217,95,2}
\definecolor{Dark2-3-B}{RGB}{217,95,2}
\definecolor{Dark2-3-3}{RGB}{117,112,179}
\definecolor{Dark2-3-C}{RGB}{117,112,179}
\definecolor{Dark2-4-1}{RGB}{27,158,119}
\definecolor{Dark2-4-A}{RGB}{27,158,119}
\definecolor{Dark2-4-2}{RGB}{217,95,2}
\definecolor{Dark2-4-B}{RGB}{217,95,2}
\definecolor{Dark2-4-3}{RGB}{117,112,179}
\definecolor{Dark2-4-C}{RGB}{117,112,179}
\definecolor{Dark2-4-4}{RGB}{231,41,138}
\definecolor{Dark2-4-D}{RGB}{231,41,138}
\definecolor{Dark2-5-1}{RGB}{27,158,119}
\definecolor{Dark2-5-A}{RGB}{27,158,119}
\definecolor{Dark2-5-2}{RGB}{217,95,2}
\definecolor{Dark2-5-B}{RGB}{217,95,2}
\definecolor{Dark2-5-3}{RGB}{117,112,179}
\definecolor{Dark2-5-C}{RGB}{117,112,179}
\definecolor{Dark2-5-4}{RGB}{231,41,138}
\definecolor{Dark2-5-D}{RGB}{231,41,138}
\definecolor{Dark2-5-5}{RGB}{102,166,30}
\definecolor{Dark2-5-E}{RGB}{102,166,30}
\definecolor{Dark2-6-1}{RGB}{27,158,119}
\definecolor{Dark2-6-A}{RGB}{27,158,119}
\definecolor{Dark2-6-2}{RGB}{217,95,2}
\definecolor{Dark2-6-B}{RGB}{217,95,2}
\definecolor{Dark2-6-3}{RGB}{117,112,179}
\definecolor{Dark2-6-C}{RGB}{117,112,179}
\definecolor{Dark2-6-4}{RGB}{231,41,138}
\definecolor{Dark2-6-D}{RGB}{231,41,138}
\definecolor{Dark2-6-5}{RGB}{102,166,30}
\definecolor{Dark2-6-E}{RGB}{102,166,30}
\definecolor{Dark2-6-6}{RGB}{230,171,2}
\definecolor{Dark2-6-F}{RGB}{230,171,2}
\definecolor{Dark2-7-1}{RGB}{27,158,119}
\definecolor{Dark2-7-A}{RGB}{27,158,119}
\definecolor{Dark2-7-2}{RGB}{217,95,2}
\definecolor{Dark2-7-B}{RGB}{217,95,2}
\definecolor{Dark2-7-3}{RGB}{117,112,179}
\definecolor{Dark2-7-C}{RGB}{117,112,179}
\definecolor{Dark2-7-4}{RGB}{231,41,138}
\definecolor{Dark2-7-D}{RGB}{231,41,138}
\definecolor{Dark2-7-5}{RGB}{102,166,30}
\definecolor{Dark2-7-E}{RGB}{102,166,30}
\definecolor{Dark2-7-6}{RGB}{230,171,2}
\definecolor{Dark2-7-F}{RGB}{230,171,2}
\definecolor{Dark2-7-7}{RGB}{166,118,29}
\definecolor{Dark2-7-G}{RGB}{166,118,29}
\definecolor{Dark2-8-1}{RGB}{27,158,119}
\definecolor{Dark2-8-A}{RGB}{27,158,119}
\definecolor{Dark2-8-2}{RGB}{217,95,2}
\definecolor{Dark2-8-B}{RGB}{217,95,2}
\definecolor{Dark2-8-3}{RGB}{117,112,179}
\definecolor{Dark2-8-C}{RGB}{117,112,179}
\definecolor{Dark2-8-4}{RGB}{231,41,138}
\definecolor{Dark2-8-D}{RGB}{231,41,138}
\definecolor{Dark2-8-5}{RGB}{102,166,30}
\definecolor{Dark2-8-E}{RGB}{102,166,30}
\definecolor{Dark2-8-6}{RGB}{230,171,2}
\definecolor{Dark2-8-F}{RGB}{230,171,2}
\definecolor{Dark2-8-7}{RGB}{166,118,29}
\definecolor{Dark2-8-G}{RGB}{166,118,29}
\definecolor{Dark2-8-8}{RGB}{102,102,102}
\definecolor{Dark2-8-H}{RGB}{102,102,102}
\definecolor{Paired-3-1}{RGB}{166,206,227}
\definecolor{Paired-3-A}{RGB}{166,206,227}
\definecolor{Paired-3-2}{RGB}{31,120,180}
\definecolor{Paired-3-B}{RGB}{31,120,180}
\definecolor{Paired-3-3}{RGB}{178,223,138}
\definecolor{Paired-3-C}{RGB}{178,223,138}
\definecolor{Paired-4-1}{RGB}{166,206,227}
\definecolor{Paired-4-A}{RGB}{166,206,227}
\definecolor{Paired-4-2}{RGB}{31,120,180}
\definecolor{Paired-4-B}{RGB}{31,120,180}
\definecolor{Paired-4-3}{RGB}{178,223,138}
\definecolor{Paired-4-C}{RGB}{178,223,138}
\definecolor{Paired-4-4}{RGB}{51,160,44}
\definecolor{Paired-4-D}{RGB}{51,160,44}
\definecolor{Paired-5-1}{RGB}{166,206,227}
\definecolor{Paired-5-A}{RGB}{166,206,227}
\definecolor{Paired-5-2}{RGB}{31,120,180}
\definecolor{Paired-5-B}{RGB}{31,120,180}
\definecolor{Paired-5-3}{RGB}{178,223,138}
\definecolor{Paired-5-C}{RGB}{178,223,138}
\definecolor{Paired-5-4}{RGB}{51,160,44}
\definecolor{Paired-5-D}{RGB}{51,160,44}
\definecolor{Paired-5-5}{RGB}{251,154,153}
\definecolor{Paired-5-E}{RGB}{251,154,153}
\definecolor{Paired-6-1}{RGB}{166,206,227}
\definecolor{Paired-6-A}{RGB}{166,206,227}
\definecolor{Paired-6-2}{RGB}{31,120,180}
\definecolor{Paired-6-B}{RGB}{31,120,180}
\definecolor{Paired-6-3}{RGB}{178,223,138}
\definecolor{Paired-6-C}{RGB}{178,223,138}
\definecolor{Paired-6-4}{RGB}{51,160,44}
\definecolor{Paired-6-D}{RGB}{51,160,44}
\definecolor{Paired-6-5}{RGB}{251,154,153}
\definecolor{Paired-6-E}{RGB}{251,154,153}
\definecolor{Paired-6-6}{RGB}{227,26,28}
\definecolor{Paired-6-F}{RGB}{227,26,28}
\definecolor{Paired-7-1}{RGB}{166,206,227}
\definecolor{Paired-7-A}{RGB}{166,206,227}
\definecolor{Paired-7-2}{RGB}{31,120,180}
\definecolor{Paired-7-B}{RGB}{31,120,180}
\definecolor{Paired-7-3}{RGB}{178,223,138}
\definecolor{Paired-7-C}{RGB}{178,223,138}
\definecolor{Paired-7-4}{RGB}{51,160,44}
\definecolor{Paired-7-D}{RGB}{51,160,44}
\definecolor{Paired-7-5}{RGB}{251,154,153}
\definecolor{Paired-7-E}{RGB}{251,154,153}
\definecolor{Paired-7-6}{RGB}{227,26,28}
\definecolor{Paired-7-F}{RGB}{227,26,28}
\definecolor{Paired-7-7}{RGB}{253,191,111}
\definecolor{Paired-7-G}{RGB}{253,191,111}
\definecolor{Paired-8-1}{RGB}{166,206,227}
\definecolor{Paired-8-A}{RGB}{166,206,227}
\definecolor{Paired-8-2}{RGB}{31,120,180}
\definecolor{Paired-8-B}{RGB}{31,120,180}
\definecolor{Paired-8-3}{RGB}{178,223,138}
\definecolor{Paired-8-C}{RGB}{178,223,138}
\definecolor{Paired-8-4}{RGB}{51,160,44}
\definecolor{Paired-8-D}{RGB}{51,160,44}
\definecolor{Paired-8-5}{RGB}{251,154,153}
\definecolor{Paired-8-E}{RGB}{251,154,153}
\definecolor{Paired-8-6}{RGB}{227,26,28}
\definecolor{Paired-8-F}{RGB}{227,26,28}
\definecolor{Paired-8-7}{RGB}{253,191,111}
\definecolor{Paired-8-G}{RGB}{253,191,111}
\definecolor{Paired-8-8}{RGB}{255,127,0}
\definecolor{Paired-8-H}{RGB}{255,127,0}
\definecolor{Paired-9-1}{RGB}{166,206,227}
\definecolor{Paired-9-A}{RGB}{166,206,227}
\definecolor{Paired-9-2}{RGB}{31,120,180}
\definecolor{Paired-9-B}{RGB}{31,120,180}
\definecolor{Paired-9-3}{RGB}{178,223,138}
\definecolor{Paired-9-C}{RGB}{178,223,138}
\definecolor{Paired-9-4}{RGB}{51,160,44}
\definecolor{Paired-9-D}{RGB}{51,160,44}
\definecolor{Paired-9-5}{RGB}{251,154,153}
\definecolor{Paired-9-E}{RGB}{251,154,153}
\definecolor{Paired-9-6}{RGB}{227,26,28}
\definecolor{Paired-9-F}{RGB}{227,26,28}
\definecolor{Paired-9-7}{RGB}{253,191,111}
\definecolor{Paired-9-G}{RGB}{253,191,111}
\definecolor{Paired-9-8}{RGB}{255,127,0}
\definecolor{Paired-9-H}{RGB}{255,127,0}
\definecolor{Paired-9-9}{RGB}{202,178,214}
\definecolor{Paired-9-I}{RGB}{202,178,214}
\definecolor{Paired-10-1}{RGB}{166,206,227}
\definecolor{Paired-10-A}{RGB}{166,206,227}
\definecolor{Paired-10-2}{RGB}{31,120,180}
\definecolor{Paired-10-B}{RGB}{31,120,180}
\definecolor{Paired-10-3}{RGB}{178,223,138}
\definecolor{Paired-10-C}{RGB}{178,223,138}
\definecolor{Paired-10-4}{RGB}{51,160,44}
\definecolor{Paired-10-D}{RGB}{51,160,44}
\definecolor{Paired-10-5}{RGB}{251,154,153}
\definecolor{Paired-10-E}{RGB}{251,154,153}
\definecolor{Paired-10-6}{RGB}{227,26,28}
\definecolor{Paired-10-F}{RGB}{227,26,28}
\definecolor{Paired-10-7}{RGB}{253,191,111}
\definecolor{Paired-10-G}{RGB}{253,191,111}
\definecolor{Paired-10-8}{RGB}{255,127,0}
\definecolor{Paired-10-H}{RGB}{255,127,0}
\definecolor{Paired-10-9}{RGB}{202,178,214}
\definecolor{Paired-10-I}{RGB}{202,178,214}
\definecolor{Paired-10-10}{RGB}{106,61,154}
\definecolor{Paired-10-J}{RGB}{106,61,154}
\definecolor{Paired-11-1}{RGB}{166,206,227}
\definecolor{Paired-11-A}{RGB}{166,206,227}
\definecolor{Paired-11-2}{RGB}{31,120,180}
\definecolor{Paired-11-B}{RGB}{31,120,180}
\definecolor{Paired-11-3}{RGB}{178,223,138}
\definecolor{Paired-11-C}{RGB}{178,223,138}
\definecolor{Paired-11-4}{RGB}{51,160,44}
\definecolor{Paired-11-D}{RGB}{51,160,44}
\definecolor{Paired-11-5}{RGB}{251,154,153}
\definecolor{Paired-11-E}{RGB}{251,154,153}
\definecolor{Paired-11-6}{RGB}{227,26,28}
\definecolor{Paired-11-F}{RGB}{227,26,28}
\definecolor{Paired-11-7}{RGB}{253,191,111}
\definecolor{Paired-11-G}{RGB}{253,191,111}
\definecolor{Paired-11-8}{RGB}{255,127,0}
\definecolor{Paired-11-H}{RGB}{255,127,0}
\definecolor{Paired-11-9}{RGB}{202,178,214}
\definecolor{Paired-11-I}{RGB}{202,178,214}
\definecolor{Paired-11-10}{RGB}{106,61,154}
\definecolor{Paired-11-J}{RGB}{106,61,154}
\definecolor{Paired-11-11}{RGB}{255,255,153}
\definecolor{Paired-11-K}{RGB}{255,255,153}
\definecolor{Paired-12-1}{RGB}{166,206,227}
\definecolor{Paired-12-A}{RGB}{166,206,227}
\definecolor{Paired-12-2}{RGB}{31,120,180}
\definecolor{Paired-12-B}{RGB}{31,120,180}
\definecolor{Paired-12-3}{RGB}{178,223,138}
\definecolor{Paired-12-C}{RGB}{178,223,138}
\definecolor{Paired-12-4}{RGB}{51,160,44}
\definecolor{Paired-12-D}{RGB}{51,160,44}
\definecolor{Paired-12-5}{RGB}{251,154,153}
\definecolor{Paired-12-E}{RGB}{251,154,153}
\definecolor{Paired-12-6}{RGB}{227,26,28}
\definecolor{Paired-12-F}{RGB}{227,26,28}
\definecolor{Paired-12-7}{RGB}{253,191,111}
\definecolor{Paired-12-G}{RGB}{253,191,111}
\definecolor{Paired-12-8}{RGB}{255,127,0}
\definecolor{Paired-12-H}{RGB}{255,127,0}
\definecolor{Paired-12-9}{RGB}{202,178,214}
\definecolor{Paired-12-I}{RGB}{202,178,214}
\definecolor{Paired-12-10}{RGB}{106,61,154}
\definecolor{Paired-12-J}{RGB}{106,61,154}
\definecolor{Paired-12-11}{RGB}{255,255,153}
\definecolor{Paired-12-K}{RGB}{255,255,153}
\definecolor{Paired-12-12}{RGB}{177,89,40}
\definecolor{Paired-12-L}{RGB}{177,89,40}
\definecolor{Accent-3-1}{RGB}{127,201,127}
\definecolor{Accent-3-A}{RGB}{127,201,127}
\definecolor{Accent-3-2}{RGB}{190,174,212}
\definecolor{Accent-3-B}{RGB}{190,174,212}
\definecolor{Accent-3-3}{RGB}{253,192,134}
\definecolor{Accent-3-C}{RGB}{253,192,134}
\definecolor{Accent-4-1}{RGB}{127,201,127}
\definecolor{Accent-4-A}{RGB}{127,201,127}
\definecolor{Accent-4-2}{RGB}{190,174,212}
\definecolor{Accent-4-B}{RGB}{190,174,212}
\definecolor{Accent-4-3}{RGB}{253,192,134}
\definecolor{Accent-4-C}{RGB}{253,192,134}
\definecolor{Accent-4-4}{RGB}{255,255,153}
\definecolor{Accent-4-D}{RGB}{255,255,153}
\definecolor{Accent-5-1}{RGB}{127,201,127}
\definecolor{Accent-5-A}{RGB}{127,201,127}
\definecolor{Accent-5-2}{RGB}{190,174,212}
\definecolor{Accent-5-B}{RGB}{190,174,212}
\definecolor{Accent-5-3}{RGB}{253,192,134}
\definecolor{Accent-5-C}{RGB}{253,192,134}
\definecolor{Accent-5-4}{RGB}{255,255,153}
\definecolor{Accent-5-D}{RGB}{255,255,153}
\definecolor{Accent-5-5}{RGB}{56,108,176}
\definecolor{Accent-5-E}{RGB}{56,108,176}
\definecolor{Accent-6-1}{RGB}{127,201,127}
\definecolor{Accent-6-A}{RGB}{127,201,127}
\definecolor{Accent-6-2}{RGB}{190,174,212}
\definecolor{Accent-6-B}{RGB}{190,174,212}
\definecolor{Accent-6-3}{RGB}{253,192,134}
\definecolor{Accent-6-C}{RGB}{253,192,134}
\definecolor{Accent-6-4}{RGB}{255,255,153}
\definecolor{Accent-6-D}{RGB}{255,255,153}
\definecolor{Accent-6-5}{RGB}{56,108,176}
\definecolor{Accent-6-E}{RGB}{56,108,176}
\definecolor{Accent-6-6}{RGB}{240,2,127}
\definecolor{Accent-6-F}{RGB}{240,2,127}
\definecolor{Accent-7-1}{RGB}{127,201,127}
\definecolor{Accent-7-A}{RGB}{127,201,127}
\definecolor{Accent-7-2}{RGB}{190,174,212}
\definecolor{Accent-7-B}{RGB}{190,174,212}
\definecolor{Accent-7-3}{RGB}{253,192,134}
\definecolor{Accent-7-C}{RGB}{253,192,134}
\definecolor{Accent-7-4}{RGB}{255,255,153}
\definecolor{Accent-7-D}{RGB}{255,255,153}
\definecolor{Accent-7-5}{RGB}{56,108,176}
\definecolor{Accent-7-E}{RGB}{56,108,176}
\definecolor{Accent-7-6}{RGB}{240,2,127}
\definecolor{Accent-7-F}{RGB}{240,2,127}
\definecolor{Accent-7-7}{RGB}{191,91,23}
\definecolor{Accent-7-G}{RGB}{191,91,23}
\definecolor{Accent-8-1}{RGB}{127,201,127}
\definecolor{Accent-8-A}{RGB}{127,201,127}
\definecolor{Accent-8-2}{RGB}{190,174,212}
\definecolor{Accent-8-B}{RGB}{190,174,212}
\definecolor{Accent-8-3}{RGB}{253,192,134}
\definecolor{Accent-8-C}{RGB}{253,192,134}
\definecolor{Accent-8-4}{RGB}{255,255,153}
\definecolor{Accent-8-D}{RGB}{255,255,153}
\definecolor{Accent-8-5}{RGB}{56,108,176}
\definecolor{Accent-8-E}{RGB}{56,108,176}
\definecolor{Accent-8-6}{RGB}{240,2,127}
\definecolor{Accent-8-F}{RGB}{240,2,127}
\definecolor{Accent-8-7}{RGB}{191,91,23}
\definecolor{Accent-8-G}{RGB}{191,91,23}
\definecolor{Accent-8-8}{RGB}{102,102,102}
\definecolor{Accent-8-H}{RGB}{102,102,102}

\definecolor{fadv}{rgb}{1.000,0.547,0.000}
\definecolor{sym}{rgb}{0.500,0.500,0.500}
\definecolor{fdel}{rgb}{0.254,0.410,0.879}

\definecolor{sym2}{rgb}{1.000,1.000,1.000}
\definecolor{fadv}{RGB}{140,81,10}
\definecolor{padv}{RGB}{216,179,101}
\definecolor{ladv}{RGB}{246,232,195}
\definecolor{sym}{RGB}{245,245,245}
\definecolor{ldel}{RGB}{199,234,229}
\definecolor{pdel}{RGB}{90,180,172}
\definecolor{fdel}{RGB}{1,102,94}

\definecolor{gro}{RGB}{27,158,119}
\definecolor{em}{RGB}{217,95,2}
\definecolor{lo}{rgb}{117,112,179}

\definecolor{f1p5}{RGB}{229,245,224}
\definecolor{f1p20}{RGB}{199,233,192}
\definecolor{f1p35}{RGB}{161,217,155}
\definecolor{f1p45}{RGB}{116,196,118}
\definecolor{f1p65}{RGB}{65,171,93}
\definecolor{f1p89}{RGB}{35,139,69}
\definecolor{f1psine}{RGB}{0,109,44}

\definecolor{f2p5}{RGB}{254,230,206}
\definecolor{f2p20}{RGB}{253,208,162}
\definecolor{f2p35}{RGB}{253,174,107}
\definecolor{f2p45}{RGB}{253,141,60}
\definecolor{f2p65}{RGB}{241,105,19}
\definecolor{f2p89}{RGB}{217,72,1}
\definecolor{f2psine}{RGB}{166,54,3}

\definecolor{f3p5}{RGB}{222,235,247}
\definecolor{f3p20}{RGB}{198,219,239}
\definecolor{f3p35}{RGB}{158,202,225}
\definecolor{f3p45}{RGB}{107,174,214}
\definecolor{f3p65}{RGB}{66,146,198}
\definecolor{f3p89}{RGB}{33,113,181}
\definecolor{f3psine}{RGB}{8,81,156}

\definecolor{cb1}{RGB}{68,1,84}
\definecolor{cb2}{RGB}{71,10,97}
\definecolor{cb3}{RGB}{72,25,107}
\definecolor{cb4}{RGB}{72,40,120}
\definecolor{cb5}{RGB}{70,52,129}
\definecolor{cb6}{RGB}{65,66,135}
\definecolor{cb7}{RGB}{57,78,140}
\definecolor{cb8}{RGB}{49,89,142}
\definecolor{cb9}{RGB}{41,101,142}
\definecolor{cb10}{RGB}{35,113,141}
\definecolor{cb11}{RGB}{30,125,140}
\definecolor{cb12}{RGB}{34,136,138}
\definecolor{cb13}{RGB}{47,146,132}
\definecolor{cb14}{RGB}{66,156,122}
\definecolor{cb15}{RGB}{90,165,110}
\definecolor{cb16}{RGB}{119,174,96}
\definecolor{cb17}{RGB}{151,180,80}
\definecolor{cb18}{RGB}{182,184,65}
\definecolor{cb19}{RGB}{211,180,58}
\definecolor{cb20}{RGB}{237,173,65}
\definecolor{cb21}{RGB}{251,161,71}
\definecolor{cb22}{RGB}{252,141,60}
\definecolor{cb23}{RGB}{240,117,46}
\definecolor{cb24}{RGB}{217,85,32}
\definecolor{cb25}{RGB}{189,55,20}
\begin{document}

\maketitle

\begin{abstract}
\footnotesize
Bubbles released from a needle show shape deformations that depend on the surfactant concentration of the surrounding liquid. We develop a model that  predicts the surfactant concentration based on experimental early-stage observations of these deformations. Using high-speed imaging, we examine bubbles within the first $144~\mathrm{ms}$ of ascent, corresponding to a vertical rise distance of $\sim40~\mathrm{mm}$ and extract the instantaneous aspect ratio ($AR$) and analyse its temporal evolution. In clean conditions, bubbles exhibit pronounced shape oscillations resulting from the periodic exchange between surface and kinetic energy. The presence of surfactants leads to an immediate damping of these oscillations, characterised by reduced $AR$ amplitudes and earlier peak deformations. This damping effect intensifies with increasing surfactant concentration until a near-saturation regime is reached, beyond which bubbles remain largely spherical and further increases in concentration produce indistinguishable $AR$ profiles within the early-stage observation window. To develop the prediction model, an aspect-ratio-based analysis methodology is proposed, which yields an empirical relationship capable of estimating surfactant concentrations between $0~\mathrm{ppm}$ and $2.9~\mathrm{ppm}$. We finally test the reliability of the model on unknown surfactant-laden bubbles. The model successfully detected the presence and relative extent of surfactant contamination as higher concentrations were introduced.
\end{abstract}

\section{Introduction\label{intro}}
Extensive research on the rise behaviours of air bubbles has been motivated by its vital role in multiple chemical engineering processes such as mixing of liquid and gas phases in bubble column reactors \citep{Martin09,Becerril02} and separation of mineral particles through froth floatation \citep{Mesa18}. In these applications, the understanding of bubble interfacial dynamics when coated with surfactants and/or particles is crucial in optimising overall process efficiencies by maximising collision and attachment probabilities \citep{Kostoglou20,Evans08} and mass transfer rates \citep{Martin09,Montes99}. However, direct measurement of interfacial properties can be challenging and invasive in such systems. Establishing reliable correlations between quantifiable bubble dynamics, such as rise velocity and shape oscillations, with surfactant levels could simplify efficiency evaluations and provide more accurate, real-time insights into interfacial phenomena.

Considerable efforts have been devoted to predicting the steady-state velocity of rising bubbles in quiescent liquid, termed as the terminal rise velocity, $U_T$, as it often serves as a characteristic parameter in evaluating system performance in engineering applications \citep{Hlawitschka22}. The rise of a bubble is driven by its buoyancy force, $F_B$, which for a given difference between liquid and gas density, $\Delta\rho=\rho_l-\rho_g$ and gravitational acceleration $g$, is a function of the bubble's volume $V$, commonly quantified by its volume equivalent diameter \citep{Fan21} as-

\begin{equation}
d_{\text{eq}} = \sqrt[3]{\frac{6V}{\pi} }
\label{eq:deq}
\end{equation}

such that 

\begin{equation}
F_B = \frac{\pi d_{eq}^3 \rho_l g}{6},
\end{equation}
where we have approximated the density difference between the liquid and gas phase $\Delta\rho\approx\rho_l$ because $\rho_l\gg\rho_g$. The buoyancy force is resisted by the total drag force, $F_D$, consisting of form drag, friction drag, and virtual or added mass force \citep{Zhang08, Nalajala14}, all combined in a drag coefficient $C_D$, giving
\begin{equation}
F_D = \frac{C_D \rho_l U^2 \pi d_{eq}^2}{8}.
\end{equation}
At steady state, the balance between these forces yields
\begin{equation}
U_T = \sqrt{\frac{4gd_{eq}}{3C_D}},
\end{equation}
which shows that the key to predicting the terminal velocity lies in the formulation of the drag coefficient, $C_D$ and its dependence on bubble and liquid properties.

Terminal rise velocities of bubbles spanning equivalent diameters $0.02~\mathrm{mm} < d_{eq} < 67.0~\mathrm{mm}$  have been systematically compiled and compared against a wide range of analytical and empirical drag-coefficient correlations \citep{Feng24}. Distinct regimes of bubble behaviour—spherical, ellipsoidal, and spherical-cap—have been identified based on bubble size, with the terminal velocity of $U_T$ exhibiting regime-dependent trends \citep{Clift78}. Bubbles with equivalent diameters in the range $1~\rm{mm}$ - $17~\rm{mm}$ which are typical of many practical applications and form the focus of the present study, lie within the ellipsoidal regime. This regime is characterised by pronounced shape oscillations and the onset of path instability for $d_{eq}>2~\rm{mm}$. Owing to these unsteady deformations, drag correlations for ellipsoidal bubbles are commonly expressed in terms of the Eötvös number, $Eo$, which quantifies the interaction between buoyancy and surface-tension forces governing bubble shape stability as-

\begin{equation}
Eo = \frac{\Delta \rho \, g \, d_{eq}^2}{\gamma},
\end{equation}
where $\gamma$ is the surface tension. In the ellipsoidal regime, increased drag associated with path instability leads to a counter-intuitive decrease in the terminal rise velocity with increasing equivalent diameter up to $7~\rm{mm}$, a behaviour not observed in the spherical or spherical-cap regimes \citep{Feng24,Mougin01,Mougin06,Cano-Lozano16}. 

Substantial effort has been devoted to extending bubble rise-velocity correlations to contaminated systems, beyond studies restricted to bubbles rising in clean water. Two primary contamination mechanisms are commonly identified in the literature: direct deposition of particles onto the bubble interface and gradual surfactant accumulation during ascent in surfactant-laden liquids.

Particle deposition onto bubble interfaces prior to release is widely reported to retard interfacial motion, resulting in increased drag and reduced terminal rise velocities compared to clean bubbles \citep{Peebles53,Wang19,Wang19Langmuir,Eskanlou18,Yan17,Yan18,Yan21,Zheng19}. The magnitude of this effect increases with the degree of surface contamination, commonly quantified by the particle-surface coverage, $\phi$. It is defined as $\phi=A_p/A_b$, where $A_p$ is the surface area covered by particles and $A_b$ the total surface area of the bubble. A pronounced reduction of 24.6\% in the terminal rise velocity has been reported as $\phi$ increased from 0\% to 50\% \citep{Yan21}. Beyond this range, further reductions in terminal velocities diminish and eventually plateau as surface contamination approaches saturation for $\phi \geq 70$\%. In addition to interfacial immobilisation, increasing particle size raises the effective density of particle-coated bubbles, reducing the density difference relative to the surrounding fluid and thereby decreasing the buoyancy force $F_B$. This reduction in buoyancy leads to lower terminal rise velocities for bubbles coated with larger particles \citep{Wang19Langmuir,Yan21}.
 
In contrast to the direct introduction of particles onto bubble surfaces as performed in the aforementioned studies, a different mechanism governs the interfacial dynamics of bubbles rising through surfactant-laden fluids. As the bubble rises, surfactant molecules accumulate on the interface and are convected from the front toward the rear, establishing an interfacial concentration gradient. This concentration gradient generates a corresponding surface-tension gradient, with lower surface tension toward the bubble rear, giving rise to tangential Marangoni stresses along the interface \citep{Peebles53,Clift78,Tomiyama02}. These stresses progressively immobilise the bubble surface, effectively transitioning the interface from a free-slip to a no-slip condition. This effect increases the frictional drag experienced by the rising bubble, reducing its rise velocity until a steady state $U_T$ is reached. Although higher surfactant concentrations have been shown to reduce the rise distance required for bubbles to reach their $U_T$, the magnitude of $U_T$ itself is insensitive to bulk surfactant concentrations \citep{Zhang01, Tomiyama02, Kracht10}. 

High-speed imaging studies have demonstrated a strong correlation between instantaneous rise velocity and bubble shape \citep{Kracht10,Wang19}. As a bubble ascends, the periodic exchange of surface energy and kinetic energy induces an oscillating shape deformation between spherical and ellipsoidal phases, which can be quantified by its aspect ratio, $AR$ defined as the ratio between the semi-major length and semi-minor length. This dynamic results in an oscillatory rise behaviour, where velocity fluctuates periodically with shape changes. Surface impurities—whether surfactant molecules or adhering particles—promote interfacial immobilisation. This immobilisation suppresses shape deformations, resulting in reduced amplitude and frequency in $AR$ oscillation profiles \citep{Chabel12, Kracht10, Wang19, Yan21}. Progressive damping of shape oscillations promotes a tendency toward near-spherical bubble geometries as interfacial contamination levels increase. This effect becomes increasingly pronounced with greater surface coverage by larger particles and has also been observed following coalescence between clean and particle-laden bubbles \citep{Wang19,Wang19Langmuir,Eskanlou18,Ata08,Wang20Langmuir}. 

Given the evident link between interfacial contamination and damped shape oscillations, this study aims to identify quantifiable indicators from $AR$ profiles of bubbles rising in fluids with varying surfactant concentrations. The present work focuses on the early-stage dynamics immediately following bubble detachment from a needle. This offers insight into the sensitivity of evolving interfacial conditions to initial shape deformations, information which is lost when analysing terminal velocity, steady-state oscillations, or path instabilities. Using high-speed imaging, the rise velocity and $AR$ oscillation profiles of bubbles in the ellipsoidal regime are analysed. The goal is to establish a reliable metric for predicting surrounding surfactant levels based on dynamic bubble behaviour. In doing so, this work addresses a critical gap in the literature: the absence of a direct correlation between early-stage bubble rise characteristics and interfacial conditions - an aspect that remains overlooked in traditional drag- and terminal velocity-based models.

\section{Methodology}
\subsection{Experimental Setup and Procedure\label{method-exp}}

\begin{figure}
  \centerline{\includegraphics[width=0.8\textwidth]{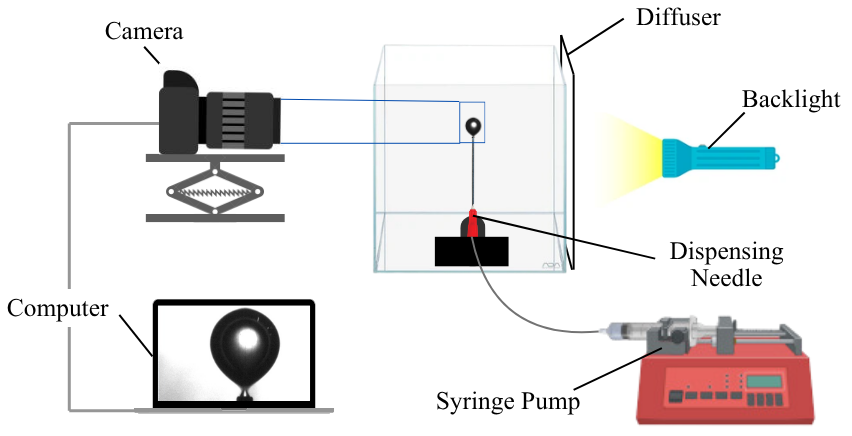}}
  \caption{Schematic diagram of experimental setup}
\label{fig:Exp. Setup}
\end{figure}

The experimental setup consists of a square tank ($180~\mathrm{mm}\times180~\mathrm{mm}\times180~\mathrm{mm}$) in which bubbles released from a dispensing needle with an inner diameter of 1.19 mm were studied, as shown in figure~\ref{fig:Exp. Setup}. The experiments were conducted under ambient laboratory conditions, with the tank filled with 5 litres of tap water and ambient air supplied to the needle using a programmable syringe pump (NE-1000). Bubble motion was recorded using a JAI Go 5000M camera operating at 125 frames per second (fps). The rectangular field of view was positioned to capture the bubble trajectory from detachment to 15 mm below the free surface, corresponding to a total vertical rise distance of approximately 50 mm at a spatial resolution of $\sim20$ pixels/mm. Illumination was provided by a focused backlight transmitted through a diffuser sheet using broadband halogen fibre-optic illuminators from Thorlabs.

\begin{table}
  \begin{center}
\def~{\hphantom{0}}
  \begin{tabular}{lccc}
    Concentration by Mass& $\mathrm{ppm}$&$\mathrm{mol/m^3}$& $\gamma~(\mathrm{mN/m})$\\[3pt]
    Clean (Tap Water)& -&-& 72.8*\\
    0.001\%       & 0.3&0.02& 49.2\\
    0.0025\%      & 0.7&0.04& 43.1\\
    0.005\%       & 1.5&0.08& 38.1\\
    0.01\%        & 2.9&0.16& 33.3\\
    0.02\%        & 5.8&0.32& 28.3\\
  \end{tabular}
  \caption{Concentrations of TX-100 solutions with surface tensions estimated using Szyszkowski equation \citep{Tagawa14}. *Surface tension of water assumed as that of pure water, subsequent values are calculated to show relative change from addition of surfactants.
  }
  \label{tab:Concentrations Summary}
  \end{center}
\end{table}

The first step of the study involved quantifying bubble rise characteristics in fluid media of varying surfactant concentrations. Precise masses of the non-ionic surfactant Triton X-100 (TX-100) were weighed using an analytical balance with a minimum resolution of 0.1 mg and progressively mixed with water to obtain the target concentrations. The prepared solutions were left to rest for 24 hours to ensure a uniform surfactant distribution and no surface foaming. The surfactant concentrations investigated, together with their estimated surface tensions calculated using the Szyszkowski equation \citep{Tagawa14}, are summarised in Table~\ref{tab:Concentrations Summary}. For clarity, all cases are hereafter referred to by their surfactant concentrations expressed in ppm.

\begin{figure}
\centering
\begin{tikzpicture}
\begin{groupplot}[
    group style={
        group size=1 by 3,
        vertical sep=12pt,
        x descriptions at=edge bottom,
    },
    width=0.7\textwidth,
    height=0.27\textwidth,
    ybar interval,
    enlarge x limits=false,
    xmin=4.0, xmax=5.75,
    ymin=0, ymax=0.4,
    xtick={4.0,4.25,4.5,4.75,5.0,5.25,5.5,5.75},
    xticklabels={4--4.25,4.25--4.5,4.5--4.75,4.75--5.0,5.0--5.25,5.25--5.5,5.5--5.75, 5.75-6.0},
    x tick label style={rotate=45, anchor=east},
    ylabel={Probability},
    tick label style={font=\small},
    legend pos=north east,
]

\nextgroupplot[
    xticklabels=\empty,
]
\addplot+[
    draw=black,
    fill=cb1
]
table[x=bin, y=probability, col sep=space, header=true]
{ArXiv/datfiles/fig2_clean_histogram.dat};
\addlegendentry{Clean}

\nextgroupplot[
    xticklabels=\empty,
]
\addplot+[
    draw=black,
    fill=cb24
]
table[x=bin, y=probability, col sep=space, header=true]
{ArXiv/datfiles/fig2_0.7ppm_histogram.dat};
\addlegendentry{0.7 ppm}

\nextgroupplot[
    xlabel={Bubble Diameter (mm)},
]
\addplot+[
    draw=black,
    fill=cb21
]
table[x=bin, y=probability, col sep=space, header=true]
{ArXiv/datfiles/fig2_5.8ppm_histogram.dat};
\addlegendentry{5.8 ppm}

\end{groupplot}
\end{tikzpicture}
\caption{Distribution of bubble diameters investigated for different concentration cases.}
\label{fig:Diameter Histograms}
\end{figure}
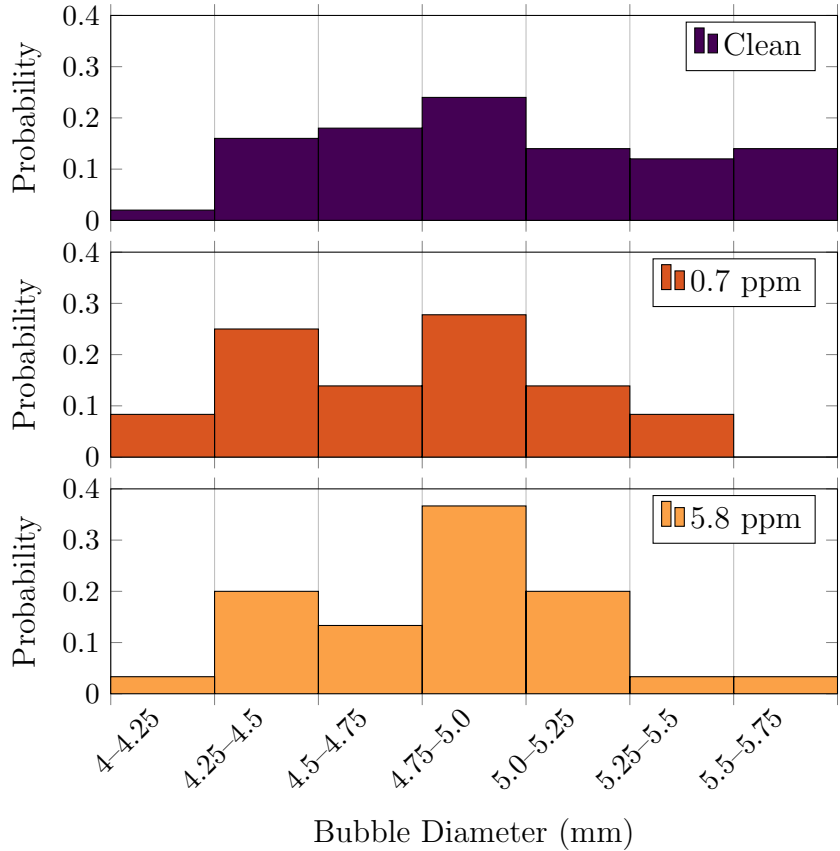

For each concentration case, individual bubbles were released into the solution through the dispensing needle. Bubble diameters in the range $4.0~\mathrm{mm} – 6.0~\mathrm{mm}$ were generated by varying the syringe-pump volumetric flow rate from $5~\mathrm{mL/min}$ to $20~\mathrm{mL/min}$. The resulting bubble-size distribution for increasing concentration cases used in this study is shown in figure~\ref{fig:Diameter Histograms}.
Minor variations in the bubble-size distribution were observed with increasing surfactant concentration. Although bubbles with diameters between $4.75~\mathrm{mm} - 5.0~\mathrm{mm}$ were most prevalent across all cases, a small downward shift in the mean diameter is apparent at higher surfactant concentrations. Accounting for bubble-to-bubble variability and experimental noise, this trend is considered negligible for the present study, and all bubbles investigated ($4.0~\mathrm{mm} \leq d_{eq} \leq 6.0~\mathrm{mm}$) remain within the same dynamical regime \citep{Tomiyama02}. 

\subsection{Surfactant Injection Setup \label{method-inj}}
\begin{figure}
  \centerline{\includegraphics[width=0.7\textwidth]{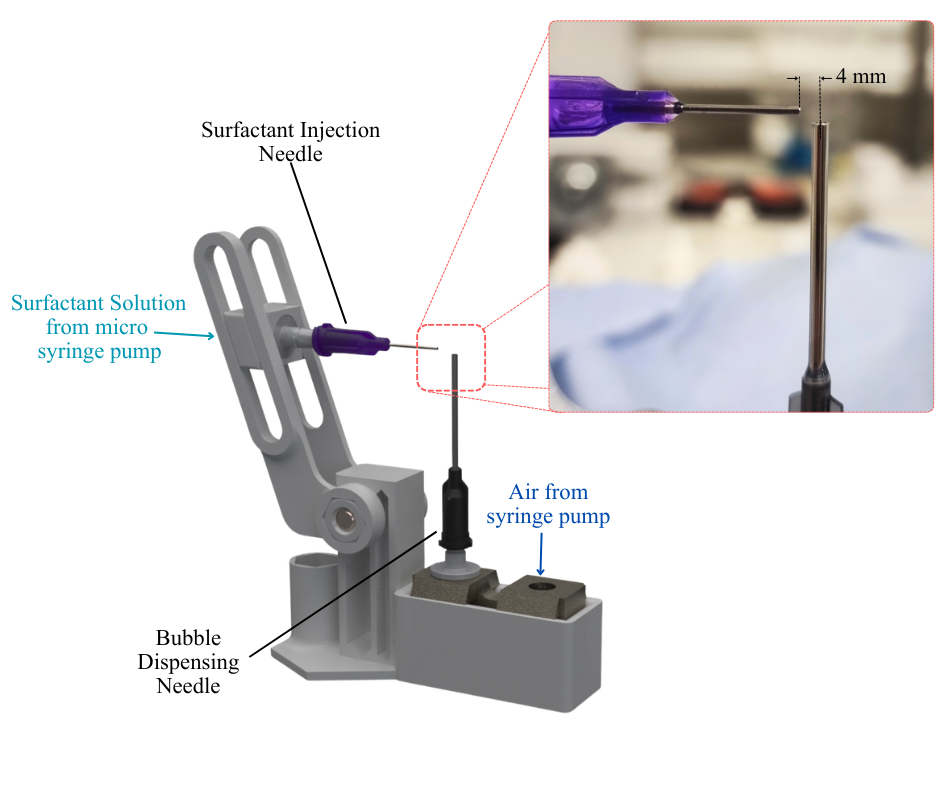}}
  \caption{Surfactant injection apparatus}
\label{fig:Injection apparatus}
\end{figure}

For a final proof-of-principle experiment that qualitatively assesses the proposed prediction framework's capability to identify surfactant-laden bubbles from a sequence of imaged bubbles, a dedicated surfactant injection apparatus was developed. The apparatus consists of a 3D-printed housing incorporating a secondary injection needle oriented perpendicular to the bubble-dispensing needle, enabling controlled pulses of premixed TX-100 solution to be delivered to the exit area of the dispensing needle, as shown in figure~\ref{fig:Injection apparatus}. To prevent interference with bubble growth prior to detachment, the tip-to-tip distance between the two needles was fixed at $4~\mathrm{mm}$ for all tests, exceeding the radius of the largest bubbles generated ($\sim 3~\mathrm{mm}$). Surfactant solutions were delivered using a stepper-motor-driven microsyringe pump controlled by an Arduino Uno, with the motor speed and step count programmed to dispense $10~\mathrm{\mu L}$ of TX-100 solution at a flow rate of $1~\mathrm{mL/min}$ on demand. 

A steady stream of bubbles in a surfactant-free water medium was generated by fixing the flow rate of the air syringe pump at $20~\mathrm{mL/min}$, producing approximately 4 bubbles per second. An imaging period of $4~\mathrm{s}$ was initiated through a manual trigger sent from the computer to the camera. The surfactant-injection pump was set to activate $1.5~\mathrm{s}$ after imaging commenced, delivering $10~\mathrm{\mu L}$ of the specified surfactant concentration. All bubbles generated within the $4~\mathrm{s}$ imaging period, including any surfactant-affected bubbles, are recorded and their rise characteristics extracted for the prediction model.

\subsection{Image Processing\label{method-img}}

Captured images were processed using a Python script to extract the bubble size, aspect ratio  $AR$, and rise velocity. Bubble sizes were quantified using the volume-equivalent diameter \( d_{\text{eq}} \)  as given in (\ref{eq:deq}), where the total bubble volume \(V\) is determined from the last image before bubble detachment (figure~\ref{fig:Bubble rise sequence}). The bubble volume was calculated using edge detection and assuming axial symmetry.

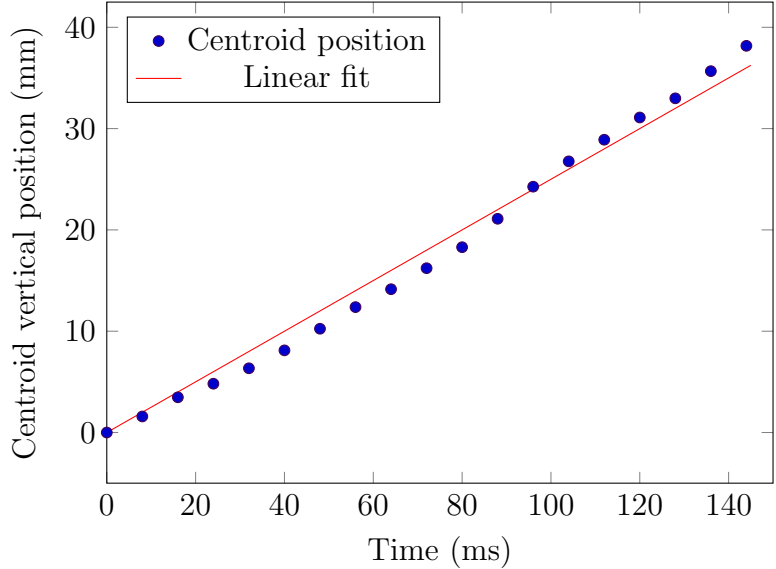
\begin{figure} 
\centering
\begin{tikzpicture}
      	\begin{axis}[width=0.65\textwidth, height=0.5\textwidth, xmin=0, xmax=150, ymin=-5, ymax==40, xlabel={Time (ms)},ylabel={Centroid vertical position (mm)}, legend pos=north west ]
      \addplot+[only marks, color=cb1] table[x=Time(ms), y=Y_centroid(mm)]{ArXiv/datfiles/Fig5_Line.dat};
      \addlegendentry{Centroid position};
      \addplot[smooth, red, domain=0:145] {0.25*x}; 
        \addlegendentry{Linear fit};
      \end{axis}
    \label{fig:centroid}
\end{tikzpicture}
      \caption{Detected centroid vertical position against time for a typical bubble rising in tap water relative to the position where bubble is first fully detached from the needle at $t = 0~\mathrm{ms}$}
    \label{fig:Vertical rise vs time}
\end{figure}

A series of images is selected spanning the interval from immediately prior to detachment until the bubble exits the camera field of view. An example of first six images in a representative sequence are shown in figure~\ref{fig:Bubble rise sequence}, where $t=0$ is defined as the first frame where the bubble has detached from the needle.  Bubble contours are detected in each frame and the corresponding centroid coordinates are extracted. With the camera operating at 125 fps, successive frames are separated by a time interval, \( \Delta t \) of $8 ms$. The centroid position is plotted as a function of time, and the gradient of a linear fit is evaluated to obtain the average rise velocity \( u \). An example of a typical clean bubble is illustrated in figure \ref{fig:Vertical rise vs time}. 

\input{ArXiv/Figures/latex/BubbleRiseSequence}

Uncertainties in the values of \( d_{\text{eq}} \) and $AR$ arise from pixel-level edge detection, where partially occupied pixels may be either fully counted or omitted during contour identification. Therefore, a maximum uncertainty of \(\pm 1\) pixel is considered. Across all bubbles recorded in this study, this translates to an approximate maximum error in \( d_{\text{eq}} \) and $AR$ of  $\pm1.4\%$ and $\pm 2.8\%$ respectively. Accounting additionally for uncertainties associated with spatial resolution and imaging frequency, the uncertainty in the mean rise velocity $u$ is estimated to lie between   $1.5\%$ and $2.6\%$.
 
\section{Results and Discussion}
\subsection{Rise Velocity\label{results-vel}}

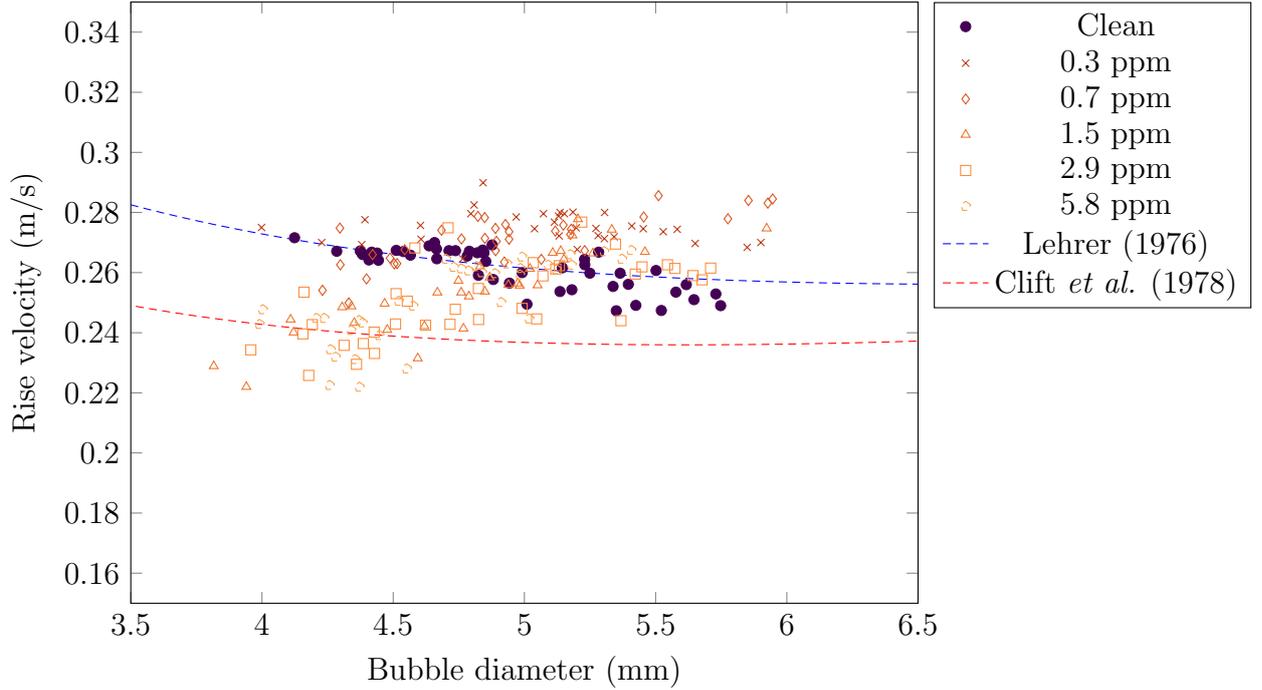
\begin{figure} 
\centering
\begin{tikzpicture}
\begin{axis}[
    width=0.75\textwidth,
    height=0.6\textwidth,
    xmin=3.5, xmax=6.5,
    ymin=0.15, ymax=0.35,
    xlabel={Bubble diameter (mm)},
    ylabel={Rise velocity (m/s)},
    legend pos=south east,
    legend style={at={(1.02,1)}, anchor=north west}
]

\addplot+[
    only marks,
    mark=*,
    mark size=1.9pt,
    color=cb1,
    mark options={fill=cb1}
] table[meta=RiseVelocity_m_per_s]
{ArXiv/datfiles/Fig6_clean_scatter.dat};
\addlegendentry{Clean};

\addplot+[
    only marks,
    mark=x,
    mark size=1.9pt,
    color=cb25
] table[meta=RiseVelocity_m_per_s]
{ArXiv/datfiles/Fig6_0.3ppm_scatter.dat};
\addlegendentry{0.3 ppm};

\addplot+[
    only marks,
    mark=diamond,
    mark size=1.9pt,
    color=cb24,
    mark options={fill=cb24}
] table[meta=RiseVelocity_m_per_s]
{ArXiv/datfiles/Fig6_0.7ppm_scatter.dat};
\addlegendentry{0.7 ppm};

\addplot+[
    only marks,
    mark=triangle,
    mark size=1.9pt,
    color=cb23,
    mark options={fill=cb23}
] table[meta=RiseVelocity_m_per_s]
{ArXiv/datfiles/Fig6_1.5ppm_scatter.dat};
\addlegendentry{1.5 ppm};

\addplot+[
    only marks,
    mark=square,
    mark size=1.9pt,
    color=cb22,
    mark options={fill=cb22}
] table[meta=RiseVelocity_m_per_s]
{ArXiv/datfiles/Fig6_2.9_scatter.dat};
\addlegendentry{2.9 ppm};

\addplot+[
    only marks,
    mark=pentagon,
    mark size=1.9pt,
    color=cb21,
    mark options={fill=cb21}
] table[meta=RiseVelocity_m_per_s]
{ArXiv/datfiles/Fig6_5.8ppm_scatter.dat};
\addlegendentry{5.8 ppm};

\addplot+[
    densely dashed,
    draw = blue,
    mark=none
] table[
    x=Diameter_mm,
    y=RiseVelocity_m_per_s
]{ArXiv/datfiles/Fig6_Lehrer_line.dat};
\addlegendentry{Lehrer (1976)}

\addplot+[
    densely dashed,
    draw = red,
    mark=none
] table[
    x=Diameter_mm,
    y=RiseVelocity_m_per_s
]{ArXiv/datfiles/Fig6_Clift_line.dat};
\addlegendentry{Clift \textit{et al.} (1978)}

\end{axis}
\end{tikzpicture}
\caption{Average rise velocity, $u$ against bubble diameter for different surfactant concentrations, and drag-based terminal rise velocity, $U_{T}$ predictions for clean bubbles by \citet{Lehrer76} and \citet{Clift78}}
\label{fig:Rise Velocity Scatter}
\end{figure}
 
The average rise velocities $u$ of individual bubbles released in solutions with varying TX-100 concentrations are presented in figure~\ref{fig:Rise Velocity Scatter}, together with predicted terminal rise velocities $U_T$ for clean bubbles obtained from theoretical drag correlations \citep{Lehrer76,Clift78}.
Close agreement is observed between the measured mean rise velocity in clean water and predictions from the drag correlation of \citet{Lehrer76}, whereas the correlation of \citet{Clift78} underpredicts the measurements by an average of approximately 10\%. Consistent across both predictions and the present measurements is a downwards trend in rise velocity as bubble diameter increases. This phenomenon has been attributed to additional drag from path oscillations experienced by bubbles in the ellipsoidal regime. However, this trend does not persist for larger bubbles $(d_{\text{eq}} > 7~\mathrm{mm})$, for which path oscillations are minimal and oscillation frequencies are near zero \citep{Lindt72,Li15}.

In general, the introduction of surfactants at concentrations below $1.5~\mathrm{ppm}$ exerts a negligible influence on the early-stage rise velocity. Interestingly, bubbles rising in the $0.3~\mathrm{ppm}$ and $0.7~\mathrm{ppm}$ solutions exhibit slightly higher rise velocities ($ \sim 5\%$) compared to those in the clean system. However, considering the significant noise in the data, no definitive trend can be established without further validations.

At higher surfactant concentrations, a reduction in \( u \) of up to $21\%$ is observed for bubbles with diameters smaller than $5~\mathrm{mm}$. This behaviour is consistent with previous observations that the distance required for bubbles to decelerate to their $U_T$ decreases in the presence of surfactants \citet{Zhang01, Tomiyama02}. Although the bubbles in the present setup have not yet reached $U_T$ within the imaged timeframe, those released in the higher concentration solutions undergo greater deceleration, which likely contributes to the lower measured values of \( u \) .

These findings suggest that at concentrations above $1.5~\mathrm{ppm}$, the accumulation of surfactant molecules at the bubble interface leads to measurable reductions in \( u \). However, this effect does not scale proportionally with further increases in TX-100 concentration, as the mean \( u \) across the range of bubble diameters remains at $ \sim 0.25~\mathrm{m/s}$ in the two highest concentration cases. Moreover, bubbles larger than $5~\mathrm{mm}$ appear largely insensitive to changes in concentration. The relatively small variability and non-monotonic changes in measured rise velocities suggest that early-stage rise velocity is not a sufficiently distinguishing parameter for characterising rise dynamics across surfactant concentrations.

\subsection{Aspect Ratio\label{results-ar}}

\input{ArXiv/Figures/latex/ARforcleanbubbles}

The temporal evolution of the aspect ratio ($AR$) of clean bubbles within the first $144~\mathrm{ms}$ after release is shown in  figure~\ref{fig:AR Clean Profile}. Immediately after detachment, the bubble remains nearly spherical ($AR\sim1$), as observed from the first bubble in figure \ref{fig:Bubble rise sequence}. Subsequently, inertial forces associated with its release initiate a deformation from a spherical bubble to an ellipsoidal disk shape, as seen in insets (a) and (b) in figure~\ref{fig:AR Clean Profile}. The $AR$ increases until it reaches an average peak value of 4.4 after approximately $60~\mathrm{ms}$. Following this peak (d), the bubble transitions back towards a near-spherical shape (e-f), and begins oscillating around its dynamic equilibrium aspect ratio, $AR_{\text{eq}}$. Although it appears that the bubble's shape is visually stabilised at this stage, smaller-amplitude oscillations are expected to continue during further ascent \citep{Kracht10,Wang19,Chabel12,Wang19Langmuir}. $AR_{\text{eq}}$ is calculated as the mean of the last four values of the $AR$ measured between 120~ms and 144~ms and yielding $AR_{\text{eq}}=1.56\pm 0.48$ for clean bubbles. Because the current research focuses on the initial rise dynamics up to $144~\mathrm{ms}$ after detachment, this equilibrium aspect ratio merely serves as a metric to quantify the final state of the bubble within the imaging period and must not be assumed to be the long-term steady state value. While it might be beneficial to investigate $AR$ profiles at later times, the largest oscillation magnitudes occur during the early stage and are captured within the initial cycle. The parameters extracted within this period, as later shown, are significantly sensitive to the tested concentrations in the prediction model. 

\input{ArXiv/Figures/latex/BubbleImages}

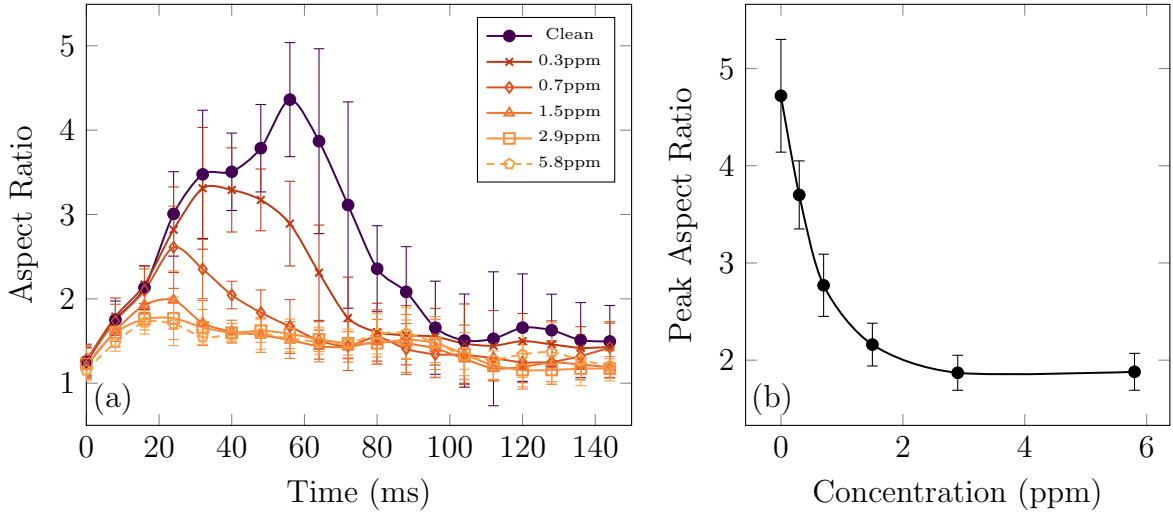
\begin{figure}
\centering
\begin{tikzpicture}

\begin{axis}[
    name=plotA,
    width=0.55\linewidth, height=0.45\linewidth,
    xmin=0, xmax=150, ymin=0.5,
    xlabel={Time (ms)}, ylabel={Aspect Ratio},
    legend pos=north east,
    legend style={font=\tiny}
]

\addplot+[smooth, color=cb1, mark=*, mark options={fill=cb1}, line width=0.75pt ] 
table[x=Time_ms, y=AR_mean]{ArXiv/datfiles/Fig9_Clean_line_std_dev.dat}; 
\addlegendentry{Clean};

\addplot+[smooth, mark=x, line width=0.75pt, color=cb25] 
table[x=Time_ms, y=AR_mean]{ArXiv/datfiles/Fig9_0.3ppm_line_std_dev.dat}; 
\addlegendentry{0.3ppm};

\addplot+[smooth, mark=diamond, line width=0.75pt, color=cb24] 
table[x=Time_ms, y=AR_mean]{ArXiv/datfiles/Fig9_0.7ppm_line_std_dev.dat}; 
\addlegendentry{0.7ppm};

\addplot+[smooth, mark=triangle, line width=0.75pt, color=cb23] 
table[x=Time_ms, y=AR_mean]{ArXiv/datfiles/Fig9_1.5ppm_line_std_dev.dat}; 
\addlegendentry{1.5ppm};

\addplot+[smooth, mark=square, line width=0.75pt, color=cb22] 
table[x=Time_ms, y=AR_mean]{ArXiv/datfiles/Fig9_2.9ppm_line_std_dev.dat}; 
\addlegendentry{2.9ppm};

\addplot+[smooth, mark=pentagon, line width=0.75pt, color=cb21] 
table[x=Time_ms, y=AR_mean]{ArXiv/datfiles/Fig9_5.8ppm_line_std_dev.dat}; 
\addlegendentry{5.8ppm};

\addplot [color=cb1, only marks] 
plot [error bars/.cd, y dir=both, y explicit] 
table[x=Time_ms,y error index=2]{ArXiv/datfiles/Fig9_Clean_line_std_dev.dat};

\addplot [color=cb25, only marks, mark=x] 
plot [error bars/.cd, y dir=both, y explicit] 
table[x=Time_ms,y error index=2]{ArXiv/datfiles/Fig9_0.3ppm_line_std_dev.dat};

\addplot [color=cb24, only marks, mark=diamond] 
plot [error bars/.cd, y dir=both, y explicit] 
table[x=Time_ms,y error index=2]{ArXiv/datfiles/Fig9_0.7ppm_line_std_dev.dat};

\addplot [color=cb23, only marks, mark=triangle] 
plot [error bars/.cd, y dir=both, y explicit] 
table[x=Time_ms,y error index=2]{ArXiv/datfiles/Fig9_1.5ppm_line_std_dev.dat};

\addplot [color=cb22, only marks, mark=square] 
plot [error bars/.cd, y dir=both, y explicit] 
table[x=Time_ms,y error index=2]{ArXiv/datfiles/Fig9_2.9ppm_line_std_dev.dat};

\addplot [color=cb21, only marks, mark=pentagon] 
plot [error bars/.cd, y dir=both, y explicit] 
table[x=Time_ms,y error index=2]{ArXiv/datfiles/Fig9_5.8ppm_line_std_dev.dat};

\end{axis}

\begin{axis}[
    name=plotB,
    at={(plotA.east)}, anchor=west, 
    xshift=1.5cm,                  
    width=0.45\linewidth, height=0.45\linewidth,
    xlabel={Concentration (ppm)}, ylabel={Peak Aspect Ratio}
]

\addplot+[smooth, mark=*, line width=0.75pt, color=black] 
table[x=Concentration_ppm, y=ARPeak]{ArXiv/datfiles/Fig9_b_line_std_dev.dat}; 

\addplot [color=black, only marks] 
plot [error bars/.cd, y dir=both, y explicit] 
table[x=Concentration_ppm, y error index=2]{ArXiv/datfiles/Fig9_b_line_std_dev.dat};

\end{axis}

\node at ($(plotA.south west)+(0.35,0.35)$) {(a)};
\node at ($(plotB.south west)+(0.35,0.35)$) {(b)};

\end{tikzpicture}

\caption{(a) Average aspect ratio profiles for different surfactant concentrations; 
(b) Peak aspect ratio vs surfactant concentration}
\label{fig:AR Profiles}

\end{figure}

The shape oscillations of bubbles in the various surfactant-laden media are presented in figure~\ref{fig:Bubble in Surfactants}. An increased damping effect was observed where bubble shapes became more spherical as surface contamination increased, confirming observations from previous studies \citep{Kracht10,Wang19,Wang19Langmuir}. This effect is most noticeable when inspecting the bubbles at a vertical position of approximately $10~\mathrm{mm}$ from the detachment point, where the presence of a distinct ellipsoidal phase diminishes at higher surfactant concentrations. The temporal $AR$ variations were quantified for a minimum of 30 bubbles in each surfactant solution. The mean $AR$ profiles with standard deviations are presented for all cases in figure~\ref{fig:AR Profiles}.

A general oscillation behaviour similar to clean bubbles is observed across all cases, where the $AR$ rises to a peak after release and gradually settles near the $AR_{eq}$. However,  systematic changes in the magnitude and timing of this peak can be deduced as surfactant concentrations increase.  The peak aspect ratio, $AR_{peak}$ decreases immediately with the introduction of surfactants, signifying a noticeable damping effect that attenuates shape oscillations even at low concentrations. This damping effect amplifies with higher surfactant concentrations, where $AR_{peak}$ values approach a minimum in the $2.9~\mathrm{ppm}$ and $5.8~\mathrm{ppm}$ cases, depicted in figure \ref{fig:AR Profiles}(b).

The increased surfactant-induced damping also shifts the time required to reach the peak aspect ratio, $t_{peak}$, to earlier values with each increasing concentration.  The diminishing shape oscillations produce a less pronounced ellipsoidal phase (figure \ref{fig:Bubble in Surfactants}). This reduces $t_{peak}$ and allows contaminated bubbles to begin relaxing towards the equilibrium $AR$ earlier. These trends observed from the $AR$ profiles in figure \ref{fig:AR Profiles}(a) are quantified and the averages for the respective concentrations are presented in table \ref{tab:AR Profile Features}.

\begin{table}
  \begin{center}
\def~{\hphantom{0}}
  \begin{tabular}{lcccc}
    Concentration ($\mathrm{ppm}$)& $AR_{\text{peak}}$ & $t_{\text{peak}}~ (\mathrm{ms})$ & $AR_{\text{eq}}$ & $t_{\text{eq}}~(\mathrm{ms})$ \\[3pt]
    Clean& ~4.72& ~59.84 & ~1.56& ~92.48 \\
    0.3& ~3.78& ~38.12 & ~1.45& ~71.76 \\
    0.7& ~2.77& ~24.66 & ~1.31& ~61.11 \\
    1.5& ~2.16& ~24.00 & ~1.21& ~48.00 \\
    2.9& ~1.87& ~29.84 & ~1.17& ~92.72 \\
    5.8& ~1.88& ~38.45 & ~1.29& ~51.55 \\
  \end{tabular}
  \caption{Average features of bubbles in different surfactant concentrations}
  \label{tab:AR Profile Features}
  \end{center}
\end{table}

At surfactant concentrations above $1.5~\mathrm{ppm}$, damping occurs almost immediately after bubble release, and a distinct ellipsoidal phase is often absent. Throughout the measurement period, the $AR$ values remain below 2, with only small fluctuations observed slightly above the final $AR_{eq}$. To distinguish the presence of an ellipsoidal bubble phase, a threshold of $AR = 2.5$ is defined between the values measured for concentrations $0.7~\mathrm{ppm}$ and $1.5~\mathrm{ppm}$. This threshold is chosen to be higher than the average $AR_{\text{peak}}$ plus one standard deviation measured at $1.5~\mathrm{ppm}$, thereby avoiding misclassification of weak oscillations as ellipsoidal deformation. Using this criterion, individual bubble images can be classified according to the presence of an ellipsoidal phase, and is later used as a distinguishing feature in the prediction model discussed in Section 5.1.

At higher concentrations, the absence of a pronounced initial peak occasionally results in maximum  $AR$ values occurring later during the ascent. This behaviour skews the average $t_{peak}$ and $t_{eq}$ producing inflated values for $2.9~\mathrm{ppm}$ and $5.8~\mathrm{ppm}$ cases reported in table \ref{tab:AR Profile Features} which deviate from the otherwise decreasing trend. Comparison of the two highest concentrations reveals nearly identical $AR$ profiles and very similar $AR_{peak}$ values. This finding suggests that the resulting accumulation of surfactant molecules on the bubble's interface approaches saturation at $2.9~\mathrm{ppm}$, beyond which further increases in concentrations produce only marginal changes to bubble shape dynamics. 

A final general observation is that clean bubbles exhibit the largest standard deviations in aspect ratio between different experiments. Although all clean bubbles are released in the same water solution, the observed variability suggests the presence of small differences in local contamination levels arising from experimental limitations in maintaining pure surfactant-free conditions. This highlights the inherent high sensitivity of clean bubbles to contamination and is in line with the highest drop in $AR_{peak}$ seen in figure \ref{fig:AR Profiles}(b) at the lowest concentrations. 

\section{Predicting Surfactant Concentrations from Aspect Ratio Variations \label{prediction model}}

Experiments conducted in surfactant-laden media with concentrations ranging from $0.3~\mathrm{ppm}$ to $5.8~\mathrm{ppm}$ demonstrate that both rise velocity and shape oscillations generally decrease for bubbles released in more concentrated media. Although measurable reductions in $u$ are observed for concentrations above $1.5~\mathrm{ppm}$, the results do not establish a consistent relationship between rise velocity and surfactant concentration across the range of bubble sizes tested. In contrast, the aspect ratio ($AR$) variations reveal a clear damping effect on shape oscillations following the introduction of surfactants, with distinct transient profiles persisting up to the $2.9~\mathrm{ppm}$ case (figure \ref{fig:AR Profiles}). Beyond this concentration the damping effect reaches a plateau, and the corresponding $AR$ profiles become largely indistinguishable.

Building on these findings, the following section introduces a workflow to derive a predictive relationship between bubble shape oscillations during initial ascent and the surrounding surfactant concentration. Because of the observed similarity in behaviour at the two highest concentrations, only empirical $AR$ data from the clean to $2.9~\mathrm{ppm}$ cases are used.

\subsection{Feature Extraction and Selection\label{predict-fea}}

Several indicators, as presented in Table~\ref{tab:AR Profile Features}, can be directly extracted to characterise the general behaviour of the $AR$ profiles, most notably the magnitude and timing of the bubble’s most pronounced ellipsoidal state, denoted by $AR_{peak}$ and $t_{peak}$.
To further characterise the oscillatory behaviour across all cases, additional quantitative features were identified and are summarised in table \ref{tab:Features Overview}. Derived from averaged and individual bubble $AR$ trends, these features serve to evaluate the intensity of shape oscillations.

\begin{table}
  \begin{center}
\def~{\hphantom{0}}
\renewcommand{\arraystretch}{1.3}
  \begin{tabular}{lp{12.5cm}}
    \textbf{Feature} & \textbf{Description} \\[6pt]
    $AR_{\text{peak}}$& Maximum $AR$ value achieved during the rise. \\[6pt]
 &$AR_{\text{peak}} = max(AR)$\\
    $t_{\text{peak}}$& Time taken to reach $AR_{\text{peak}}$.\\[6pt]
    $AR_{\text{eq}}$& Mean $AR$ value after 120 ms. \\[6pt]
 &$\displaystyle AR_{eq} =  \frac{1}{n_{t>120ms}} \sum_{t_i \ge 120ms} \text{AR}(t_i)$\\
 &where $n$ is the number of recorded $AR$ values.\\
    $t_{\text{eq}}$& Time instance where $AR_{\text{eq}}$ is first reached.\\[6pt]
    $\bar{AR}$& Mean $AR$ value throughout rise. \\[6pt]
 &$\displaystyle \bar{AR} = \frac{1}{n} \sum_{i} \text{AR}_{i} $\\
    $\sigma_{\text{$AR$}}$& 
    Standard deviation of $AR$. \\
 &$ \displaystyle \sigma_{\text{$AR$}} = \sqrt{\frac{1}{n} \sum_{i} (AR_i - AR_{\text{mean}})^2}$\\
    $t_{ellip}$& Time period where $AR$ is above a preset elliptical threshold of \\ & $AR$ = 2.5 (discussed in section \ref{results-ar})\\[6pt]
    $\dot{AR}_{max}$& 
    Maximum rate of change of $AR$. \\
 &$\displaystyle \dot{AR}_{max} = \max\left(\frac{AR(t_{i+1}) - AR(t_i)}{\Delta t}\right)$\\ 
    & where $\Delta t = 8$ ms based on the camera’s capture rate of 125 fps.\\
    $\Delta_{pt}$& 
    Peak to trough ratio. \\[6pt]
 &$\displaystyle \Delta_{pt} = \frac{\max(AR)}{\min(AR)}$\\
    $AUC_{AR}$& 
    Area under curve of $AR$ variation against time. \\
 &$\displaystyle AUC_{AR} = \sum_{i} \frac{AR(t_{i+1}) + AR(t_i)}{2} (t_{i+1} - t_i)$\\
  \end{tabular}
  \caption{Summary of extracted features from $AR$ variation profiles.}
  \label{tab:Features Overview}
  \end{center}
\end{table}

As higher surfactant concentrations induce stronger damping effects, the corresponding feature values are expected to decrease. The extent to which these features respond to varying concentrations determines their effectiveness in distinguishing between cases. To assess this sensitivity, the Pearson correlation coefficient, $r$ is employed to quantify the relationship between each feature and surfactant concentration. 

\begin{equation}
r_{x,s} = \frac{\sum_{i=1}^{n} (x_i - \bar{x})(s_i - \bar{s})}
{\sqrt{\sum_{i=1}^{n} (x_i - \bar{x})^2 \sum_{i=1}^{n} (s_i - \bar{s})^2}}
\end{equation}

where $r_{x,s}$ is the correlation between the value of a particular feature $x$ and surfactant concentration $s$, and $\bar{x}$ and $\bar{s}$ are their respective mean values across all measured cases.

\begin{table}
  \begin{center}
\def~{\hphantom{0}}
\renewcommand{\arraystretch}{1.3}
  \begin{tabular}{l>{\centering\arraybackslash}p{6cm}}
    Feature & Pearson correlation coefficient, $r$ with respect to surfactant concentration \\[6pt]
    $t_{\text{eq}}$ & 0.008\\[6pt]
    $t_{\text{peak}}$ & -0.445\\[6pt]
    $AR_{\text{eq}}$ & -0.587\\[6pt]
    $AUC_{AR}$& -0.762\\[6pt]
    $\Delta_{pt}$& -0.763\\[6pt]
    $\bar{AR}$& -0.765\\[6pt]
    $t_{ellip}$& -0.791\\[6pt]
    $\sigma_{AR}$ & -0.805\\[6pt]
    $AR_{\text{peak}}$ & -0.833\\[6pt]
    $\dot{AR}_{max}$& -0.835\\
  \end{tabular}
  \caption{Correlation of extracted features with surfactant concentration}
  \label{tab:Feature r values}
  \end{center}
\end{table}
The computed $r$ values for all identified features are tabulated in table \ref{tab:Feature r values}. The Pearson coefficient quantifies the strength and direction of a linear relationship between two parameters, where values close to $1$ or $-1$ indicate strong positive or negative linear correlations, respectively, while a value near zero suggests no linear relationship. Apart from $t_{eq}$, all features exhibit negative correlations of varying strengths with surfactant concentration. To further illustrate the variation between features of high and low $r$ values, individual features are examined across the tested concentrations.

\begin{figure}
\centering
    \begin{tikzpicture}
         \begin{axis}[width=0.6\textwidth, axis line style={black},every axis label/.append style ={fadv},
    every tick label/.append style={fadv},  xmin=-0.1, xmax=3, ymin=0.07, ymax=0.145, axis y line*= right, ylabel={\textcolor{fadv}{$t_{eq}~(\mathrm{s})$}}]
          \addplot+[smooth, line width=0.75pt, color=fadv, mark=square*, mark color=fadv] table[x=Concentration_ppm, y=Mean]{ArXiv/datfiles/Fig11_b_line_std_dev.dat};
            \addplot [color=fadv, only marks,legend image post style={mark=[correct mark]}]
 plot [error bars/.cd, y dir = both, y explicit]
 table[x=Concentration_ppm,y error index=2]{ArXiv/datfiles/Fig11_b_line_std_dev.dat};
        \end{axis}
        \begin{axis}[width=0.6\textwidth, axis line style={black},every axis label/.append style ={fdel},
    every tick label/.append style={fdel} , xmin=-0.1, xmax=3, ymin=35, ymax=300, axis y line*=left, xlabel={\textcolor{black}{Surfactant Concentration (ppm)}}, ylabel={\textcolor{fdel}{$\dot AR_{max}~(\mathrm{s^{-1}})$}}]
\addplot+[smooth, mark=*, line width=0.75pt, color=fdel] table[x=Concentration_ppm, y=Mean]{ArXiv/datfiles/Fig11_a_line_std_dev.dat}; 
    \addplot [color=fdel, only marks,legend image post style={mark=[correct mark]}]
 plot [error bars/.cd, y dir = both, y explicit]
 table[x=Concentration_ppm,y error index=2]{ArXiv/datfiles/Fig11_a_line_std_dev.dat};
        \end{axis}

    \end{tikzpicture}
    \caption{Variation of features extracted from measured $AR$ profiles with the strongest and weakest correlations to surfactant concentrations up to $2.9ppm$: (a) Max rate of change, $\dot{AR}_{max}$; (b) Time to dynamic equilibrium, $t_{eq}$}
    \label{fig:feature comparison}
\end{figure}
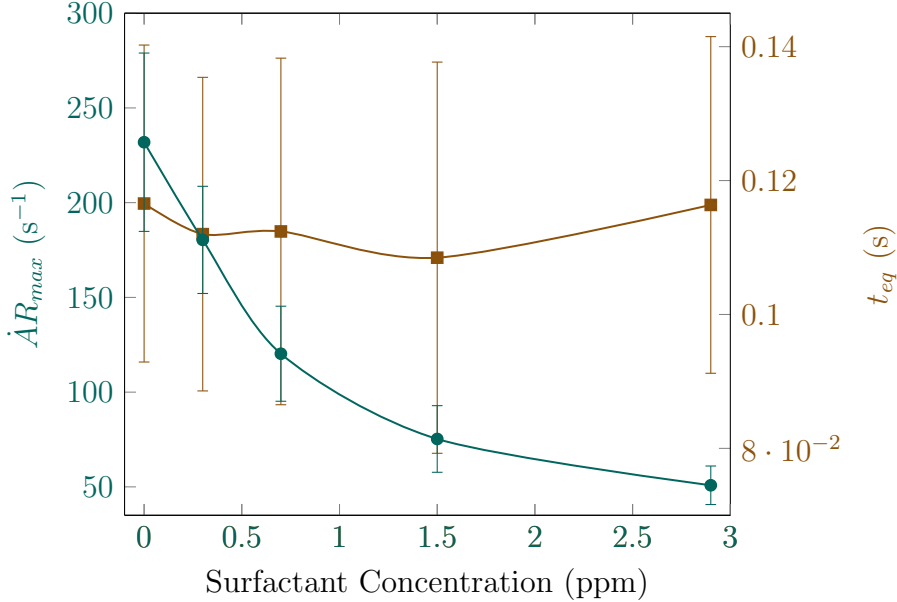

The mean and standard deviation of two representative features across all concentration cases— the most strongly correlated feature, the maximum rate of change in aspect ratio, $\dot{AR}_{max}$, and the weakest correlated feature $t_{eq}$— are presented in Figure~\ref{fig:feature comparison}. A clear decrease in $\dot{AR}_{max}$ is observed with increasing surfactant concentration, reflecting the progressive suppression of peak $AR$ values. The separation of feature distributions across concentrations further highlights its potential as a reliable predictor. In contrast, $t_{eq}$ exhibits no obvious trend and shows substantial scatter in all cases, further emphasising its limited sensitivity to damping effects. These comparisons highlight the contrast between features that robustly capture concentration-dependent behaviour and those that remain less informative. Consequently, this makes feature selection essential, ensuring that only features with strong and consistent relationships are retained for model development.

A feature selection procedure was implemented to retain features with meaningful predictive influence while discarding those contributing to noise or inaccuracies. All extracted features are first ranked by their absolute correlation coefficient values, $|r|$, with the ordering reflected in table \ref{tab:Feature r values}. A variable threshold was then introduced and systematically increased to exclude features with $|r|$ values below the threshold. The threshold increments were selected so that only one additional feature is excluded at a time. This creates a series of candidate feature sets, each representing a progressively stricter selection criteria. For each set, the features were used to form a cumulative index that yields an interpretable trend with surfactant concentrations. Details on the derivation of this index is described in the following subsections. Once obtained, the index is used to estimate  predicted concentration value for all tested bubbles, which are then compared with the experimentally prescribed concentrations.

To determine the threshold that yields the optimal feature set, two performance metrics were evaluated from the predictions: error and uncertainty. The error metric was computed as the mean percentage deviation from the experimentally prescribed surfactant concentrations. The uncertainty metric was derived from the spread of predicted values at each concentration. Using the standard deviation of the computed index, upper and lower bounds were obtained in addition to the central prediction, and the difference between these bounds was taken as a measure of prediction confidence.

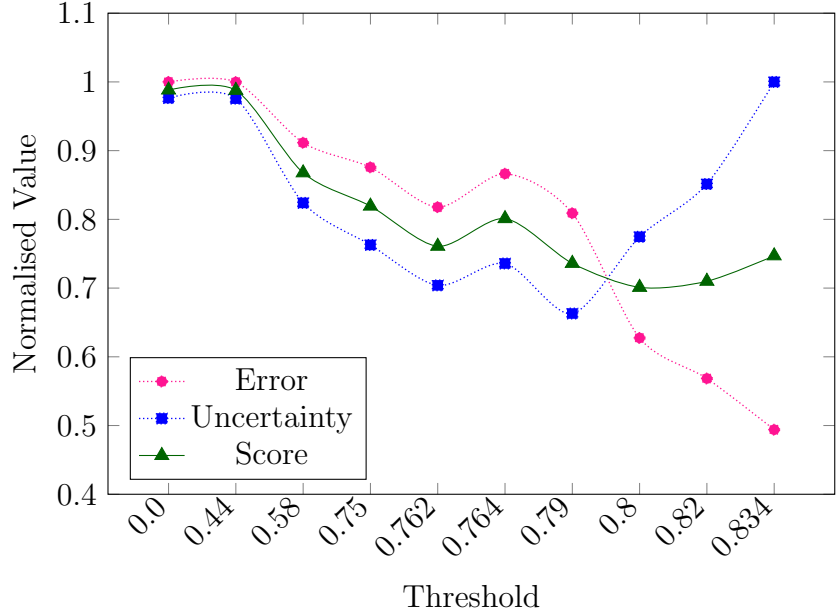
\begin{figure}
\centering
\begin{tikzpicture}
\begin{axis}[
    width=0.7\textwidth,
    height=0.5\textwidth,
    ymin=0.4, ymax=1.1,
    xlabel={Threshold},
    ylabel={Normalised Value},
    legend pos=south west,
    xtick=data,
    xticklabels={0.0,0.44,0.58,0.75,0.762,0.764,0.79,0.8,0.82,0.834},
    x tick label style={rotate=45, anchor=east},
]

\addplot+[
    densely dotted,smooth,line width= 0.5pt,
    color=deeppink, mark=*,
    mark options={fill=deeppink}
] table[
    x expr=\coordindex,   
    y=Error
]{ArXiv/datfiles/Fig12_error_line.dat};

\addlegendentry{Error}

\addplot+[
    densely dotted, line width=0.5pt, smooth,
    color=blue, mark=square*,
    mark options={fill=blue}
] table[
    x expr=\coordindex,   
    y=Uncertainty
]{ArXiv/datfiles/Fig12_uncertainty_line.dat};

\addlegendentry{Uncertainty}

\addplot+[mark=triangle*,
    mark options={fill=darkgreen}, mark size=3 pt,
    smooth,
    color=darkgreen
] table[
    x expr=\coordindex,   
    y=Score
]{ArXiv/datfiles/Fig12_score_line.dat};

\addlegendentry{Score}

\end{axis}
\end{tikzpicture}
\caption{Variation of prediction error, uncertainty and score with imposed threshold values}
\label{fig:Threshold score}
\end{figure}

For each candidate feature set, these metrics were evaluated across all concentration cases. Both were normalised by their maximum values to allow comparability, and subsequently combined into a single performance score defined as:

\begin{equation}
\text{Score} = 0.5 \cdot \text{Error} + 0.5 \cdot \text{Uncertainty}
\end{equation}
By assigning both metrics an equal weight, the score provides a balanced measure of predictive accuracy and reliability. Figure \ref{fig:Threshold score} presents the trend of both normalised metrics and the final scores for each threshold value imposed to give the individual candidate feature sets.

The trends clearly indicate that when up to a $|r|$ threshold of 0.44 is applied, and all weakly correlated features are included, the derived prediction model exhibits the largest errors and uncertainties. When only the strongest correlated feature is retained with the threshold set at 0.83, minimum error is observed. This outcome is expected, as predictions is only driven solely by the most-sensitive feature, eliminating any noise from weaker correlating features. However, as shown in figure \ref{fig:feature comparison}, variations in feature values between bubbles persist at each concentration, and relying on a single feature amplifies this variability, leading to higher uncertainties. Incorporating multiple features with relatively high $|r|$ values introduces an averaging effect, improving robustness and confidence in predictions. Ultimately, the optimal set of features yielding the minimum score was identified at a threshold of $|r|$ = 0.8. 

\subsection{Aspect Ratio Damping Index\label{predict-ardi}}

While individual high-correlated features provide useful insight into the damping behaviour induced by increasing surfactant concentrations, relying on any one in isolation limits the robustness of predictions. The Aspect Ratio Damping Index ($ARDI$) condenses the selected features into a single quantitative indicator, establishing a link between bubble shape oscillation dynamics and surfactant concentrations. We define the index as:
\begin{equation}
ARDI = \sum_{i=1}^{N} X_i \cdot W_i
\label{ARDI eqn}
\end{equation}
where $N$ is the total number of selected features, $X_i$ is the standardised value of feature $x_i$, and $W_i$ is its assigned weight. Standardisation is performed across all recorded bubbles as:
\begin{equation}
X_i = \frac{x_i - \bar{x}}{\sigma_x}
\end{equation}
With $\bar{x}$ and $\sigma_x$ denoting the mean and standard deviation of the feature across all concentrations, respectively. The weight of each feature is determined from its relative $r$ value listed in table \ref{tab:Feature r values} as:
\begin{equation}
    W_i = \frac{|\: r_i \:|}{\sum |\: r_{\text{selected}} \:|}    
\end{equation}
The final set of selected features from imposing the determined threshold of $|r|$ = 0.8 and their corresponding weights alongside their initial $r$ values are summarised in table \ref{tab:Selected feature weights}. It is noted that the $W$ is nearly identical across the selected features owing to their similar $r$ values.

\begin{figure}
\centering
\begin{tikzpicture}
\begin{axis}[
    width=0.7\textwidth,
    height=0.5\textwidth,
    xmin=-0.1, xmax=3,
    ymin=-1.5, ymax=2,
    xlabel={Surfactant Concentration (ppm)},
    ylabel={$ARDI$},
    legend pos=north east,
]

\addplot+[
    smooth,
    draw=cb1,
    mark=none
] table[
    x=Concentration_ppm,
    y=ARDI_mean_fit
]{ArXiv/datfiles/Fig13_mean_fit_line.dat};
\addlegendentry{Mean Fit}

\addplot+[
    densely dashed,
    mark=none
] table[
    x=Concentration_ppm,
    y=ARDI_upper_std_fit
]{ArXiv/datfiles/Fig13_upper_std_fit_line.dat};
\addlegendentry{Upper std Fit}

\addplot+[
    densely dashed,
    draw = blue,
    mark=none
] table[
    x=Concentration_ppm,
    y=ARDI_lower_std_fit
]{ArXiv/datfiles/Fig13_lower_std_fit_line.dat};
\addlegendentry{Lower std Fit}

\addplot+[
    only marks,
    mark=*,
    mark size=2.2pt,
    draw=black,
    fill=black,
    error bars/.cd,
        y dir=both,
        y explicit,
        error bar style={line width=0.6pt},
        error mark options={mark size=2pt},
] table[
    x=Concentration_ppm,
    y=Mean_ARDI,
    y error=yerr
]{ArXiv/datfiles/Fig13_mean_points_scatter_std_dev.dat};
\addlegendentry{Mean $\pm$ std}

\end{axis}
\end{tikzpicture}
\caption{$ARDI$ against surfactant concentration with mean fit and $\pm\sigma$ envelopes}
\label{fig:ARDI derived}
\end{figure}
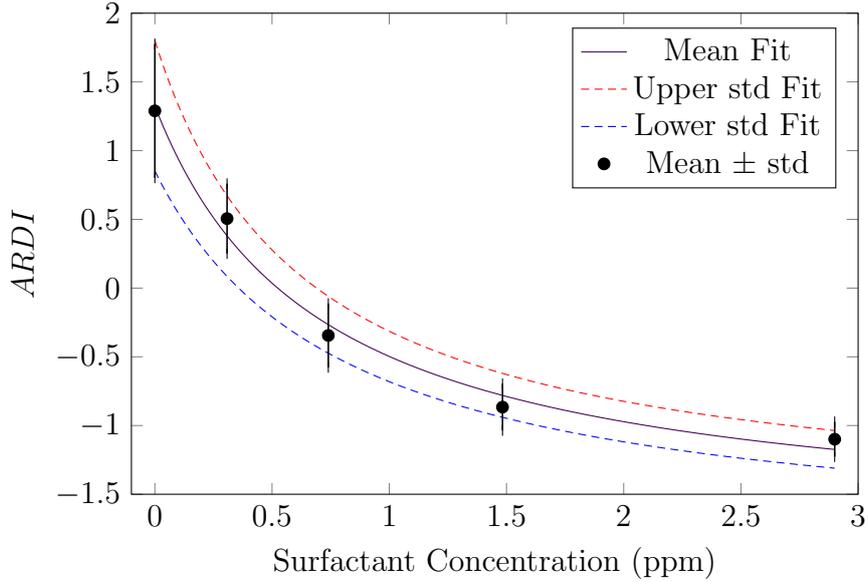

\begin{table}
  \begin{center}
\def~{\hphantom{0}}
  \begin{tabular}{lcc}
    Selected Feature& Pearson correlation coefficient, $r$ & Assigned Weight, $W$\\[3pt]
    $\dot{AR}_{max}$& -0.835& 0.338\\
    $AR_{\text{peak}}$& -0.833& 0.337\\
    $\sigma_{AR}$& -0.805& 0.325\\
  \end{tabular}
  \caption{$W$ and $r$ values of selected features}
  \label{tab:Selected feature weights}
  \end{center}
\end{table}
Using these tabulated parameters, the $ARDI$ of every bubble can be evaluated by extracting the features from their measured $AR$ profiles and using eq. \ref{ARDI eqn}. The resulting average $ARDI$ values across the tested concentrations are presented in figure \ref{fig:ARDI derived}.

The derived relationship between $ARDI$ and surfactant concentration provides a direct and interpretable means of estimating concentration from the transient $AR$ behaviour of rising bubbles. Confidence bounds are obtained from the upper and lower fitted curves, quantifying the uncertainty in predictions based on variability within the dataset. The fitted expressions of $ARDI$ as a function of surfactant concentration, $s$, for each of these curves are presented below:
\begin{equation}
ARDI(s)_{mean}  = -1.781 + \frac{2.189}{0.704 + s}
\label{ARDI_mean}
\end{equation}

\begin{equation}
ARDI(s)_{upper} = -1.657 + \frac{2.196}{0.635 + s}
\label{ARDI_upper}
\end{equation}

\begin{equation}
ARDI(s)_{lower} = -1.906 + \frac{2.204}{0.799 + s}
\label{ARDI_lower}
\end{equation}
The value of $ARDI$ decays with higher concentrations, mirroring the damping effect on shape oscillations. At high concentrations, the index appears to be approaching a plateau. This behaviour reflects the diminishing changes in $AR$ variation as surfactant concentrations exceed $2.9~\mathrm{ppm}$, as observed in figure \ref{fig:AR Profiles}  and discussed in section \ref{results-ar}. This saturation effect highlights a practical limitation of the current model, as predictions become less reliable when $AR$-based features converge and lose distinctiveness. Owing to the standardisation procedure, the computed $ARDI$ values lie predominantly within the range of $-1$ to $+1$. Bubbles released in clean water yield $ARDI$ values close to $+1$, where the largest deviations are also observed, demonstrating the sensitivity of interfacial dynamics to even trace levels of contamination. In contrast, $ARDI$ values near $-1$ correspond to bubbles in highly surfactant-rich media, which in the present study is capped at $2.9~\mathrm{ppm}$ surfactant concentration.

\subsection{Quality of Predictions\label{predict-quality}}

The prescribed methodology aims to provide a systematic empirical approach to derive the relationship between surfactant levels and a quantifiable index representative of bubble shape damping effects. This subsection analyses the performance of the derived model by producing predicted concentrations of the bubbles recorded in the experiments. Unlike in the feature selection process, where a global average of error and uncertainty were used to guide the choice of features, the present analysis reports these quantities individually for each tested concentration. This provides a clearer picture of how prediction accuracy varies with surfactant level, and highlights the relative strengths and limitations of the model under different conditions.

\begin{figure}
\centering
\begin{tikzpicture}
\begin{axis}[ybar, 
    width=0.7\linewidth,     
    height=0.55\linewidth,
    xmin=-0.2, xmax=4.2, 
	xlabel={Concentrations Tested},  xtick={0,1,2,3,4},xticklabels={0ppm,0.3ppm ,0.7ppm,1.5ppm,2.9ppm},
	ylabel=Surfactant Concentration (ppm),
	enlargelimits=0.05, 
    legend pos=north west,
]
\addplot coordinates {(0,0)
        (1,0.3)
		 (2,0.7) 
         (3,1.5) 
         (4,2.9)};
\addplot+[error bars/.cd, y dir=both,y explicit]coordinates{(0, 0.058408) +- (0,0.097624)
(1, 0.26526) +- (0, 0.11737)
(2, 0.85868) +-(0, 0.25456)
(3, 1.777) +-(0, 0.50163)
(4, 2.6162) +-(0,0.60598)  
};
\legend {Actual,
Predicted}

\node[font=\scriptsize, text=red,xshift=-10pt] at (axis cs:1,0.45) {-13.9\%};
\node[font=\scriptsize, text=red,xshift=-15pt] at (axis cs:2,0.82) {+16.0\%};
\node[font=\scriptsize, text=red,xshift=-15pt] at (axis cs:3,1.65) {+19.8\%};
\node[font=\scriptsize, text=red,xshift=-10pt] at (axis cs:4,3.05) {-9.8\%};

\end{axis}
\end{tikzpicture}
\caption{Predicted surfactant concentrations using $ARDI(s)_{mean}$ (eq. \ref{ARDI_mean}) of recorded bubbles compared against actual experimental concentrations; with standard deviation for each concentration case and relative error in percentage annotated in red.}
\label{fig:Act vs pred}
\end{figure}
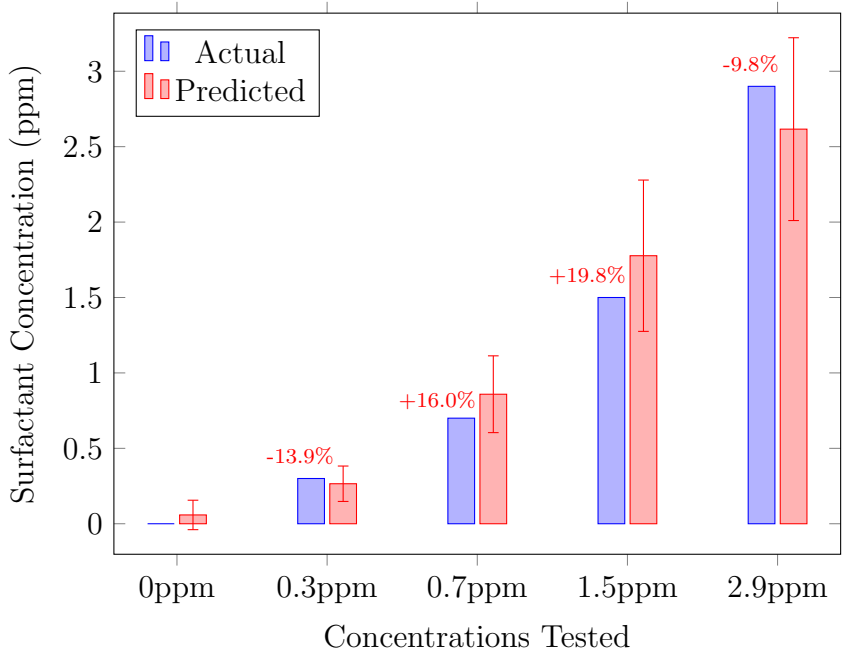

From $ARDI$ values computed for each recorded bubble, surfactant concentration predictions can be made by inverting the obtained relationships (\ref{ARDI_mean}) - (\ref{ARDI_lower}). The error is then defined as the percentage deviation between the predicted value using the mean $ARDI$ fit and actual experimental concentrations. Figure \ref{fig:Act vs pred} presents the predicted values averaged across all bubbles in each tested surfactant concentration with their relative error.

The results demonstrate that the derived model is capable of distinguishing bubbles released in both high and low surfactant concentration media, reinforcing the robustness of $ARDI$ as a predictive indicator. 
At low concentrations, bubble dynamics are highly sensitive to even small changes in surfactant concentration, resulting in significant variations in shape oscillation profiles and, consequently, in the extracted features. This strong behavioural contrast benefits the model, where predictions are less sensitive to small uncertainties in $ARDI$ as seen from the steep slope at lower concentrations in figure \ref{fig:ARDI derived}. At intermediate concentrations between $0.3~\mathrm{ppm}$ and $1.5~\mathrm{ppm}$, bubbles continue to exhibit relatively large variations in oscillation behaviour but also become more sensitive to $ARDI$, leading to moderate increases in prediction error. The model's ability to discriminate intermediate concentrations is evident from the standard deviation bands in figure \ref{fig:Act vs pred}, where they remain disjoint between cases up till $2.9~\mathrm{ppm}$. Here, where oscillations are heavily damped and changes in concentration produce only minute variations in $AR$ profiles, high  $ARDI$ sensitivity diminishes the model's reliability. Compounding errors from measurements and experimental uncertainties are amplified, leading to larger deviations in predicted values. 
Although a relatively low prediction error is observed for bubbles released in the $2.9~\mathrm{ppm}$ solution, it does not reflect the model's performance for highly surfactant-laden bubbles. The reported error is only meaningful when the surfactant concentration is known to lie within the model’s calibrated range ($\leq2.9~\mathrm{ppm}$). In general, for bubbles released in higher-concentration media, predictions will collapse towards the upper training limit, regardless of the true value. This loss of reliability arises from the increasingly indistinguishable shape oscillation behaviour of near-saturated bubbles, ultimately restricting the current model’s applicability to concentrations up to $2.9~\mathrm{ppm}$.

\section{Application of Surfactant Concentration Prediction Model}

In this final section, we evaluate the predictive capability of the surfactant-concentration model by applying it to bubbles subjected to a controlled, localised surfactant perturbation. Using the injection apparatus described in section \ref{method-inj}, a fixed volume of surfactant solution is dispensed onto a rising bubble stream at a prescribed moment after imaging begins. A $10~\mathrm{\mu L}$ pulse is released at $t=1.5~\mathrm{s}$, allowing a single or small number of bubbles to encounter the injected solution as it rises. The aim is to assess whether the derived model can first repeatedly identify the presence of surfactant-laden bubbles, and secondly determine the relative magnitude of contamination across different injected concentrations.


\begin{table}
  \begin{center}
    \def~{\hphantom{0}}
    \begin{tabular}{lc}
      Concentration by mass& Concentration (ppm)\\[3pt]
      0.02\%        & 6.2\\
      0.03\%        & 9.0\\
      0.04\%        & 11.7\\
      0.05\%        & 14.4\\
    \end{tabular}
    \caption{Surfactant concentrations used in injection experiments}
    \label{tab:Injected Concentrations}
  \end{center}
\end{table}

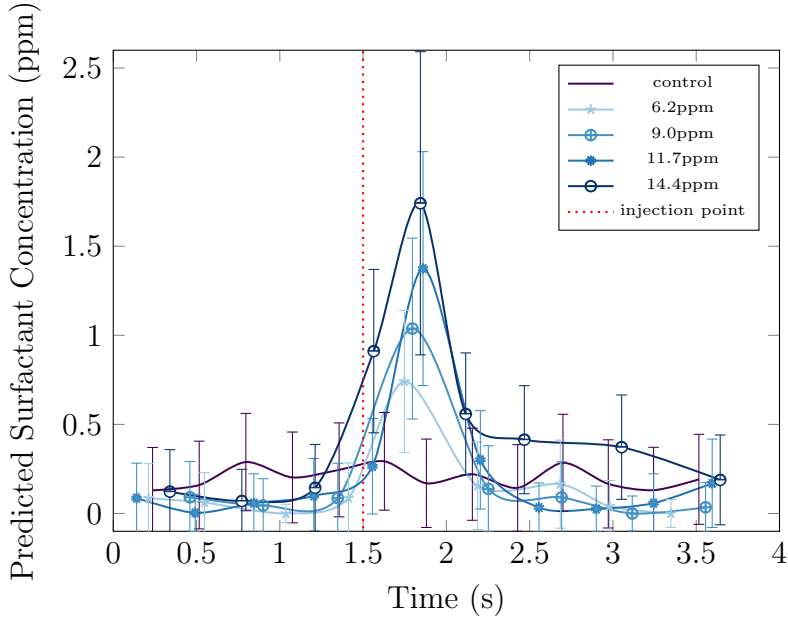
\begin{figure}
    \centering
    \begin{tikzpicture}
   \begin{axis}[width=0.65\textwidth, height=0.5\textwidth, xmin=0, xmax=4, ymin=-0.1, ymax=2.6, xlabel={Time (s)},ylabel={Predicted Surfactant Concentration (ppm)}, legend pos=north east, legend style={font=\tiny} ]

\addplot+[smooth, mark=plus, line width=0.75pt,solid,color=cb1] table[x=time(s), y=pred_mean(ppm)]{ArXiv/datfiles/Fig15_control_line_std_dev.dat}; 
\addlegendentry{control};
\addplot+[smooth, mark=star, line width=0.75pt, color=Blues-6-3] table[x=time(s), y=pred_mean(ppm)]{ArXiv/datfiles/Fig15_6.2ppm_line_std_dev.dat}; 
\addlegendentry{6.2ppm};
\addplot+[smooth, mark=oplus,line width=0.75pt, solid, color=Blues-7-5] table[x=time(s), y=pred_mean(ppm)]{ArXiv/datfiles/Fig15_9.0ppm_line_std_dev.dat}; 
\addlegendentry{9.0ppm};
 \addplot+[smooth, mark=10-pointed star,line width=0.75pt, solid, color=Blues-8-7] table[x=time(s), y=pred_mean(ppm)]{ArXiv/datfiles/Fig15_11.7ppm_line_std_dev.dat}; 
 \addlegendentry{11.7ppm};
\addplot+[smooth, mark=halfcircle, line width=0.75pt, solid,  color=Blues-9-9] table[x=time(s), y=pred_mean(ppm)]{ArXiv/datfiles/Fig15_14.4ppm_line_std_dev.dat}; 
\addlegendentry{14.4ppm};
\addplot[color=red,dotted,thick] coordinates {(1.5,-1) (1.5,3)};
\addlegendentry{injection point}

    \addplot [color=cb1, only marks,mark=plus, legend image post style={mark=[correct mark]}]
 plot [error bars/.cd, y dir = both, y explicit]
 table[x=time(s),y error index=2]{ArXiv/datfiles/Fig15_control_line_std_dev.dat};
    \addplot [color=Blues-6-3, only marks,mark=star,legend image post style={mark=[correct mark]}]
 plot [error bars/.cd, y dir = both, y explicit]
 table[x=time(s),y error index=2]{ArXiv/datfiles/Fig15_6.2ppm_line_std_dev.dat};
    \addplot [color=Blues-7-5, only marks,  mark=oplus, legend image post style={mark=[correct mark]}]
 plot [error bars/.cd, y dir = both, y explicit]
 table[x=time(s),y error index=2]{ArXiv/datfiles/Fig15_9.0ppm_line_std_dev.dat};
    \addplot [color=Blues-8-6, only marks, mark=10-pointed star, legend image post style={mark=[correct mark]}]
 plot [error bars/.cd, y dir = both, y explicit]
 table[x=time(s),y error index=2]{ArXiv/datfiles/Fig15_11.7ppm_line_std_dev.dat};
    \addplot [color=Blues-9-9, only marks, mark=halfcircle, legend image post style={mark=[correct mark]}]
 plot [error bars/.cd, y dir = both, y explicit]
 table[x=time(s),y error index=2]{ArXiv/datfiles/Fig15_14.4ppm_line_std_dev.dat};
 
   \end{axis}
    \end{tikzpicture}
    \caption{Predicted surfactant concentrations using derived $ARDI_{mean}$ relationship (eq. \ref{ARDI_mean}, with uncertainties depicted for different injected concentrations}
    \label{fig:Injection Predictions}
\end{figure}

Four surfactant concentrations, ranging from $0.02\%$ to $0.05\%$ by mass were injected into a stream of bubbles rising in clean tap-water environment. Their equivalent concentrations of these injected solutions in ppm are listed in table \ref{tab:Injected Concentrations}. It should be emphasised that these values correspond to the concentrations inside the syringe pump. After injection, the surfactant immediately mixes and dilutes within the tank. A control run with no injection was also performed. For each case, the $AR$ profiles for bubbles observed within the $4~\mathrm{s}$ imaging window were extracted and processed using the prediction model, following the same procedure detailed in section \ref{prediction model}. The resulting predicted concentrations are presented in figure~\ref{fig:Injection Predictions}.

 

The results demonstrate that the model reliably detects the onset of surfactant contamination. In the control case, the model consistently predicts low-level concentrations, with small variations reflecting the background presence of weak surface-active impurities in tap water. When surfactant is injected, a distinct and rapid increase in predicted concentration occurs immediately after $t=1.5~\mathrm{s}$. The magnitude of this spike increases with injected concentration, confirming the model's ability to distinguish between different levels of contamination. As previously discussed in figure \ref{fig:Act vs pred}, the prediction uncertainties amplifies with more heavily contaminated bubbles, which is reflected in the larger error bars for higher injected concentrations.

Following the injection, bubbles in concentrations $11.7~\mathrm{ppm}$ and below return to $AR$ values comparable to the control case within approximately $0.5~\mathrm{s}$ after the peak. However, for the highest injected concentration case of $14.4~\mathrm{ppm}$, elevated predictions persist for nearly $2~\mathrm{s}$, indicating that stronger injections leave measurable residual surfactant near the needle. This residual continues to be absorbed by subsequent bubbles, delaying the system's return to a clean state. 

A discrepancy naturally arises between the nominal injected concentration and the concentration predicted for the affected bubble. The immediate dilution of the injected solution prevents the bubble from experiencing the full syringe-pump concentration. Moreover, due to the stochastic nature of bubble pinch-off and detachment, the bubble's vertical position at the moment of injection is not perfectly controllable. This leads to some pulses being delivered after majority of the bubble of interest has already risen past the injection point. In such cases, the injection will be shared with the next bubble in the sequence, skewing the actual peak prediction magnitude for the tested injection concentration. Despite this variability, results in figure \ref{fig:Injection Predictions} demonstrate that in all trials, a bubble released within $0.5~\mathrm{s}$ of the injection event exhibits a clearly altered dynamic behaviour and can be consistently identified by the model as surfactant-laden.

Overall, these results indicate that the prediction model is well suited for detecting localised, transient surfactant perturbations and for differentiating between relative concentration levels, even in the presence of dilution and timing variability. Its limitations primarily arise from experimental constraints such as imperfect synchronisation and instantaneous mixing rather than from deficiencies in the predictive framework itself. 

\section{Conclusion}

The early-stage dynamics of bubbles rising in water containing varying concentrations of TX-100 surfactant (up to $5.8~\mathrm{ppm}$) were investigated experimentally to identify quantitative indicators capable of predicting interfacial contamination conditions.  High-speed imaging of the first $144~\mathrm{ms}$ of ascent (corresponding to a rise distance of approximately $40~\mathrm{mm}$) enabled the extraction of each bubble’s instantaneous $AR$  and $u$. 

The results reveal that $u$ extracted at this early stage exhibits inconclusive and non-monotonic behaviour with increasing surfactant concentration within the tested range, limiting its suitability as a reliable prediction parameter. In contrast, the temporal $AR$ profiles demonstrate a high sensitivity to even trace amounts of surfactant, responding rapidly to interfacial contamination. A clear damping effect on bubble shape oscillations is observable at concentrations as low as $0.3~\mathrm{ppm}$ and becomes increasingly significant with further increases in concentration up to $2.9~\mathrm{ppm}$, when it approaches saturation. This behaviour is evident from the progressive suppression of the ellipsoidal deformation phase during ascent and is most notably quantified by the reduction in peak aspect ratio, $AR_{peak}$. Further correlation analysis of the $AR$ profiles across concentrations identified that the maximum rate of change of $AR$ and standard deviation of $AR$ across bubbles recorded in the same condition are also highly sensitive metrics to surfactant concentration with computed Pearson correlation coefficient magnitudes above 0.8. 

Our study combines these strongly correlated parameters derived from temporal $AR$ measurements into a single quantitative indicator, termed $ARDI$, which characterises the bubble’s deformation dynamics. An empirical relationship between $ARDI$ and surfactant concentration is established, providing a direct and practical link between measurable shape characteristics and the surrounding contamination level. The predictive performance of the model was assessed using bubbles from known conditions, yielding an average error of $14.9\%$ across more than $170$ bubbles. A key limitation arises at concentrations exceeding the calibrated upper bound of $2.9~\mathrm{ppm}$, where prediction reliability deteriorates. The largely similar $AR$ profiles exhibited by heavily damped bubbles in concentrations above  $2.9 ~\mathrm{ppm}$ leads to potential inaccuracies, where predictions tend to converge toward the upper limit regardless of the true concentration. 

Finally, the predictive framework was applied to assess its capability in detecting and quantifying surfactant contamination from a stream of unknown bubbles. The model demonstrated strong functionality and repeatability, successfully identifying transient increases in surfactant concentration within $0.5~\mathrm{s}$ following the introduction of a controlled surfactant pulse. Beyond detection, the model consistently differentiated between varying contamination levels, predicting higher concentrations in response to more concentrated pulses. These results confirm the potential of the proposed approach as a rapid and sensitive tool for real-time monitoring of interfacial contamination in bubble-driven systems.

\bibliographystyle{apalike}
\bibliography{ArXiv/references}

@Article{Martin09,
  author = 	 {M. Martín and F.~J. Montes and M.~A. Galán},
  title = 	 {Mass transfer from oscillating bubbles in bubble column reactors},
  journal = 	 {Chem.~Eng.~J.},
  year = 	 {2009},
  volume = 	 {151},
  number =     {1--3},
  pages = 	 {79--88},
  month = 	 aug,
}

@Article{Becerril02,
  author = 	 {E.~L. Becerril and A. Cockx and A. Line},
  title = 	 {Effect of bubble deformation on stability and mixing in bubble columns},
  journal = 	 {Chem.~Eng.~Sci.},
  year = 	 {2002},
  volume = 	 {57},
  number =     {16},
  pages = 	 {3283--3297},
  month = 	 aug,
}

@Article{Mesa18,
  author = 	 {D. Mesa and P.~R. Parada},
  title = 	 {Scale-up in froth flotation: a state-of-the-art review},
  journal = 	 {Sep.~Purif.~Technol.},
  year = 	 {2018},
  volume = 	 {210},
  pages = 	 {950--962},
  month = 	 aug,
}

@Article{Kostoglou20,
  author =       {M. Kostoglou and T.~D. Karapantsios and O. Oikonomidou},
  title =        {A critical review on turbulent collision frequency/efficiency models in flotation: unravelling the path from general coagulation to flotation},
  journal =      {Adv.~Colloid Interface Sci.},
  year =         {2020},
  volume =       {279},
  month =        may,
}

@Article{Evans08,
  author =       {G.~M. Evans and E. Doroodchi and G.~L. Lane and P.~T.~L. Koh and M.~P. Schwarz},
  title =        {Mixing and gas dispersion in mineral flotation cells},
  journal =      {Chem.~Eng.~Res.~Des.},
  year =         {2008},
  volume =       {86},
  number =       {12},
  pages =        {1350--1362},
  month =        dec,
}

@Article{Montes99,
  author =       {F.~J. Montes and M.~A. Galan and R.~L. Cerro},
  title =        {Mass transfer from oscillating bubbles in bioreactors},
  journal =      {Chem.~Eng.~Sci.},
  year =         {1999},
  volume =       {54},
  number =       {15--16},
  pages =        {3127--3136},
  month =        jul,
}

@Article{Nalajala14,
  author =       {V.~S. Nalajala and N. Kishore},
  title =        {Drag of contaminated bubbles in power-law fluids},
  journal =      {Colloids Surf.~A Physicochem.~Eng.~Aspects},
  year =         {2014},
  volume =       {443},
  pages =        {240--248},
  month =        feb,
}

@Article{Zhang08,
  author =       {L. Zhang and C. Yang and Z.~S. Mao},
  title =        {Unsteady motion of a single bubble in highly viscous liquid and empirical correlation of drag coefficient},
  journal =      {Chem.~Eng.~Sci.},
  year =         {2008},
  volume =       {63},
  number =       {8},
  pages =        {2099--2106},
  month =        apr,
}

@Article{Yan17,
  author =       {X. Yan and Y. Jia and L. Wang and Y. Cao},
  title =        {Drag coefficient fluctuation prediction of a single bubble rising in water},
  journal =      {Chem.~Eng.~J.},
  year =         {2017},
  volume =       {316},
  pages =        {553--562},
}

@Article{Yan18,
  author =       {X. Yan and K. Zheng and Y. Jia and Z. Miao and L. Wang and Y. Cao and J. Liu},
  title =        {Drag coefficient prediction of a single bubble rising in liquids},
  journal =      {Ind.~Eng.~Chem.~Res.},
  year =         {2018},
  volume =       {57},
  number =       {15},
  pages =        {5385--5393},
}

@Article{Zheng19,
  author =       {K. Zheng and C. Li and X. Yan and H. Zhang and L. Wang},
  title =        {Prediction of bubble terminal velocity in surfactant aqueous solutions},
  journal =      {Can.~J.~Chem.~Eng.},
  year =         {2019},
  volume =       {98},
  number =       {2},
  pages =        {607--615},
  month =        jul,
}

@Article{Yan21,
  author =       {X. Yan and K. Zheng and W. Su and L. Wang and H. Zhang and Y. Cao and C. Guo},
  title =        {Predictions of terminal rising velocity, shape and drag coefficient for particle-laden bubbles},
  journal =      {Minerals Eng.},
  year =         {2021},
  volume =       {173},
  month =        nov,
}

@Article{Peebles53,
  author =       {F.~N. Peebles and H.~J. Garber},
  title =        {Studies on the motion of gas bubbles in liquids},
  journal =      {Chem.~Eng.~Prog.},
  year =         {1953},
  volume =       {49},
  pages =        {88--97},
}

@Article{Tomiyama02,
  author =       {A. Tomiyama and G.~P. Celata and S. Hosokawa and S. Yoshida},
  title =        {Terminal velocity of single bubbles in surface tension force dominant regime},
  journal =      {Int.~J.~Multiphase Flow},
  year =         {2002},
  volume =       {28},
  pages =        {1497--1519},
}

@Article{Kracht10,
  author =       {W. Kracht and J.~A. Finch},
  title =        {Effect of frother on initial bubble shape and velocity},
  journal =      {Int.~J.~Mineral Processing},
  year =         {2010},
  volume =       {94},
  number =       {3--4},
  pages =        {115--120},
  month =        apr,
}

@Article{Wang19,
  author =       {P. Wang and J.~J. Cilliers and S.~J. Neethling and P.~R. Parada},
  title =        {The behaviour of rising bubbles covered by particles},
  journal =      {Chem.~Eng.~J.},
  year =         {2019},
  volume =       {365},
  pages =        {111--120},
  month =        feb,
}

@Article{Chabel12,
  author =       {N.~A. Chabel and J. Vejrazka and O. Masbernat and F. Risso},
  title =        {Shape oscillations of an oil drop rising in water: effect of surface contamination},
  journal =      {J.~Fluid Mech.},
  year =         {2012},
  volume =       {702},
  pages =        {533--542},
  month =        apr,
}

@Article{Wang19Langmuir,
  author =       {P. Wang and J.~J. Cilliers and S.~J. Neethling and P.~R. Parada},
  title =        {Effect of Particle Size on the Rising Behavior of Particle-Laden Bubbles},
  journal =      {Langmuir},
  year =         {2019},
  volume =       {35},
  number =       {10},
  pages =        {3680--3687},
}

@Article{Eskanlou18,
  author =       {A. Eskanlou and M. Khalesi and M. Mirmogaddam and M. Chegeni and B. Hassas},
  title =        {Investigation of trajectory and rise velocity of loaded and bare single bubbles in flotation process using video processing technique},
  journal =      {Sep.~Sci.~Technol.},
  year =         {2018},
  volume =       {54},
  number =       {11},
  pages =        {1795--1802},
  month =        oct,
}

@Article{Ata08,
  author =       {S. Ata},
  title =        {Coalescence of bubbles covered by particles},
  journal =      {Langmuir},
  year =         {2008},
  volume =       {24},
  number =       {12},
  month =        may,
}

@Article{Wang20Langmuir,
  author =       {H. Wang and P.~R. Parada},
  title =        {Coalescence dynamics of particle-laden bubbles},
  journal =      {Langmuir},
  year =         {2020},
  volume =       {36},
  number =       {19},
  month =        apr,
}

@Book{Clift78,
  author =       {R. Clift and J.~R. Grace and M.~E. Weber},
  title =        {Bubbles, drops and particles},
  publisher =    {Academic Press},
  year =         {1978},
  address =      {New York},
}

@Article{Mougin01,
  author =       {G. Mougin and J. Magnaudet},
  title =        {Path Instability of a Rising Bubble},
  journal =      {Phys.~Rev.~Lett.},
  year =         {2001},
  volume =       {88},
  month =        dec,
}

@Article{Mougin06,
  author =       {G. Mougin and J. Magnaudet},
  title =        {Wake-induced forces and torques on a zigzagging/ spiralling bubble},
  journal =      {J.~Fluid Mech.},
  year =         {2006},
  volume =       {567},
  pages =        {185--194},
  month =        jul,
}

@Article{Zhang01,
  author =       {Y. Zhang and J.~A. Finch},
  title =        {A note on single bubble motion in surfactant solutions},
  journal =      {J.~Fluid Mech.},
  year =         {2001},
  volume =       {429},
  pages =        {63-66},
}

@Article{Tagawa14,
  author =       {Y. Tagawa and S. Takagi and Y. Matsumoto},
  title =        {Surfactant effect on path instability of a rising bubble},
  journal =      {J.~Fluid Mech.},
  year =         {2014},
  volume =       {738},
  pages =        {124-142},
}

@Article{Lehrer76,
  author =       {I.~H. Lehrer},
  title =        {A rational terminal velocity equation for bubbles and drops at intermediate and high reynolds numbers},
  journal =      {J.~Chem Eng.of Japan},
  year =         {1976},
  volume =       {9},
  pages =        {237-240},
}

@Article{Feng24,
  author =       {Y. Feng and L. Sun and Z. Mo and M. Du and C. Zhu and W. Yang and X. Xu},
  title =        {An evaluation of predictive correlations for the terminal rising velocity of a single bubble in quiescent clean liquid},
  journal =      {Int.~J.~Multiphase Flow},
  year =         {2024},
  volume =       {173},
}

@Article{Lindt72,
  author =       {J.~T. Lindt},
  title =        {On the periodic nature of the drag on a rising bubble},
  journal =      {Chem.~Eng.~Sci.},
  year =         {1972},
  volume =       {27},
  pages =        {1775-1781},
}

@Article{Li15,
  author =       {Z. Li and X. Song and S. Jiang and J. Yu and M. Ishii},
  title =        {Experimental investigation on motion characteristics of relative large bubble},
  journal =      {Nucl. Power Eng.},
  year =         {2015},
  volume =       {36},
  pages =        {161-164},
}

@Article{Hlawitschka22,
  author =       {M.~W. Hlawitschka and P. Kovats and B. Donmez and K. Zahringer and H.~J. Bart},
  title =        {Bubble motion and reaction in different viscous liquids},
  journal =      {Exp.~Comp.~Multiphase Flow},
  year =         {2022},
  volume =       {4},
  pages =        {26-38},
}

@Article{Fan21,
  author =       {Y. Fan and J. Fang and I. Bolotnov},
  title =        {Complex bubble deformation and break-up dynamics studies using interface capturing approach},
  journal =      {Exp.~Comp.~Multiphase Flow},
  year =         {2021},
  volume =       {3},
  pages =        {139-151},
}

@Article{Cano-Lozano16,
  author =       {J.~C, Cano-Lozano and C. Martinez-Bazan},
  title =        {Paths and wakes of deformable nearly spheroidal rising bubbles close to the transition to path instability},
  journal =      {Phys.~Rev.~Fluids},
  year =         {2016},
  volume =       {1},
  pages =        {053604},
  numpages =     {30},

}

\end{document}